%% file: coleman.tex
\documentclass[
  ,final            
  ,numberedheadings 
  ]{aipprocx}

\usepackage{epsfig}
\usepackage{epsf}
\layoutstyle{6x9}
\newcommand \beq  {\begin{equation}}
\newcommand \eeq  {\end{equation}}
\newcommand \bea {\begin{eqnarray} }
\newcommand \eea {\end{eqnarray}}
\renewcommand \hbar{{\mathchar'26\mkern-9muh}}
\newlength{\spear}

\newlength{\fight}
\newcommand{\fg}[3]
{\begin{figure}[tb]
\resizebox{\fight}{!} {\includegraphics{#1}}
\caption{#2}\label{#3}\end{figure}}
\newcommand{\fgb}[3]
{\begin{figure}[b]
\resizebox{\fight}{!} {\includegraphics{#1}}
\caption{#2}\label{#3}\end{figure}}

\newlength{\bxwidth}\bxwidth=1.5 truein
\newcommand\frm[1]{\epsfig{file=#1,width=\bxwidth}}
\newcommand \boxit[1]{\noindent\marginpar{\mbox{}}
\fbox{\rule{0pt}{2 \baselineskip}\parbox{0.98\textwidth}{#1}}}
\newenvironment{Quote}%
  {\begin{list}{}{%
	\setlength{\leftmargin}{1truein}
       \setlength{\rightmargin}{0.5\leftmargin}}
           \item[]\ignorespaces}
  {\end{list}}
\newenvironment{problems}{\begin{enumerate}\small}
{\end{enumerate}}
\newlength{\wx}
\newlength{\wy}
\newlength{\wz}
\def\3he{{$^3${\rm He}}}

\def\etal{{\it et al.}}

\def\slD{\raise.15ex\hbox{$/$}\kern-.57em\hbox{$D$}}
\def\dsl{\raise.15ex\hbox{$/$}\kern-.57em\hbox{$\Delta$}}
\def\slp{{\raise.15ex\hbox{$/$}\kern-.57em\hbox{$\partial$}}}
\def\nsl{\raise.15ex\hbox{$/$}\kern-.57em\hbox{$\nabla$}}
\def\sla{\raise.15ex\hbox{$/$}\kern-.57em\hbox{$\rightarrow$}}
\def\slla{\raise.15ex\hbox{$/$}\kern-.57em\hbox{$\lambda$}}
\def\gtwid{\raise.3ex\hbox{$>$\kern-.75em\lower1ex\hbox{$\sim$}}}
\def\ltwid{\raise.3ex\hbox{$<$\kern-.75em\lower1ex\hbox{$\sim$}}}

\def\12{{1\over2}}

\def\part{\partial}
\def\la{\lambda}

\def\bethlogo{\vbox{\bf \line{\hrulefill} 
    \kern-.5\baselineskip 
    \line{\hrulefill\phantom{ ELIZABETH A. MASON }\hrulefill} 
    \kern-.5\baselineskip 
    \line{\hrulefill\hbox{ ELIZABETH A. MASON }\hrulefill} 
    \kern-.5\baselineskip 
    \line{\hrulefill\phantom{ 1411 Chino Street }\hrulefill} 
    \kern-.5\baselineskip 
    \line{\hrulefill\hbox{ 1411 Chino Street }\hrulefill} 
    \kern-.5\baselineskip 
    \line{\hrulefill\phantom{ Santa Barbara, CA 93101 }\hrulefill} 
    \kern-.5\baselineskip 
    \line{\hrulefill\hbox{ Santa Barbara, CA 93101 }\hrulefill}
    \kern-.5\baselineskip 
    \line{\hrulefill\phantom{ (805) 962-2739 }\hrulefill} 
    \kern-.5\baselineskip 
    \line{\hrulefill\hbox{ (805) 962-2739 }\hrulefill}}}
\def\lisalogo{\vbox{\bf \line{\hrulefill} 
    \kern-.5\baselineskip 
    \line{\hrulefill\phantom{ LISA R. GOODFRIEND }\hrulefill} 
    \kern-.5\baselineskip 
    \line{\hrulefill\hbox{ LISA R. GOODFRIEND }\hrulefill} 
    \kern-.5\baselineskip 
    \line{\hrulefill\phantom{ 6646 Pasado }\hrulefill} 
    \kern-.5\baselineskip 
    \line{\hrulefill\hbox{ 6646 Pasado }\hrulefill} 
    \kern-.5\baselineskip 
    \line{\hrulefill\phantom{ Santa Barbara, CA 93108 }\hrulefill} 
    \kern-.5\baselineskip 
    \line{\hrulefill\hbox{ Santa Barbara, CA 93108 }\hrulefill}
    \kern-.5\baselineskip 
    \line{\hrulefill\phantom{ (805) 962-2739 }\hrulefill} 
    \kern-.5\baselineskip 
    \line{\hrulefill\hbox{ (805) 962-2739 }\hrulefill}}}

\def\la{{\lambda}}

\def\low#1{\lower.5ex\hbox{${}_#1$}}
\def\ltwid{\raise.3ex\hbox{$<$\kern-.75em\lower1ex\hbox{$\sim$}}}

\def\om{{\omega}}

\def\psl{\raise.15ex\hbox{$/$}\kern-.57em\hbox{$\partial$}}
\def\partt{\raise.15ex\hbox{$\widetilde$}{\kern-.37em\hbox{$\partial$}}}
\def\parts{\raise.15ex\hbox{$/$}{\kern-.6em\hbox{$\partial$}}}
\def\nablas{\raise.15ex\hbox{$/$}{\kern-.6em\hbox{$\nabla$}}}
\def\oprod{\hbox{$\rm O$}{\kern -0.8em\hbox{$\Pi$}}}
\def\partw#1{\raise.15ex\hbox{$/$}{\kern-.6em\hbox{${#1}$}}}

\def\si{{\sigma}}

\def\gtappr{{{\lower4pt\hbox{$>$} } \atop \widetilde{ \ \ \ }}}
\def\ltappr{{{\lower4pt\hbox{$<$} } \atop \widetilde{ \ \ \ }}}

\def\topppageno1{\global\footline={\hfil}\global\headline
={\ifnum\pageno<\firstpageno{\hfil}\else{\hss\twelverm --\ \folio
\ --\hss}\fi}}

\def\toppageno2{\global\footline={\hfil}\global\headline
={\ifnum\pageno<\firstpageno{\hfil}\else{\rightline{\hfill\hfill
\twelverm \ \folio
\ \hss}}\fi}}

\def\ltdash{\raise-1.8pt\hbox{$\scriptscriptstyle |$}}

\def \ra{\rangle}
\def\la{\langle}

\def\dg{{^
{\dag}}}

\def\ra{\rangle}
\def\la{\langle}

\def\1{{\bf 1}}
\def\2{{\bf 2}}

\def\rarrow{\rightarrow}

\def\vk{\vec k}

\def\ell{{\it l } {\rm n}}

\def\si{\sigma}

\def\cx2{\sqrt{c^2_x+c^2_y}}

\def\gkk{\gamma _{\vec k}}
\def\gk2{\gkk ^2}
\def\dw{\downarrow}
\def\up{\uparrow}
\def\gtappr{{{\lower4pt\hbox{$>$} } \atop \widetilde{ \ \ \ }}}
\def\ltappr{{{\lower4pt\hbox{$<$} } \atop \widetilde{ \ \ \ }}}

\def\dsp{\displaystyle}
\def\pbar{{\partial\kern-1.2ex\raise0.25ex\hbox{/}}}

\def\up{\uparrow}
\def\dw{\downarrow}

\def\dsp{\displaystyle}

\def\dg{{^{\dag}}}

\def\ra{\rangle}
\def\la{\langle}

\def\1{{\bf 1}}
\def\2{{\bf 2}}

\def\rarrow{\rightarrow}

\def\vk{\vec k}

\def\ell{{\it l } {\rm n}}

\def\si{\sigma}

\def\cx2{\sqrt{c^2_x+c^2_y}}

\def\gkk{\gamma _{\vec k}}
\def\gk2{\gkk ^2}
\def\gtappr{{{\lower4pt\hbox{$>$} } \atop \widetilde{ \ \ \ }}}
\def\ltappr{{{\lower4pt\hbox{$<$} } \atop \widetilde{ \ \ \ }}}

\def\thickra{\hbox{\raise0.2pt\hbox{{$\bf >\mkern-13mu>\mkern-13mu>$}}}}
\def\thickrarrow{\hbox{\raise0.28pt\hbox{{$\bf >\mkern-13mu>\mkern-13mu>$}}}}

\begin{document}
\newlength{\upit}\upit=0.1truein
\setlength{\unitlength}{1mm}
\newcommand{\nmat}[4]{\left[{
\hbox{${
{{\dsp \raisebox{4pt}{$#1$}
 \above1pt \dsp  \raisebox{-4pt}{$#3$} }}}$}
\left| 
{\bf
{{\dsp \raisebox{0pt}{$#2$}
 \above1pt \dsp\raisebox{-0pt}{$#4$} }}}
\right.
} \right]}

\title{Local moment physics in heavy electron systems}

\author{P. Coleman}{
  address={Center for Materials Theory, Department of Physics and Astronomy,
Rutgers University, 136 Frelinghausen Road, Piscataway, NJ 08855, USA
}
}


\begin{abstract}
 This set of lectures describes the physics of moment formation, 
  the basic physics of the Kondo effect and the development of 
a coherent heavy electron fluid in the dense Kondo lattice. 
The last lecture discusses the open problem of quantum criticality
in heavy electron systems. 
\end{abstract}

\maketitle
\tableofcontents

\section{Local moment Formation}
\subsection{Introduction}

The last two decades have seen a growth of interest 
in ``strongly
correlated electron systems'': materials where the
electron interaction energies dominate  the electron kinetic
energies, becoming so large that  they qualitatively transform 
the physics of the medium. \cite{physicsworld}

Examples
of strongly correlated systems include
\begin{itemize}
\item Cuprate superconductors, \cite{hitcbook}
where interactions amongst
electrons in localized 3d-shells form 
an antiferromagnetic  Mott insulator, which 
develops high temperature superconductivity
when doped. 

\item Heavy electron compounds, where 
localized magnetic moments formed by rare earth or actinide
ions transform the metal in which they are immersed, generating 
quasiparticles with masses in excess of 1000 bare electron masses.\cite{hewson}

\item Fractional Quantum Hall systems, where the 
interactions between electrons in the lowest Landau level
of a semi-conductor heterojunction 
generate a new electron fluid, described by the
Laughlin ground-state, with quantized fractional Hall constant
and quasiparticles with fractional charge and statistics. \cite{fqhe}

\item ``Quantum Dots'', which are 
tiny pools of electrons in semiconductors that act as artificial atoms. 
As the gate voltage is changed, the
Coulomb repulsion between electrons in the dot leads to the so-called
``Coulomb Blockade'', whereby electrons can be added
one by one to the quantum dot. \cite{qdots}

\end{itemize}

Strongly interacting materials
develop ``emergent'' properties: properties which 
require a new language\cite{physicsworld} and new
intellectual building blocks for their understanding.  
This chapter will illustrate and discuss one area of
strongly correlated electron physics in which localized magnetic
moments form the basic driving force of strong correlation. 
When electrons localize, they can form
objects whose low energy excitations involve spin degrees of moment. 
In the simplest case, such 
``localized magnetic moments'' are represented by a single,
neutral spin  operator
\[
\vec{ S}= \frac{\hbar }{2}\vec{\si}
\]
where $\vec{\sigma }$ denotes the Pauli matrices of the localized
electron.   Localized moments develop within highly localized atomic
wavefunctions. The most severely localized  wavefunctions in nature
occur inside the partially filled $4f$ shell of rare earth compounds
(Fig.~\ref{fig01})
such
as cerium ($Ce$) or Ytterbium ($Yb$).
Local moment formation also occurs in the localized $5f$ levels of
actinide atoms as uranium and the slightly more delocalized 
$3d$ levels of first row transition metals(Fig.~\ref{fig01}).
\fight=5truein
\fg{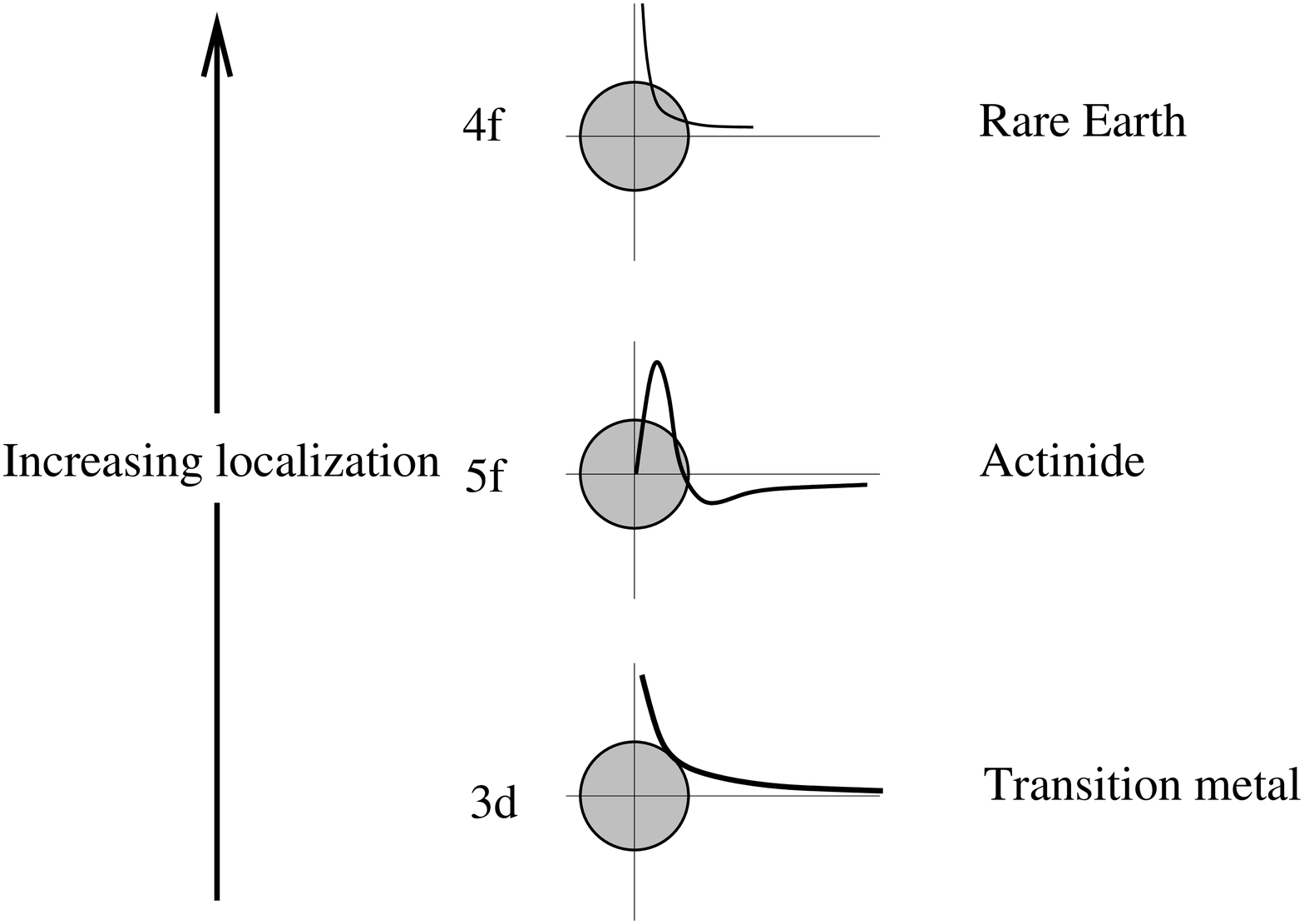}{Depicting localized 
$4f$, $5f$ and $3d$ atomic wavefunctions. 
}{fig01}
Localized moments are the origin of magnetism in
insulators, and in metals their interaction with the mobile charge carriers
profoundly changes the nature of the metallic state via a mechanism
known as the ``Kondo effect''.

In the past decade, the physics of local moment formation
has also reappeared in connection with quantum dots, where it gives rise
to the Coulomb blockade phenomenon and the non-equilibrium Kondo
effect. 

\subsection{Anderson's Model of Local Moment Formation} Though the concept 
of localized moments was employed 
in the earliest applications
of quantum theory to condensed matter\footnote{
Landau and N{\'e}el
invoked the notion of the localized moment in their 1932 papers on
antiferromagnetism, and in 1933, Kramers used this idea again in his
theory of magnetic superexchange.}, a theoretical understanding of the
{\sl mechanism} of moment formation did not develop until the early
sixties, when experimentalists began to systematically study 
impurities in metals. \footnote{
It was not until the sixties that materials physicist could
control the concentration of magnetic impurities in the parts per million
range required for the study of individual impurities.  Such control of
purity evolved during the  1950s, with the development of new techniques
needed for semiconductor physics, such as zone refining. }

In the early 1960s, Clogston,  Mathias and collaborators\cite{earlymagrefs}
showed that when small concentrations $n_{i}$ of 
magnetic ions, such as iron are added to a 
metallic host, they develop 
a Curie component to the magnetic susceptibilty
\begin{equation}\label{curie}
\chi =n_{i}{  M^2\over 3T } \qquad \qquad M^2 = g_J^2\mu _{B}^{2}J(J+1), 
\end{equation}
indicating the formation of a local moment. 
However, the local moment does not always develop, depending 
on the metallic host in which the magnetic ion was embedded.   For
example, iron  dissolved at $1\%$ concentration in pure $Nb$ does not
develop a local moment, but in the alloy $Nb_{1-x}Mo_{x}$ a local
moment
develops for $x>0.4$, rising to $2.2 \mu_{B}$ above $x=0.9$.
What is the underlying physics behind this phenomenon?

Anderson\cite{anda}  was the first to 
identify interactions between localized
electrons as the driving force for local moment formation. Earlier
work by Friedel\cite{Friedel} and Blandin\cite{blandin} had
already identified part of the essential physics of local moments
with the development of resonant bound-states.  Anderson now included
interactions to this picture. 
Much of
the basic physics can be understood by considering an isolated 
atom with a localized $S=1/2$ atomic state which we shall refer to as
a localized ``d-state''. 
In isolation, the atomic bound state is stable and can 
be modeled in terms of a single level
of energy $E_d$ and a Coulomb interaction 
\begin{equation}\label{}
U= \frac{1}{2}\int d^{3}x
d^{3}x' V (\vec{x}- \vec{ x}')\vert \psi (\vec{x})\vert ^{2}
\vert \psi (\vec{x}')\vert ^{2},
\end{equation}
where 
$V (\vec{x}- \vec{x}')={e^{2}}/{4 \pi \epsilon_{0} \vert
\vec{x}-\vec{ x'}\vert } $ is the Coulomb
potential. 

The  Anderson model for a localized impurity atom is given by 
\\

\boxit{
\begin{equation}\label{}
H=H_{c}+H_{mix}+\overbrace {H_{d}+ H_{U}}^{H_{atomic}}
\end{equation}}
\\

where $H_d= E_d\sum_{\sigma}\hat n_{d\sigma }$
describes an isolated atomic d-state of energy $E_{d}$
and
occupancy $n_{d\sigma}$  in the 
``up'' and ``down'' state. 
$H_{U} = U\hat n_{d\up}\hat n_{d\dw}$
is the inter-atomic interaction
between the up and down d-electrons.  
The term 
\[
H_{c}= \sum_{\vec{ k}\si}\epsilon_{\vec{k}}
c\dg _{\vk  \sigma  }c _{\vk  \sigma  }
\]
describes the dispersion of electrons in the conduction sea which
surrounds the ion, 
where $c\dg _{\vk  \sigma  }
$ creates an electron of momentum $\vec{k}$, spin $\sigma $ and
energy $\epsilon_{\vec{k}}$. When the ion is embedded within a metal,
the energy of the d-state is degenerate with band-electron states,
and the term 
\[
H_{mix}= 
 \sum_{j \si}
\bigl[V_{\vec k}c\dg_{k  \si}d_{\si} + {\rm H. c.}\bigr]  
\]
describes the hybridization that then takes place 
with the conduction electron sea, where
$d\dg _{\si}$ describes the creation of a d-electron. 
The  matrix  element of the ionic potential
between a plane wave conduction state and the d-orbital is
\begin{equation}\label{}
V_{\vec k}
= 
\la \vec k \si \vert\hat V \vert d^1 \si'\ra =
\int d^{3}r e^{-i\vec{k}\cdot \vec{r}}V_{ion} (r) \psi (\vec{r})
\delta_{\si \si'}. 
\end{equation}
\noindent where  $\psi (\vec{r})$ is the wavefunction of 
the localized orbital and $V_{ion} (r)$ is the ionic potential.This matrix element will have the same symmetry as the localized
orbital-
a matter of some importance for real d-states, or f-states
\footnote{A direct calculation shows that
\begin{equation}\label{}
V(k) = 4 \pi i^{-l}\int r^2 dr j_l(kr) V(r) R_{\Gamma}(r)        \qquad\qquad(l=2)
\end{equation}
is the overlap of the radial wavefunctions $R_{\Gamma} (r)$
of the d-state and the
$l=2$ partial wave state of the conduction electron, with the ionic potential.}. However, for the discussion that 
follows, 
the detailed $\vec{k}$ dependence of
this object can essentially
be ignored 
.

Let us first focus on the atomic part of H, 
\[\label{}
H_{atomic}=H_{d}+ H_{U}= E_d\sum_{\sigma}\hat n_{d\sigma }
+U n_{d\uparrow}n_{d\downarrow } .
\]
The four states of this
ion are 
\begin{eqnarray}\label{}
\begin{array}{rl}
\vert d^2 \ra \qquad\qquad & E (d^{2}) = 2E_d+U\cr
\vert d^0 \ra \qquad\qquad &E (d^{0}) =0\cr
\vert d^1 \up\ra \quad \vert d^1\dw\ra \qquad\qquad & E (d^{1}) =E_d.
\end{array}
\end{eqnarray}
To obtain a magnetic doublet as the ground-state, the 
excitation energies out of the doublet state must be greater than
zero, i.e 
\begin{eqnarray}\label{}
E (d^{2})-E (d^{1})=E_{d}+U>0&\Rightarrow& E_{d}+U/2 > - U/2\cr
E (d^{0})-E (d^{1})=-E_{d}>0&\Rightarrow& U/2 > E_{d}+U/2
\end{eqnarray}
so that for 
\[
U/2 >  \vert E_d+U/2\vert,
\]
the isolated 
ion has a doubly degenerate magnetic ground-state, as illustrated 
in \ref{fig1}. We see that provided the Coulomb interaction $U$
is large enough compared with the level spacing, the ground-state
of the ion becomes magnetic. The d-excitation spectrum of the 
ion will involve two sharp levels, one at energy $E_{d}$, the other at
energy $E_{d}+U$. 
\fight=3truein
\fg{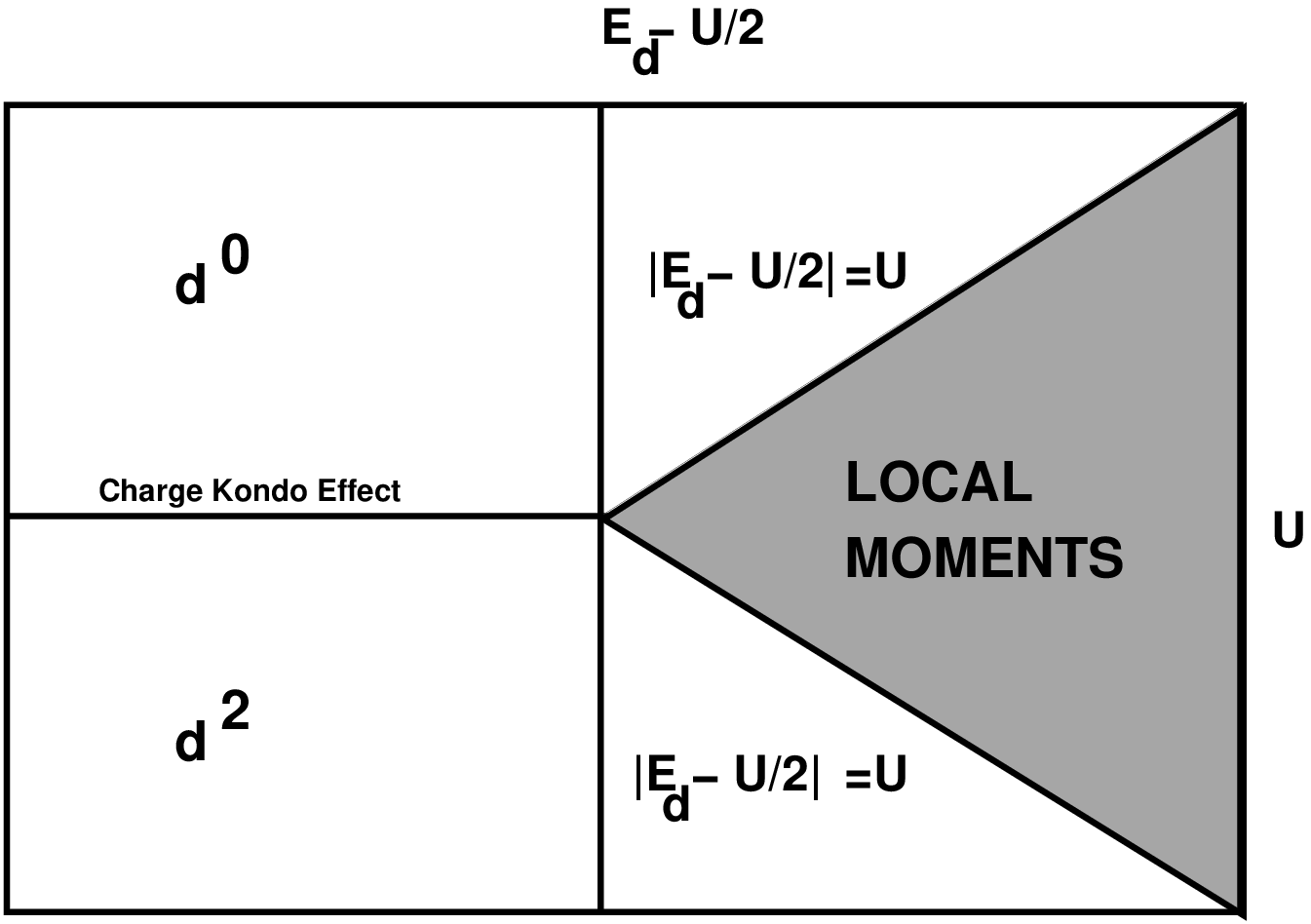}{Phase diagram for Anderson Impurity Model in the
Atomic Limit.
}{fig1}
\fight=3truein

Suppose this ion is embedded in a metal: the free electron continuum
is then pulled downwards by the  work function of the metal so that 
now the  d-level  energy is degenerate with conduction electron energy
levels.  In this situation 
we expect the d-level to hybridize with the conduction electron
states, broadening the  sharp 
d-level into a resonance with a width 
$\Delta= \Delta (E_{d})$, where $\Delta (\epsilon )$ is given by 
Fermi's Golden Rule.  
\begin{equation}\label{golden}
\Delta(\epsilon)=
\pi \sum_{\vec k } \vert V(\vk ) \vert^2 \delta( \epsilon_{\vk} 
- \epsilon )
= \pi \overline{N(\epsilon)V^{2}(\epsilon)}
\end{equation}
where $N(\epsilon) = \sum_{\vk} \delta(\epsilon_{\vk} - \epsilon)$
is the electron density of states (per spin). In the discussion that
follows, let us assume that over the energy width of the resonance,
$V(\epsilon)$ and $N(\epsilon)\sim N(0)$ are essentially constant.

When this hybridization is small compared with $U$, we expect the
ground-states of the ion to be essentially that of the atomic limit.
For weak interaction strength $U$
the hybridization with the
conduction sea will produce a single d-resonance of width $\Delta $
centered around $E_{d}$. In Anderson's model for moment formation, 
when $U>U_{c}\sim \pi\Delta $ the single resonance splits into two, so
that for large $U\gg U_{c}$, there are two d-resonances centered
around
$E_{d}$ and $E_{d}+U$, 
as shown in Fig. \ref{fig2}. 
\fight=\textwidth
\fg{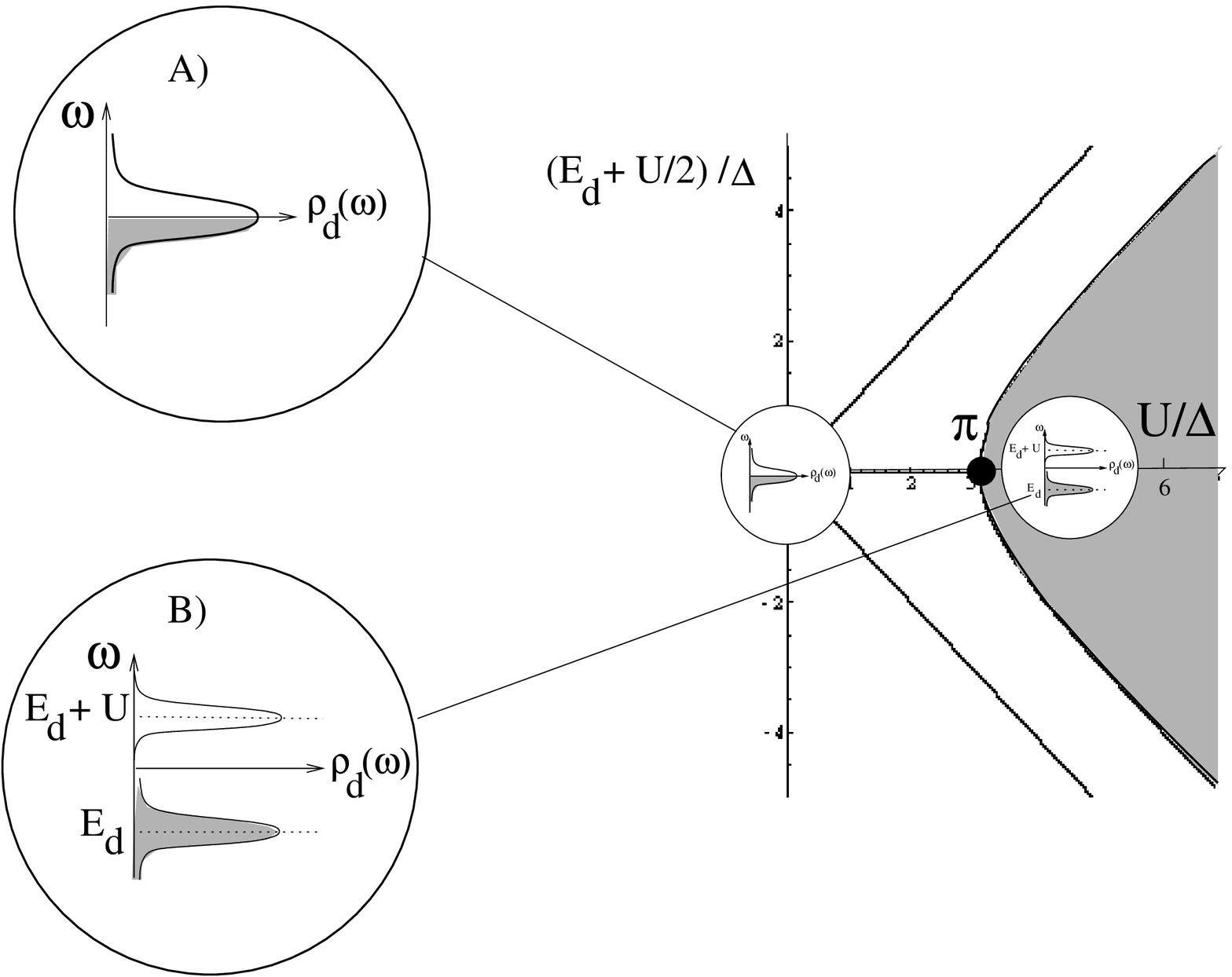}{Illustrating how the d-electron resonance splits to
form a local moment. A) $U<\pi \Delta $, single half-filled
resonance. B) $U>\pi \Delta $, up and down components of the resonance
are split by an energy $U$. }{fig2}
\fight=3truein
To illustrate the calculations that lead to this conclusion,  let us use
a Feynman diagram approach.  We shall treat $H_{I}= H_{mix }+ H_{U}$ as a
perturbation to the non-interacting
part of the Hamiltonian to be
$H= H_{b}+H_{d}$. The Green's functions of the bare d-electron and
conduction electron are  then denoted by
\bxwidth=1.5truein
\upit=-0.05truein
\newcommand{\raiser}[1]{\raisebox{\upit}[0cm][0cm]{#1}}
\begin{eqnarray}\label{}
\raiser{\frm{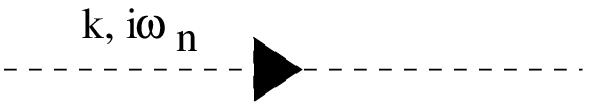}}&\qquad&
G^{0} (k,i\omega _{n})= [i\omega _{n}-\epsilon
_{k}]^{-1}\cr&\quad &\cr
\raiser{\frm{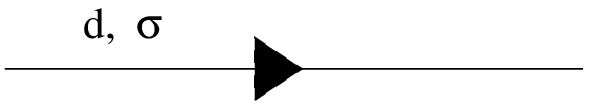}}&\qquad& G^{0}_{d}=[ (i\omega _{n})-E_{d}]^{-1}\nonumber
\end{eqnarray}
whilst the Feynman diagrams for the hybridization and the interaction
terms are then
\bxwidth=0.7truein
\upit=-0.05truein

\begin{eqnarray}\label{}
\raiser{\frm{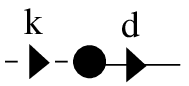}}&\qquad&
V^{*} (k)
\cr&\qquad&\cr
\raiser{\frm{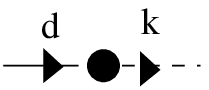}}&\qquad&
V (k)
\cr&\qquad&
\cr
\upit=-0.15truein
\raiser{\frm{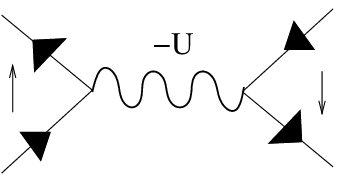}}&\qquad& -U
\end{eqnarray}
Quite generally, the propagator for the d-electrons can be written
\begin{eqnarray}\label{}
G_{d\sigma } (\omega )= [\omega -E_{d}-\Sigma_{d\sigma
}
(\omega )]^{-1}
\end{eqnarray}
where $\Sigma_{d\sigma } (\omega)$ is the the self-energy of the
d-electron
with spin $\sigma $. We delineate between ``up'' and ``down'', anticipating
Anderson's broken symmetry description of a local moment as a
resonance immersed in a self-consistently determined 
Weiss field.
The density of states associated with the d-resonance
is determined by  the imaginary part of the d-Green function:
\begin{equation}\label{dos}
\rho _{d\sigma } (\omega )= \frac{1}{\pi }\hbox{Im}G_{d\sigma } (\omega -i\delta ).
\end{equation}
The Anderson model for local moment formation is equivalent to the Hartree
approximation to the d-electron self-energy, denoted by
\bxwidth=3 truein
\upit=-0.4truein
\begin{eqnarray*}
&\raiser{\frm{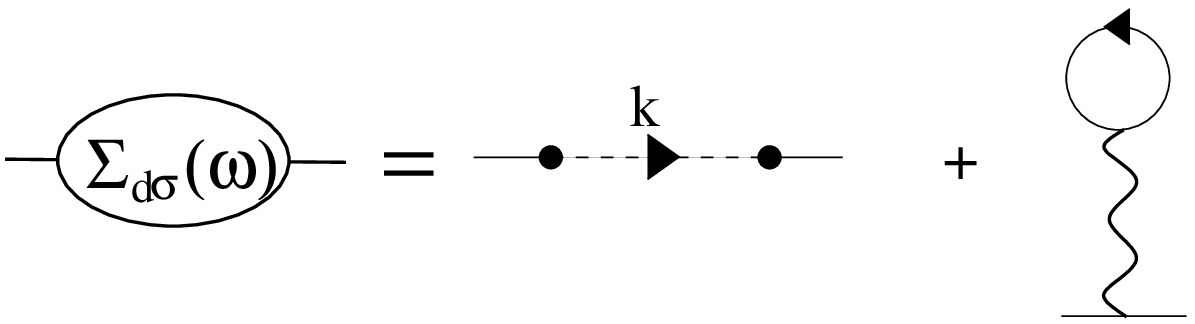}}\cr
&\qquad \cr
&\qquad \cr
&\qquad \cr
&\qquad \quad \Sigma _{d\sigma } (\omega )\qquad =\  \ \qquad   \Sigma _{1} (\omega )+\quad
\qquad U n_{d-\sigma }
\end{eqnarray*}
The first term in this expression derives from the hybridization of the
d-electrons with the conduction sea.  Notice that the d-state
fluctuates
into \underline{all} k-states of the conduction sea, so that there
is a sum over $k$ inside $\Sigma_{1} (\omega )$.
The second term is the Hartree
approximation to the interaction self energy. We can identify 
the fermion loop here as the occupancy of the $-\sigma $ d-state, so that 
\begin{equation}\label{}
G_{d\sigma } (\omega )= [\omega - ( E_{d}+ U n_{d -\sigma })-
\Sigma_{1}(\omega )]^{-1}
\end{equation}
so the Hartree approximation is equivalent to replacing
$E_{d}\rightarrow E_{d\sigma } = E_{d} + U n_{d, - \sigma }$. The
hybridization part of the self energy  is
\[
\Sigma_{1} (\omega+ i \delta  )= \sum_{\vec{k} }\frac{\vert V (\vec{ k})\vert ^{2}
}{\omega- \epsilon_{\vec{k}}-i\delta }
\]
Notice that
since $1/ (x\mp i\delta ) = P (1/x)\pm i \pi \delta (x)$, it follows that
$Im\Sigma_{d} (\omega\pm i \delta )= \mp \Delta (\omega)$, so the imaginary
part of this quantity has a discontinuity along the real axis
equal to the hybridization width. Using 
(\ref{golden}), you can verify that we can now rewrite this 
as 
\begin{eqnarray}\label{}
\Sigma_{1}(\omega+i \delta ) 
&=& \int \frac{d\epsilon}{\pi}
{{ \pi \sum_{\vec k } \vert V(\vk ) \vert^2 \delta( \epsilon_{\vk} 
- \epsilon)}
\over { \omega -
\epsilon- i \delta }}\cr
&=& \int \frac{d\epsilon 
}{\pi}
{  \Delta (\epsilon)\over { \omega -
\epsilon}- i \delta }
\end{eqnarray}
Typically $\Delta (\epsilon )$ will only vary substantially on
energies
of order the bandwidth, so that over the width of the resonance we can
replace
$\Delta
(\epsilon)\rightarrow \Delta $. Moreover, for a broad band of width
$D$,  the real
part of $\Sigma (\omega)\sim \frac{1}{\pi }\Delta (0) \ln [(\omega -D)/ (\omega+D)]$ 
is of order $\omega/D$ and can be ignored, or absorbed into 
into a small  renormalization of 
$E_d$. 
This allows us to 
make the replacement
$$
\Sigma_{1}(\omega\pm i \delta)=\mp i \Delta {\rm sgn \delta}
\eqno(24)$$
so that 
\begin{equation}\label{}
 G_{d\sigma }( \omega-i\delta)= {1\over (\omega-E_{d \si}-i \Delta) }. 
\end{equation}
The density of  states described by the Green-function is a Lorentzian
centered around energy $E_{d\sigma }$:
\[
\rho_{d\sigma } (\omega)= \frac{1}{\pi}Im G_{d\sigma } (\omega - i \delta )= 
\frac{\Delta }{(\omega-E_{d\sigma })^{2} + \Delta ^{2}}, 
\]
moreover, the occupancy of the d-state is given by 
the d-occupation at zero temperature is
\begin{equation}\label{}
\begin{array}{rcl}
n_{d \si}  &=&
\int_{-\infty}^{0} d\om \rho_{d\si}(\omega)\cr
&=&\frac{1}{\pi}\cot ^{-1}\left(\frac{E_{d}+Un_{d-\sigma } }{\Delta } \right)
\end{array}
\end{equation}
This equation defines Anderson's  mean-field theory.  \footnote{The quantity 
$\delta _{\sigma }= \cot ^{-1}\left(\frac{E_{d}+Un_{d-\sigma }
}{\Delta } \right)$ is actually the phase shift for scattering an
electron off the d-resonance (see exercise), and the identity
$n_{d\sigma }= \frac{1}{\pi }\delta _{\sigma }$  is a particular
realization of the ``Friedel sum rule'', which relates the charge
bound in an atomic potential to the number of nodes ($=\sum_{\sigma}\frac{\delta
_{\sigma }}{\pi }$) introduced into the scattering state wavefunction. }
It is convenient
to
introduce an occupancy $n_{d}= \sum_{\sigma }n_{d\sigma }$ and
magnetization
$M=n_{d\uparrow}- n_{d\downarrow}$, so that $n_{d\sigma }= \frac{1}{2}
(n_{d}+ \sigma M)$ ($\sigma =\pm 1$). The mean-field equation for the occupancy and
magnetization are then
\begin{eqnarray}\label{mf}
n_{d}&=&\frac{1}{\pi}\sum_{\sigma =\pm 1}
\cot^{-1}\left(\frac{E_{d}+U/2 (n_{d}-\sigma M)}{\Delta } \right) 
\cr
M&=&\frac{1}{\pi}\sum_{\sigma =\pm 1}\sigma 
\cot^{-1}\left(\frac{E_{d}+U/2 (n_{d}-\sigma M)}{\Delta } \right) 
\end{eqnarray}
To find the critical size of the interaction strength where a local moment develops, 
set $M\rightarrow 0^{+}$  (replacing the second equation by its
derivative w.r.t. $M$), 
which gives 
\begin{eqnarray}\label{}
n_{d}&=&\frac{2}{\pi}\cot ^{-1}\left(\frac{E_{d}+ U_{c} n_{d}/2}{\Delta }
\right)\cr
1&=& \frac{U_{c}}
{\pi \Delta }
\frac{1}{1 + \left(\frac{E_{d}+U
n_{d}/2}{\Delta } \right)^{2}}
\end{eqnarray}
which can be written parametrically as 
\begin{eqnarray}\label{par}
E_{d}+ \frac{U_{c}}{2} &=& \Delta 
\left( c + 
\frac{\pi}{2}
( 1-n_{d})(1+c^{2}) \right)
\cr
U_{c} &=& \pi \Delta ( 1 + c^{2})
\end{eqnarray}
where $c\equiv \cot \left(\frac{\pi n_{d}}{2} \right).
$ The critical curve described by these equations is shown in
Fig. \ref{fig2}.  

From the mean-field equations, it is easily seen that for $n_{d}=1$,
when the d-levels are half filled, the critical value $U_{c}= \pi
\Delta $. This enables us to qualitatively understand the
experimentally observed formation of local moments. 
When dilute magnetic ions are dissolved into a metallic host, the formation
of a local moment is dependent on whether the ratio $U/\pi \Delta $ is larger
than, or smaller than zero.  When iron is dissolved in  pure niobium, 
the failure of the moment to form reflects the higher density of states
and larger value of $\Delta $ in this alloy.  When iron is dissolved in
molybdenum, the lower density of states causes $U> U_{c}$, and local
moments
form. \cite{earlymagrefs}

\subsubsection{The Coulomb Blockade}

A modern context for the physics of local moments is found within
quantum dots.  A quantum dot is a  tiny electron pool in a doped semi-conductor,
small enough so that the electron states inside the dot are quantized,
loosely resembling the electronic states of  an atom.  Unlike a
conventional atom, the separation of the electronic states is of the
order of milli-electron volts, rather than volts. The 
overall position of the quantum dot energy levels can be changed by
applying a gate voltage to the dot. It is then possible to pass a
small 
current through the dot by 
placing it between two leads. The differential conductance $G= dI/dV
$
is directly proportional to the density of states $\rho (\omega )$
inside the dot $G \propto \rho (0) $. Experimentally, when G is measured
as a function of gate voltage $V_{g}$, the differential conductance is
observed to develop a periodic structure, with a period of a few
milli-electron volts. \cite{qdots}

This phenomenon is known as the ``Coulomb blockade'' and it results
from precisely the same physics that is responsible for moment
formation. A simple model for a quantum dot considers it as a sequence
of single particle levels at energies $\epsilon _{\lambda }$, interacting
via a single Coulomb potential $U$, according to the model
\begin{equation}\label{qdot}
H_{dot}= \sum_{\lambda } (\epsilon _{\lambda } -e V_{g})n_{\lambda
\sigma }+ 
\frac{U}{2}N (N-1) 
\end{equation}
where $n_{\lambda \sigma }$ is the 
occupancy of the spin $\sigma $ state of the  $\lambda $ level, 
$N=\sum_{\lambda\sigma }n_{\lambda \sigma }$ is the total number
of electrons in the dot and $V_{g}$ the gate voltage. 
This is a simple generalization of the 
single atom part of the Anderson model. Notice that the capacitance of the
dot is  $C= e^{2}/U$.

Provided that $U$ is far greater than the energy separation of the
individual levels, $U>> \epsilon _{\lambda }-\epsilon _{\lambda '}$,
the energy difference between the $n$ electron and $n+1$ electron 
state of the dot is
given by $E (n+1)-E (n) = n U- e V_{g}$. 
As the gate voltage is raised, the quantum dot
fills each level sequentially, as illustrated in Fig. \ref{fig3},
and when $eV_{g}= U$, the n-th level becomes degenerate with the Fermi
energy of each lead. At this point, electrons can pass coherently through
the resonance giving rise to a sharp peak in the conductance. At maximum
conductance, the transmission and reflection of electrons is unitary,
and the conductance of the quantum dot will reach a substantial fraction
of the quantum of conductance, $e^{2}/h $ per spin. A simple calculation of
the zero-temperature conductance through a single non-interacting
resonance coupled symmetrically to two leads gives
\begin{equation}\label{cond}
G (eV_{g})  = \frac{2e^{2}}{h}\frac{\Delta ^{2}}{(E_{\lambda}-e
V_{g})^{2}+\Delta ^{2}}
\end{equation}
where the factor of two derives from two spin channels. 
At finite temperatures, the resonance becomes broadened by
thermal excitation effects, giving
\[
G (eV_{g},T)  = \frac{2e^{2}}{h}\int d\epsilon \left(-\frac{df
(\epsilon -eV_{g})}{d\epsilon } \right)
\frac{\Delta ^{2}}{(E_{\lambda}-\epsilon )^{2}+\Delta ^{2}}
\]
where $f (\epsilon )= 1/ (e^{\beta \epsilon }+1)$
is the Fermi function. 
When interactions are included, we must sum over the n-levels, giving (See Fig. \ref{fig3}. ) 
\[
G (eV_{g},T)  = \frac{2e^{2}}{h}\int d\epsilon \left(-\frac{df
(\epsilon -eV_{g})}{d\epsilon } \right)
\frac{\Delta ^{2}}{(nU-\epsilon )^{2}+\Delta ^{2}}
\]
\fight=3.6truein
\fg{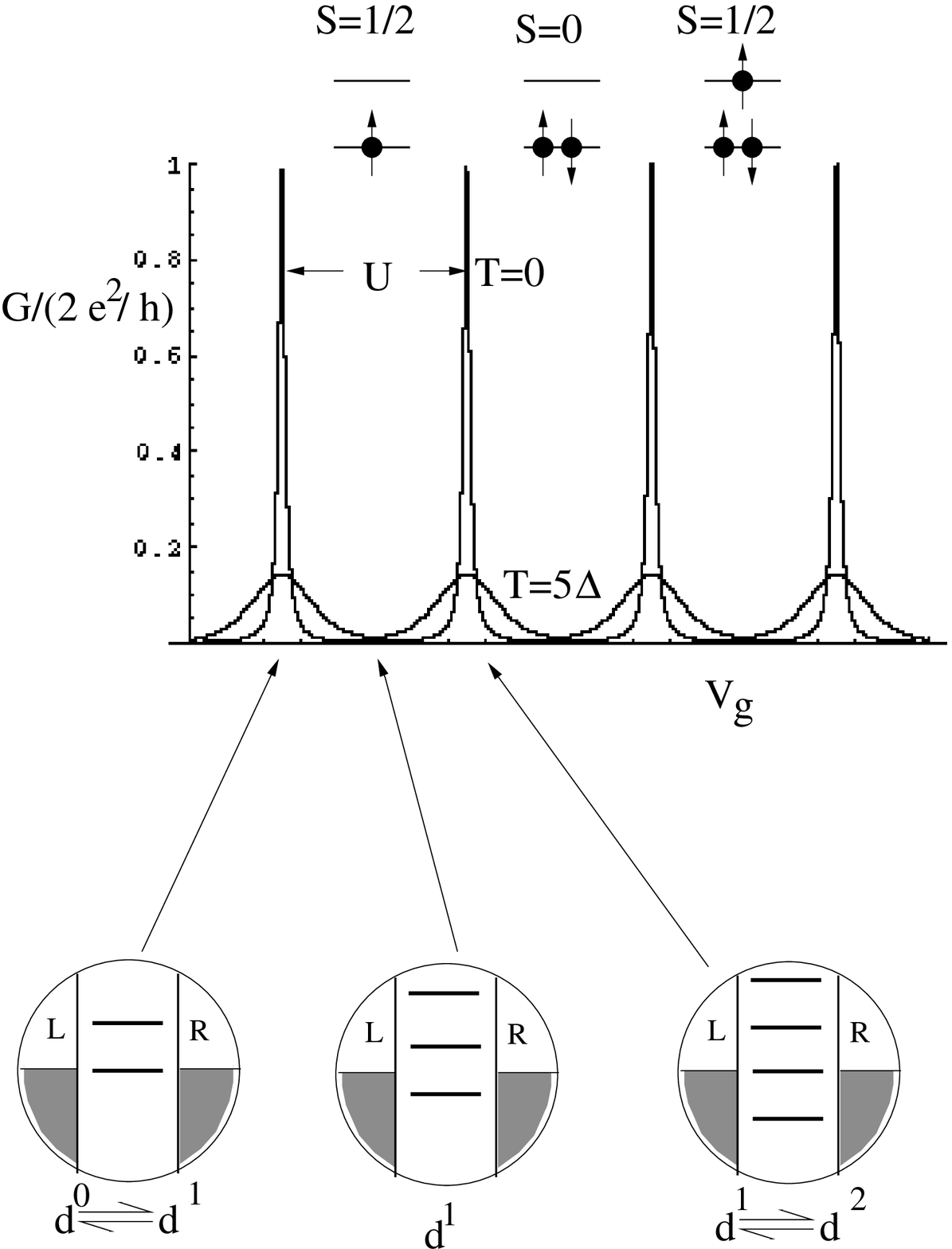}{Variation of zero bias 
conductance $G=dI/dV$ with gate voltage in a quantum dot. 
Coulomb interactions mean that for each additional electron in the
dot, the energy to add one electron increases by $U$. When  the
charge on the dot is integral, the Coulomb interaction blocks
the addition of electrons and the conductance is suppressed. 
When the energy to add an electron is degenerate with the
Fermi energy of the leads, unitary transmission occurs, and
for symmetric leads, $G=2e^{2}/h$. 
}{fig3}

The effect of field on these results is interesting.  When the number of
electrons in the dot is even, the quantum dot
is in a singlet state.  When the number of electrons is odd, the
quantum dot forms a local moment.  In a magnetic field,  the energy
of the odd-electron dot is reduced, whereas the energy of the even
spin dot is unchanged, with the result that at low temperatures
\begin{eqnarray}\label{}
E (2n+1)-E (2n)&=& 2nU - \mu _{B}B\cr
E (2n+2)-E (2n+1)&=& (2n+1)U + \mu _{B}B
\end{eqnarray}
so that the voltages of the odd and even numbered peaks in the 
conductance develop an alternating field dependence.  

It is remarkable that the physics of moment formation and the
``Coulomb blockade'' operate in both artificial mesoscopic devices and
naturally occurring magnetic ions. 

\subsection{Exercises }

\begin{problems}

\item By expanding a plane wave state in terms of spherical harmonics:
\[
\langle \vec{r}\vert \vec{k}\rangle = 
e^{i\vec{k}\cdot \vec{ r}}= 4\pi \sum_{l,m} i^{l}j_{l} (kr)Y^{*}_{lm }
(\hat k)Y_{lm} (\hat r)
\]
show that the overlap between a state $\vert \psi \rangle $
with wavefunction $\langle \vec{x}\vert \psi \rangle = R (r)
Y_{lm} (\hat r)$ with a plane wave is given by $V (\vec{k})=
\langle \vec{k}\vert V\vert \psi \rangle = V (k)Y_{lm} (\hat{k})$ where
\begin{equation}\label{}
V (k)= 4\pi 	i^{-l} \int dr r^{2} V (r) R (r)j_{l } (kr)
\end{equation}

\item 
\begin{itemize}
\item [(i)]
Show that $\delta = \cot ^{-1}\left(\frac{E_{d}}{\Delta } \right)$ is the scattering phase shift for scattering
off a resonant level at position $E_{d}$. 

\item [(ii)] Show that the
energy  of states in the continuum  is shifted by an amount $-\Delta
\epsilon \delta (\epsilon )/\pi $, where $\Delta \epsilon $ is the
separation  of states in the continuum.

\item [(iii)]Show that the increase in density of states is given by $\partial
\delta /\partial E= \rho _{d} (E)$. (See chapter~3.)
\end{itemize}

\item Derive the formula (\ref{cond}) for the conductance of a single isolated
resonance.  

\end{problems}

\section {The Kondo Effect}

Although Anderson's mean-field theory
provided a mechanism for moment formation, it raised many new
questions.  One of its inadequacies is that
of the magnetic moment is regarded as a broken symmetry order
parameter. Broken symmetry is possible when the object that breaks the
symmmetry involves a macroscopic number of degrees of freedom, but
here, we are dealing with a single spin.  There
will always be a certain quantum mechanical amplitude for the
spin to flip between an up and down configuration. 
This 
tunneling rate $\tau ^{-1}$ defines a temperature scale
\[
k_{B} T_{K} = \frac{\hbar }{\tau }
\]
called the Kondo temperature, which sets the dividing line between
local moment behavior, where the spin is free, and the low temperature
limit, where the spin becomes highly correlated with the
surrounding electrons. Experimentally, this temperature marks
the low temperature limit of a Curie susceptibility. 
The physics by which the local moment
disappears or ``quenches'' at low temperatures is closely analagous
to the physics of quark confinement and it is named the ``Kondo
effect'' after the Japanese physicist Jun Kondo. \cite{kondo}

The Kondo effect has a wide range of manifestations in condensed
matter physics: not only does it govern the quenching of magnetic moments
inside a metal, but it also is responsible for the formation of 
heavy fermion metals, where the local moments transform into 
composite quasiparticles with masses sometimes in excess of a thousand
bare electron masses.\cite{stewart} Recently, the Kondo effect has also been observed
to take place in quantum dots that carry a local moment.  (Typically
quantum dots with an odd number of electrons).  \cite{qdots}

In this section we will first derive the Kondo model from the Anderson model,
and then discuss the properties of this model in the language of the
renormalization group . 

\subsection{Adiabaticity}

Let us discuss some of the properties of the Anderson model at low temperatures
using the idea of adiabaticity.  We suppose that the interaction
between electrons in the Anderson model is increased continuously
to values $U>>\Delta $, whilst maintaining the occupancy of the d-state
equal to unity $n_{d}=1$. The requirement that $n_{d}=1 $ ensures that the d-electron density of
states is  particle-hole symmetric, which implies that $E_{d}=0$ and
$\Sigma' (0)=0$.  

When $U>> \Delta $, we expect that the d-electron spectral function
$\rho _{d}= \frac{1}{\pi }\hbox{Im} G_{d} (\omega -i\delta )$ will contain
two peaks at $\omega= \pm U/2$. Since the total spectral weight
integrates to unity, $\int d\omega \rho (\omega )=1$, we expect
that the weight under each of these peaks is approximately $1/2$.
Remarkably, as we shall now see, 
the spectral function at $\omega=0$ is unchanged
by the process of increasing the interaction strength and remains
equal to its non-interacting value
\[
\rho _{d} (\omega =0)= \frac{1}{\Delta }
\]
This means that the d-spectral function must contain a narrow peak,
of vanishingly small spectral weight $Z<<1$, height $\frac{1}{\Delta
}$  and hence width $\Delta^{*}= Z\Delta <<\Delta $. This peak
in the d-spectral function is associated with the Kondo effect,
and is known as the Abrikosov-Suhl, or the ``Kondo''  resonance. 
\fight=3in
\fg{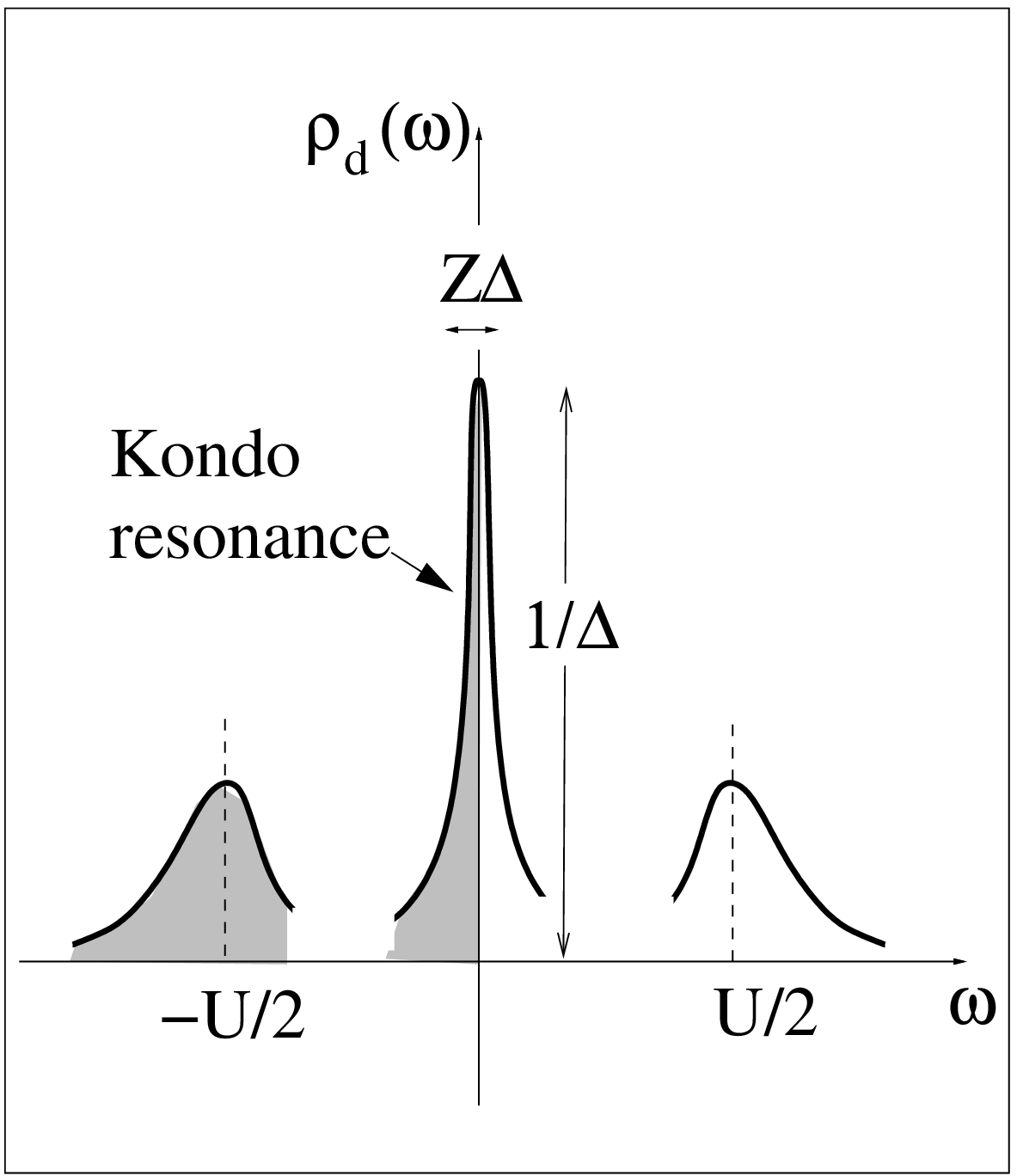}{``Kondo resonance'' in the d-spectral function. 
At large $U$ for the particle hole-symmetric case where
$n_{d}=1$,  the d-spectral function contains two peaks about
$\omega\sim \pm U/2$, both of weight approximately $1/2$.
 However, since the spectral function is constrained
by the requirement that $\rho _{d} (\omega =0)= \frac{1}{\Delta }$, 
the spectral function must preserve a narrow peak of fixed height, but
vanishingly small weight $Z<<1$. 
}{figsuhl}
\fight=2in
Let us see how this comes about as a consequence of adiabaticity.
For a single magnetic ion, we expect that the 
interactions between electrons can be increased continuously,
without any risk of instabilities, so that the excitations of
the strongly interacting case remain in one-to-one correspondence
with the excitations of the non-interacting case $U=0$, forming a
``local Fermi liquid''. 

In this local Fermi liquid, one can divide the d-electron self-energy
into two components- the first derived from hybridization, the second
derived from interactions:
\begin{eqnarray}\label{}
\Sigma (\omega-i \delta  ) &=&  i \Delta  + \Sigma _{I} (\omega -i\delta
)\cr
\Sigma_{I} (\omega -i\delta ) &=& (1 - Z^{-1})\omega + i A \omega ^{2}.
\end{eqnarray}
The ``wavefunction'' renormalization $Z$ is less
than unity. 
The quadratic energy dependence of $\Sigma´´_{I} (\omega )\sim \omega ^{2}$
follows from the quadratic energy dependence of the phase space
for producing particle-hole pairs. Using this result, 
the form of the d-electron propagator for $n_{d}$
at low energies is 
\begin{eqnarray}\label{}
G_{d} (\omega-i\delta  ) &=& \frac{1}{\omega -i\Delta - \Sigma_{I}
(\omega )}\cr
&=& \frac{Z}{\omega -iZ\Delta - i O (\omega ^{2}).
}
\end{eqnarray}
This corresponds to a renormalized resonance of reduced 
weight $Z<1$, renormalized width $Z\Delta
$. 
One of the remarkable results of this line of reasoning, is the
discovery that d-spectral weight 
\[
\rho _{d} (\omega \sim 0) = \frac{1}{\pi}{\rm Im }G_{d} (\omega -i \delta
)\vert _{\omega =0}= \frac{1}{\Delta }
\] 
is independent of the strength of $U$. 
This result, first discovered by Langreth\cite{langreth}
guarantees a peak in the 
d-spectral function at low energies, no matter how large $U$ becomes. 
Since we  also expect a peak in the d-spectral function around
$\omega\sim \pm U/2  $, this line of reasoning suggests that 
the structure of the d-spectral function at large $U$, contains three peaks.

\subsection{Schrieffer-Wolff transformation}

If a local moment forms within an atom, 
the object left behind is a pure
quantum top- a quantum mechanical object with purely spin degrees of
freedom.  
\footnote{In the simplest version of the Anderson model, the local moment is a
$S=1/2$, but in more realistic atoms much large moments can be
produced.
For example, an electron in a Cerium $Ce^{3+}$ ion atom lives in a $4f^{1}$
state. Here spin-orbit coupling combines orbital and spin angular momentum
into a total angular moment $j= l-1/2= 5/2$. The Cerium ion that forms
thus has a spin $j=5/2$ with a spin degeneracy of $2j+1=6$. 
In multi-electron atoms, the situation can become still more complex,
involving Hund's coupling between atoms. }

These spin degrees of freedom do interact with the surrounding
conduction sea. In particular 
virtual charge fluctuations, in which
an electron briefly migrates off, or onto the ion
lead to spin-exchange between the local moment and the conduction
sea.  This induces an antiferromagnetic interaction between the
local moment and the conduction electrons. 
To see this 
consider the 
two possible spin exchange processes 
\begin{eqnarray}\label{}
e_{\uparrow}+ d^{1}_{{\downarrow }}&\leftrightarrow& d^{2}
\leftrightarrow e_{\downarrow }+d^{1}_{{\uparrow}}\qquad \Delta E_{I} \sim
U + E_{d}\cr
e_{\uparrow}+ d^{1}_{{\downarrow }}&\leftrightarrow& 
e_{{\uparrow}}+e_{\downarrow }
\leftrightarrow e_{\downarrow }+d^{1}_{{\uparrow}}\qquad \Delta E_{II} \sim
-E_{d}
\end{eqnarray}
The first process passes via a doubly
occupied singlet d-state, so it can only take place
if the incoming conduction electron and d-electron are in a mutual 
$S=0$ state.  In the second process, 
in order that the conduction electron can hybridize with the d-state,
it has to arrive and depart 
in a state with precisely the same d- orbital symmetry. This means that
the intermediate state formed in the second process must be spatially
symmetric, and must therefore be a spin-antisymmetric singlet $S=0$ state. 
From these arguments, we see that spin exchange only takes place in
the singlet channel, lowering the energy of the singlet configurations
by an amount of order
\begin{eqnarray}
J & \sim &  V^{2}\left[\frac{1}{\Delta E_{1}}+ \frac{1}{\Delta E_{2}} \right] \\ 
& = & \left[\frac{1}{-E_{d}}+ \frac{1}{E_{d}+U} \right]
\end{eqnarray}
where $V$ is the size of the hybridization matrix element near
the Fermi surface. If we introduce the electron spin density operator
$\vec{S} (0)= \frac{1}{N}\sum_{k,k'}c\dg _{k\alpha }\vec{\sigma
}_{\alpha \beta }c _{k'\beta }
$, where $N$
is the number of sites in the lattice, then we expect that
the effective interaction induced by the virtual charge fluctuations will
have the form
\[
H_{eff}= J \vec{S} (0)\cdot \vec{S}_{d}
\]
where $\vec{S}_{d}$ is the spin of the localized moment.  Notice that
the sign of $J$ is \underline{antiferromagnetic}. This kind of
heuristic argument was ventured in Anderson's paper on local moment
formation in 1961.  The antiferromagnetic sign in this interaction was
quite unexpected, for it had been tacitly assumed by the community
that exchange forces would induce a ferromagnetic interaction between
the conduction sea and local moments. This seemingly innocuous sign
difference has deep consequences for the physics of local moments at
low temperatures, as we shall see in the next section.

Let us now  carry out the transformation a little more
carefully, using the method of canonical transformations introduced by
Schrieffer  and Wolff\cite{swolf,coqblin}.  The  Schrieffer-Wolff  transformation is  very
close to  the idea of the  renormalization group and will  help set up
our renormalization  group discussion.  When a local  moment forms,
the  hybridization with the conduction sea 
induces virtual charge fluctuations.  It is
therefore  useful to  consider dividing  the Hamiltonian into two terms
\newcommand{\diagmatrix}[4]{
\left[\matrix{
\left. \matrix{&&\cr
&#1&\cr
&&}\right| 
& 
\left. \matrix{&&\cr
&#2&\cr
&&}
\right.
\cr
\overline{\left. \matrix{&&\cr
&#3 &\cr
&&}
\right|}
&
\overline{\left.\matrix{&&\cr
&#4&\cr
&&}
\right.}
}
\right]
}
\newcommand{\diager}[4]{
\left[\matrix{
{\left. #1\right| #2}\cr
\overline{
{\left. #3 \right| #4}} 
}
\right]
}
\[
H=H_{1}+ \lambda {\mathcal{V}}
\]
where $\lambda$ is an expansion parameter. Here, 
\[
H_{1}= H_{band} + H_{atomic}=\nmat{H_{L}}{\ 0}{\ \ 0\ \ \   }{H_{H}}
\]
is diagonal in the low energy 
$d^{1}$ ($H_{L}$) and  the high energy $d^{2}$ or $d^{0}$ ($H_{H}$)
subspaces, whereas  the hybridization term 
\[
\mathcal{V}= H_{mix}= \sum_{j \si}
\bigl[V_{\vec k}c\dg_{k  \si}d_{\si} + {\rm H. c.}\bigr]  
=\nmat{\ \ 0\ }{V\dg }{V}{0}
\]
provides the off-diagonal matrix elements between these
two subspaces. 
The idea of the  
Schrieffer Wolff transformation is to carry out a canonical transformation
that returns the Hamiltonian to block-diagonal form, as follows:
\begin{equation}\label{ug}
U\nmat{H_{L}}{\lambda V\dg}{\lambda V}{H_{H}}U\dg =
\nmat{H^{*}}{0\phantom{\dg }}{0}{H'}.
\end{equation}
This is a ``renormalized'' Hamiltonian, and the block-diagonal part
of this matrix $H^{*}= P_{L}H'P_{L}$ in the low energy subspace provides an {\sl effective}
Hamiltonian for the low energy physics and low temperature
thermodynamics. 
If we set $U=e^{S}$, where $S=-S\dg $ is anti-hermitian 
and expand 
S in a power series
\[
S = \lambda S_{1} + \lambda ^{2} S_{2}+ \dots, 
\]
then expanding (\ref{ug}) using the identity
$e^{A}Be^{-A}= B + [A,B] + \frac{1}{2!}[A,[A,B]]\dots $ 
\[
e^{S} ( H_{1}+\lambda {\mathcal{V}})e^{-S} = H_{1} + \lambda
\left(\mathcal{V}+[S_{1},H_{1}]\right) 
+ \lambda ^{2}\left( \frac{1}{2}[S_{1},[S_{1},H]]+
[S_{1},\mathcal{V}] + [S_{2},H_{1}] \right) + \dots 
\]	
so that to leading order 
\begin{equation}\label{basic}
[S_{1},H_{1}]=-\mathcal{V},
\end{equation}
and to second order
\[
e^{S} ( H_{1}+\lambda {\mathcal{V}})e^{-S}= H_{1} 
+ \lambda ^{2}\left( \frac{1}{2}[S_{1},\mathcal{V}]+
[S_{2},H_{1}] \right) + \dots .
\]	
Since $[S_{1},\mathcal{V}]$ is block-diagonal, we can satisfy
(\ref{ug} ) to second order by requiring 
$S_{2}=0$, so that to this order, the renormalized Hamiltonian
has the form (setting $\lambda =1$)
\[
H^{*} = H_{L} + H_{int}
\]
where
\[
H_{int}=\frac{1}{2}P_{L}[S_{1},\mathcal{V}]P_{L}+\dots 
\]
is an interaction term induced by virtual fluctuations into the
high-energy manifold. 
Writing 
\[
S=\nmat{0}{-s\dg }{s}{0}
\]
and substituting into (\ref{basic}), we obtain $V=-sH_{L}+
H_{H}s$. Now since $( H_{L})_{{ab}}=
E^{{L}}_{a}\delta _{ab}$  and 
$( H_{H})_{{ab}}=
E^{{H}}_{a}\delta _{ab}$  are diagonal, it follows that 
\begin{equation}\label{thes}
s_{ab}= \frac{V_{ab}}{E_{a}^{H}-E_{b}^{L}}, \qquad 
-s\dg _{ab}= \frac{V\dg _{ab}}{E_{a}^{L}-E_{b}^{H}}, \qquad .
\end{equation}
From (\ref{thes}), we obtain
\[
( H_{int})_{ab}=- \frac{1}{2}(V\dg s + s\dg  V)_{ab}= \frac{1}{2}
\sum_{\lambda \in \vert H\rangle }\left[\frac{V\dg _{a\lambda
}V_{\lambda b}}{E_{a}^{L}-E_{\lambda }^{H}}+
\frac{V\dg _{a\lambda
}V_{\lambda b}}{E_{b}^{L}-E_{\lambda }^{H}}
 \right]
\]
Some important points about this result
\begin{itemize}

\item We recognize this result as a simple generalization of second-order 
perturbation theory which encompasses off-diagonal matrix elements.

\item 
$H_{int}$ can also be written 
\[
H_{int}= \frac{1}{2} [T(E_{a}) + T (E_{b})]
\]
where $T$ is given by
\begin{eqnarray}\label{tmatrix}
\hat T (E)&=& P_{L}{\mathcal{V}}\frac{P_{H}}{E-H_{1}}{\mathcal{V}}P_{L}\cr
T_{ab} (E)&=& \sum_{\lambda \in \vert H\rangle }\left[\frac{V\dg _{a\lambda
}V_{\lambda b}}{E-E_{\lambda }^{H}}
 \right]
\end{eqnarray}
is the leading order expression for the scattering T-matrix induced by
scattering off $\mathcal{V}$.
We can thus relate $H_{int}$ to a
scattering
amplitude, and schematically represent it by a Feynman diagram,
illustrated in Fig. \ref{fig4}.

\fight=2in
\fg{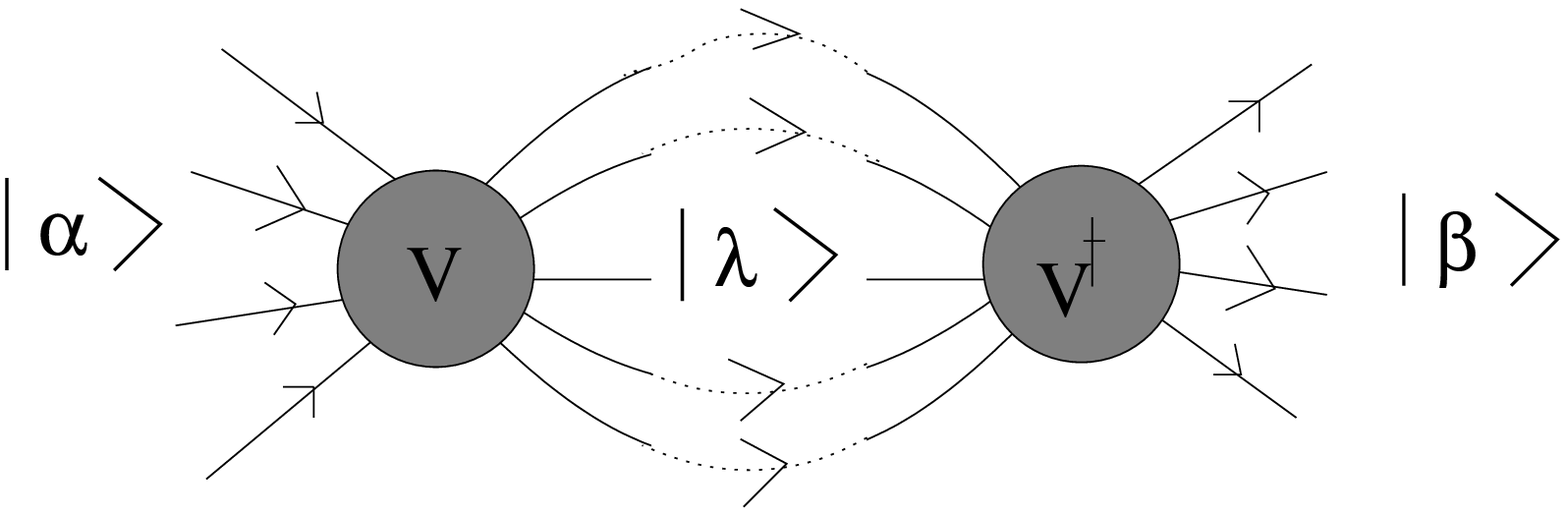}{T-matrix representation of interaction induced by
integrating out high-energy degrees of freedom}{fig4}

\item If the separation of the low and high energy subspaces is large, 
then the energy denominators in the above expression will not depend
on the initial and final states $a$ and $b$, so that this expression
can be simplified to the form 
\[
H_{int}= -\sum_{\lambda\in \vert H\rangle }\frac{V\dg P_{\lambda
}V}{\Delta E_{\lambda }}
\]
where $\Delta E_{\lambda }= E^{H}_{\lambda }-E^{L}$ is the excitation
energy in the high energy subspace labeled by $\lambda $, and 
the projector $P_{\lambda }= \sum_{\vert a\rangle
\in\vert \lambda\rangle
}\vert a\rangle \langle a\vert $ .

\end{itemize}

If we apply this method to the Anderson model, we have two high-energy
subspaces, with excitation energies $\Delta E (d^{1}\rightarrow
d^{0})= -E_{d}$ and $\Delta E (d^{1}\rightarrow d^{2})= E_{d}+U$, 
so that the renormalized
interaction is
\[
H_{int}=-\sum_{k\sigma ,k'\sigma '}V^{*}_{k'}V_{k}\Biggl[
\overbrace{ 
\frac{
( c\dg _{k\sigma }d_{\sigma})
( d\dg _{\sigma '}c_{k'\sigma '})
}{E_{d}+U}}
^{d^{1}+e^{-}\leftrightarrow d^{2}}
+
\overbrace{\frac{( d\dg _{\sigma '}c_{k'\sigma '})
( c\dg _{k\sigma }d_{\sigma})
}{-E_{d}}}^{d^{1}\leftrightarrow d^{0}+e^{-}} \Biggr]
\]
Using the identity
$\delta _{ab}\delta _{cd}+\vec{\sigma }_{ab}\cdot
\vec{\sigma }_{cd}= 2\delta _{ad}\delta _{bc}$
we may cast the renormalized Hamiltonian
in the form
\begin{eqnarray}\label{konfin}
H_{int}&=& \sum_{k\alpha ,k'\beta }
J_{k,k'}c\dg _{k\alpha }\vec{\sigma
}c_{k'\beta }\cdot \vec{S}_{d}+H'\cr
J_{k,k'}&=&
V^{*}_{k'}V_{k}\Biggl[\overbrace{\frac{1}{E_{d}+U}}^{{d^{1}+e^{-}\leftrightarrow
d^{2}}}+\overbrace {\frac{1}{-E_{d}}}^{d^{1}\leftrightarrow d^{0}+e^{-}} \Biggr]
\end{eqnarray}
where 
\begin{equation}\label{abrikosov}
\vec{S}_{d}\equiv d\dg _{\sigma }\left(\frac{\vec{\sigma }_{\alpha
\beta }}{2} \right)d_{\beta }, \qquad (n_{d}=1)
\end{equation}
where we have replaced $n_{d}=1$ in the low energy subspace. Apart
from a constant, the
second term 
\[
H'= -\frac{1}{2}\sum_{k,k'\sigma }
V^{*}_{k'}V_{k}\left[ \frac{1}{E_{d}+U}+\frac{1}{E_{d}}\right]
c\dg _{k\sigma }c_{k'\sigma} 
\]
is a residual potential scattering term off the local moment. This
term vanishes for the particle-hole symmetric
case $E_{d}=- (E_{d}+U)$ and 
will be dropped, since it does not involve the internal dynamics
of the local moment. 
 Summarizing, the effect of the high-frequency
valence fluctuations is to induce an antiferromagnetic
coupling between the local spin density of the conduction electrons and
the local moment:\\

\boxit{\begin{equation}\label{kondomodel}
H= \sum_{k\sigma }\epsilon _{k}c\dg _{k\sigma }c_{k\sigma } + \sum
_{k,k'}J_{k,k'}c\dg _{k\alpha }\vec{\sigma
}c_{k'\beta }\cdot \vec{S}_{d}
\end{equation}}
This is the infamous ``Kondo model''. For many purposes, the $k$
dependence of the coupling constant can be dropped. 
In this case, 
the Kondo interaction can be written 
$H_{int}=J
\psi \dg (0)\vec{\sigma }\psi (0)
\cdot \vec{S}_{d}$, where $\psi_{\alpha }
(0)=\frac{1}{\sqrt{N}}\sum c_{k\alpha }$ is the electron operator
at the origin and $\psi \dg (0)\vec{\sigma }\psi (0)
$ is the spin density at the origin. 
In this simplified form, the Kondo model takes the deceptively simple form\\

\boxit{\begin{equation}\label{kondomodel2}
H= \sum_{k\sigma }\epsilon _{k}c\dg _{k\sigma }c_{k\sigma } +\overbrace{J
\psi \dg (0)\vec{\sigma
}\psi (0)\cdot \vec{S}_{d}}^{{H_{int}}}.
\end{equation}}\\
\vskip 0.1in

\noindent In other words, there is a simple point-interaction between the spin
density of the metal at the origin and the local moment. 
Notice how all reference to the fermionic character of the
d-electrons has gone, and in their place, is a $S=1/2$ spin operator. 
The fermionic representation (\ref{abrikosov}) of the spin operator
proves to be very useful in the case where the Kondo effect takes
place. 

\subsection{Renormalization concept}\label{}

To make further progress, we need to make use of the 
concept of renormalization.
In a general sense, physics occurs
on several widely
spaced energy scales in condensed matter systems. 
We would like to distill the  essential effects
of 
the high energy  atomic physics at electron volt scales
on the low energy physics at millivolt scales without getting
caught up in the fine details.
An essential tool for this task is the ``renormalization group''. 
\cite{yuval,poor,wilson,Noz}

The concept of the renormalization group 
permits us to describe  complex condensed matter
systems using simple  models that reproduce only the
relevant low energy physics of the system.
The idea here is that only certain gross features
of the
high energy physics are relevant to the low energy excitations.
The continuous family of model Hamiltonians with the same low energy
excitation spectrum constitute a ``universality class'' of models.
(Fig. \ref{fig9})
Suppose we parameterize each model Hamiltonian $H(D)$
by its cutoff energy
scale, $D$, the energy of the largest 
excitations. 
The scaling procedure, involves rescaling the cutoff 
$D\rightarrow D'=D/b$ where $b>1$, integrating out the 
excitations $E\in [D',D]$ to obtain
an effective Hamiltonian $\tilde{H_{L}}$for the remaining low-energy degrees of
freedom.  The energy scales are then rescaled, to obtain a new $H
(D')=b\tilde{H_{L}}$
Generically, the Hamiltonian will have the
block-diagonal
form
\begin{equation}\label{bdiag}
H=\nmat{H_{L}}{V\dg }{V}{H_{H}}
\end{equation}
where $H_{L}$ and $H_{H}$ act on states 
in the low-energy and high-energy subspaces respectively, and $V$ and
$V\dg $ provide the matrix elements between them. 
The high energy degrees of freedom may be 
``integrated out'' \footnote{The term ``integrating out'' is
originally derived from the path integral formulation
of the renormalization group, in which high energy
degrees of freedom are removed by integrating over 
these variables inside the path integral.
}
by carrying out a  canonical transformation 
and projecting  out the low-energy
component
$\tilde{H}_{L}$
\begin{equation}\label{cantrans}
H (D)\rightarrow  U H (D) U \dg 
= \nmat{\tilde{H_{L}}} {0}{0}{\tilde{H_{H}}}
\end{equation}
By rescaling 
\begin{equation}\label{rescale}
H(D')= b \tilde{H_{L}}
\end{equation}
one arrives at a new Hamiltonian describing the physics on the 
reduced scale. The transformation from $H (D)$ to $H (D')$
is referred to as a ``renormalization group'' (RG)
transformation. This term was coined long ago, even though 
the transformation does not form a real group, since there is no
inverse transformation.
Repeated application of the RG procedure
\fight=3.5in
\fg{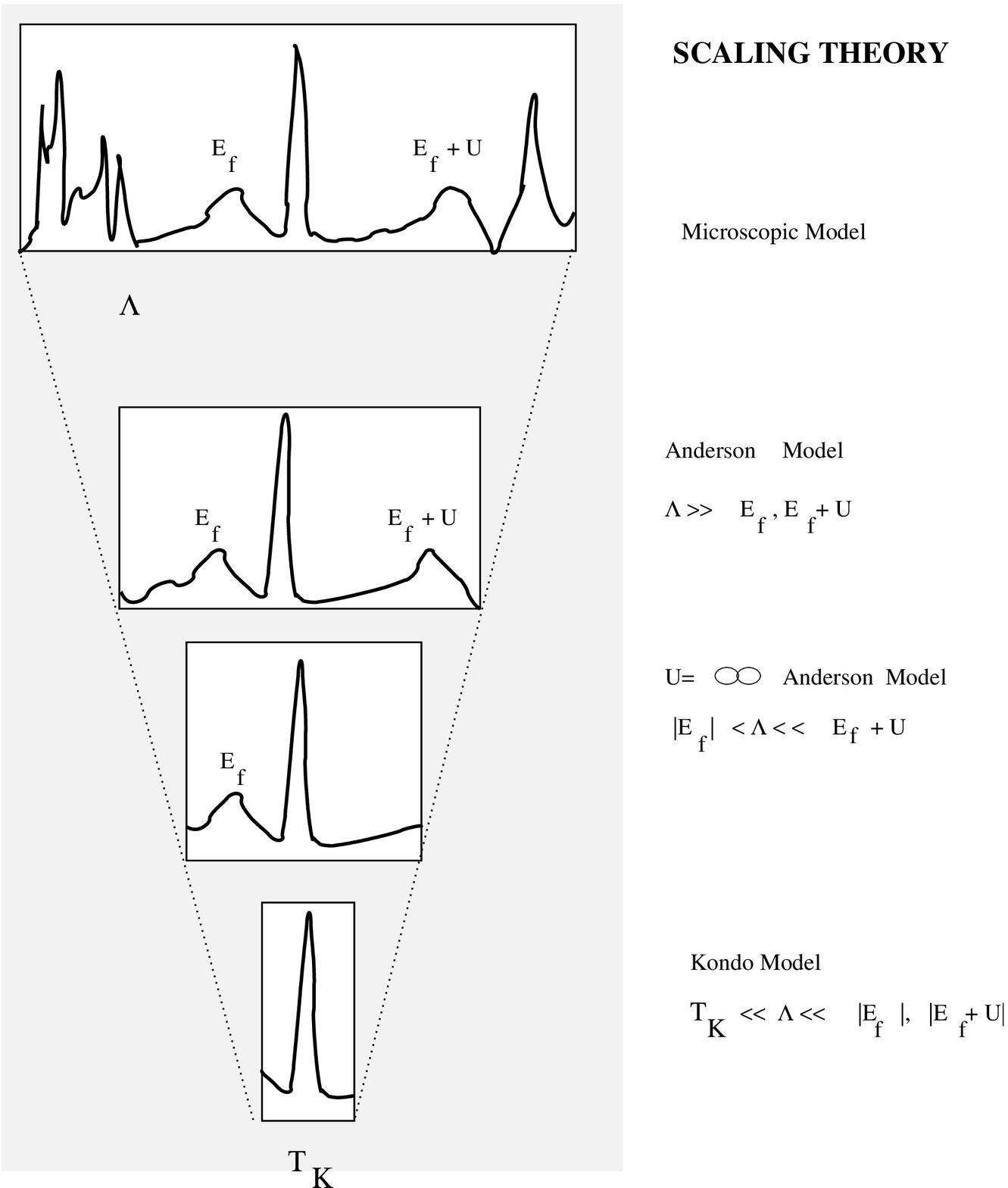}{Scaling concept. Low energy model Hamiltonians are
obtained from the detailed original model by integrating out
the high energy degrees of freedom. At each stage, the physics
described by the model spans a successively lower frequency window 
in the excitation spectrum.}{fig9}
leads to a family of Hamiltonians $H (D)$. By taking
the  limit  $b\rightarrow 1$, these Hamiltonians evolve
continuously with $D$. 
Typically, $H$ will contain a series of
dimensionless 
coupling constants $\{g_{i} \}$ which denote the strength of various interaction
terms in the Hamiltonian. The evolution of these 
coupling constants with cut-off is given by a scaling equation, so
that for the simplest case
\[
\frac{\partial g_{j}}{\partial \ln  D} = \beta_{j} (\{g_{i} \})
\]
A negative $\beta $ function denotes a ``\underline{relevant}''
coupling constant which grows as the cut-off is reduced. 
A positive $\beta $ function denotes  an 
``\underline{irrelevant}'' coupling constant which diminishes as the
cut-off is reduced. 
There are two types of event that can occur in such a scaling
procedure (Fig.~\ref{fig10}):

\begin{itemize}
\item{i)} A {\sl crossover}. When the cut-off energy scale $D$
becomes smaller than the characteristic energy scale of a
particular class of high frequency excitations, 
then at lower energies, these excitations may only occur
via a virtual process. To accommodate this change, the Hamiltonian
changes its structure, acquiring additional terms that simulate the
effect of the high frequency virtual fluctuations on the low energy 
physics. The passage from the Anderson to the Kondo model is
an example of one such  cross-over.  In the renormalization group
treatment of the Anderson model,  when the band-width of the
conduction electrons becomes smaller than the energy to produce a valence
fluctuation, a cross-over takes place in which real charge fluctuations
are eliminated, and the physics at all lower energy scales is described
by the Kondo model.  

\item{ii)} {\sl Fixed Point}. If the cut-off energy scale drops
below the lowest energy scale in the problem, then there are no
further changes to occur in the Hamiltonian, which will now remain
invariant under 
the scaling procedure (so that the $\beta $ function of all remaining
parameters in the Hamiltonian must vanish). 
This {\sl ``Fixed Point Hamiltonian''}
describes the essence of the low energy physics.
\end{itemize}

\subsection{``Poor Man'' Scaling}

We shall now apply the scaling concept to the Kondo model. This was
originally carried out by Anderson and Yuval using a method formulated
in the time, rather than energy domain. The method presented here
follows Anderson's `` Poor Man's'' scaling approach, in which 
the evolution of the coupling constant is followed as the band-width
of the conduction sea is reduced. The 
Kondo model is written
\begin{eqnarray}\label{kondomodel3}
H&=& \sum_{\vert \epsilon_{k}\vert <D }\epsilon _{k}c\dg _{k\sigma
}c_{k\sigma } +H^{(I)}\cr
H^{(I)}&=& J (D)\sum_{\vert \epsilon_{k}\vert,\vert \epsilon_{k'}\vert  <D}
c\dg _{k\alpha }\vec{\sigma}_{\alpha \beta }
c_{k'\beta }\cdot \vec{S}_{d}
\end{eqnarray}
where the density of conduction electron states $\rho (\epsilon ) $ 
is taken  to
be constant. 
The Poor Man's renormalization procedure 
follows the evolution of $J (D)$ that results from reducing $D$ by
progressively
integrating out the electron states at the edge of the conduction band.
In the Poor Man's procedure, the band-width is not rescaled to its
original size after each renormalization, which avoids the need
to renormalize the electron operators so that 
instead of Eq. (\ref{rescale}), $H (D')=\tilde{H_{L}}$. 

To carry out the renormalization procedure, we integrate out the
high-energy spin fluctuations using the t-matrix formulation
for the induced interaction $H_{int}$, derived in the last section.
Formally, the induced interaction is given by
\[
\delta H^{int}_{ab} = \frac{1}{2}[T_{ab} (E_{a})+T_{ab} (E_{b})]
\]
where
\[
T_{ab} (E) = \sum_{\lambda \in \vert H\rangle }\left[\frac{H^{(I)}_{a\lambda
}H^{(I)}_{\lambda b}}{E-E_{\lambda }^{H}}\right] 
\]
where the energy of state $\vert \lambda\rangle $ lies in the range
$[D',D]$.  There are two possible intermediate states that can be 
produced by the action of $H^{(I)}$ on a one-electron state: (I) either
the electron state is scattered directly, or (II) a virtual electron hole-pair
is created in the intermediate state.  
In process (I), the T-matrix can be represented by the Feynman diagram\\

\centerline {\frm{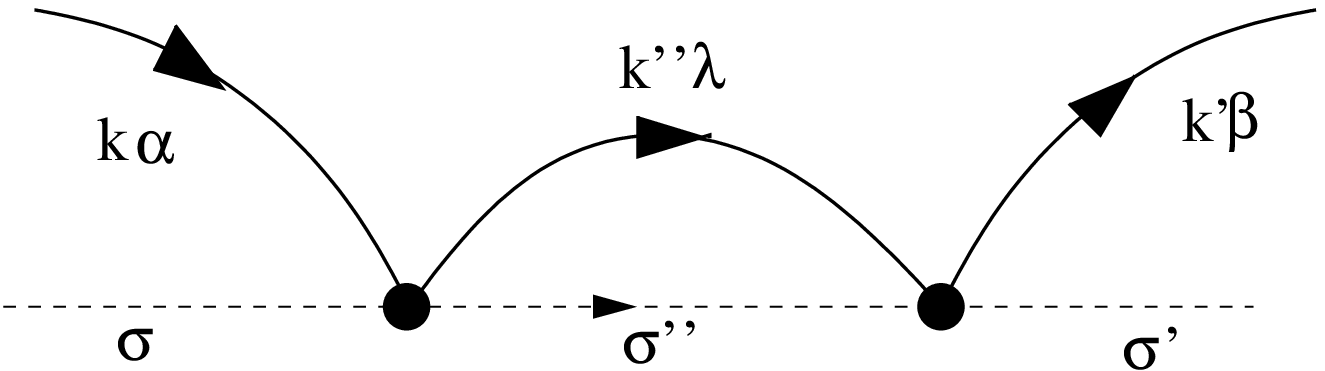}}

\noindent 
for  which the T-matrix for scattering into a high energy electron
state is
\begin{eqnarray}\label{proca}
T^{(I)} (E)_{k'\beta \sigma' ;k\alpha \sigma } &=& 
\sum_{\epsilon _{k''} \in [D-\delta D,D]}
\left[\frac{1}{E-\epsilon _{k''}}
\right]
J^{2 }(\sigma ^{a}\sigma ^{b})_{\beta \alpha }
(S^{a}S^{b})_{\sigma' \sigma }\cr
&\approx &J^{2}\rho  \delta D
\left[\frac{1}{E-D}
\right]
(\sigma ^{a}\sigma ^{b})_{\beta \alpha }
(S^{a}S^{b})_{\sigma' \sigma }
\end{eqnarray}
In process (II), \\

\centerline{\frm{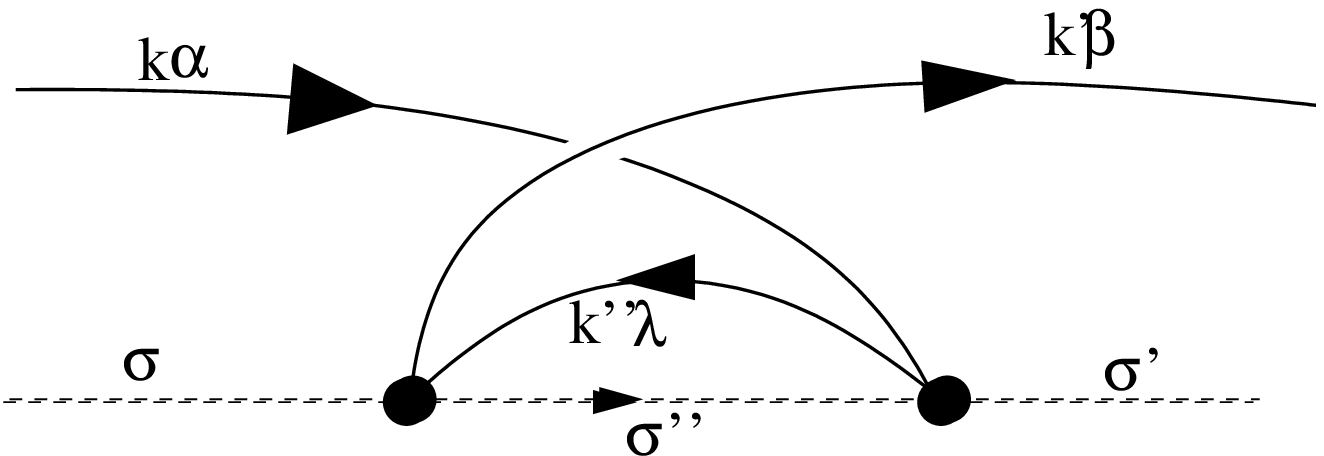}}

\noindent 
the formation of a particle-hole pair involves a
conduction electron line that crosses itself, leading to a negative
sign. 
Notice how the spin operators of the conduction sea and antiferromagnet
reverse their relative order in process II, so that the T-matrix
for scattering into a high-energy hole-state is given by
\begin{eqnarray}\label{procb}
T^{(II)} (E)_{k'\beta \sigma' ;k\alpha \sigma } &=& 
-\sum_{\epsilon _{k''} \in [-D,-D+\delta D]}
\left[\frac{1}{E- (\epsilon _{k}+\epsilon_{k'}-\epsilon _{k''})}
\right]
J^{2 }(\sigma ^{b}\sigma ^{a})_{\beta \alpha }
(S^{a}S^{b})_{\sigma' \sigma }\cr
&=&-J^{2}\rho  \delta D
\left[\frac{1}{E-D}
\right]
(\sigma ^{a}\sigma ^{b})_{\beta \alpha }
(S^{a}S^{b})_{\sigma' \sigma }
\end{eqnarray}
where we have assumed that 
the energies $\epsilon _{k}$ and $\epsilon _{k'}$  are negligible
compared with $D$. 
Adding (Eq. \ref{proca}) and
(Eq. \ref{procb}) gives 
\begin{eqnarray}\label{}
\delta H^{int}_{k'\beta \sigma';k\alpha \sigma }&=& \hat T^{I} + T^{II}= 
-\frac{J^{2}\rho \delta D}{D}
 [\sigma ^{a}, \sigma ^{b}]_{\beta \alpha } S^{a}S^{b}\cr
&=&\frac{J^{2}\rho \delta D}{D}
\vec{ \sigma }_{\beta \alpha }\vec{ S}_{\sigma
'\sigma }.
\end{eqnarray}
In this way we see that the virtual emission of a high
energy electron and hole generates an antiferromagnetic correction
to the original Kondo coupling constant
\[
J (D')= J (D) + 2 J^{2}\rho \frac{\delta D}{D}
\]
High frequency spin fluctuations  thus \underline{{\sl antiscreen}}
the antiferrromagnetic interaction. 
If we introduce the coupling constant $g = \rho J$, we see that it
satisfies
\[
\frac{\partial g}{\partial \ln  D} =\beta (g)= - 2 g^{2} + O (g^{3}).
\]
This is an example of a \underline{negative} $\beta $ function:  
a signature of an interaction which is weak at high frequencies,
but which grows as the energy scale is reduced. The local moment
coupled to the conduction sea is said to be \underline{asymptotically
free}.
The solution to this scaling equation is
\begin{equation}\label{}
g (D')= \frac{g_{o}}{1 - 2 g_{o} \ln (D/D')}
\end{equation}
and if we introduce the scale
\begin{equation}\label{}
T_{K} = D\exp\left[-\frac{1}{2 g_{o} } \right]
\end{equation}
we see that this can be written
\[
2g (D')= \frac{1}{\ln  (D'/T_{K})}
\]
This is an example of a running coupling constant- a coupling constant
whose strength depends on the scale at which it is measured. 
(See Fig. \ref{fig10}). 

\fight=0.8\textwidth
\fg{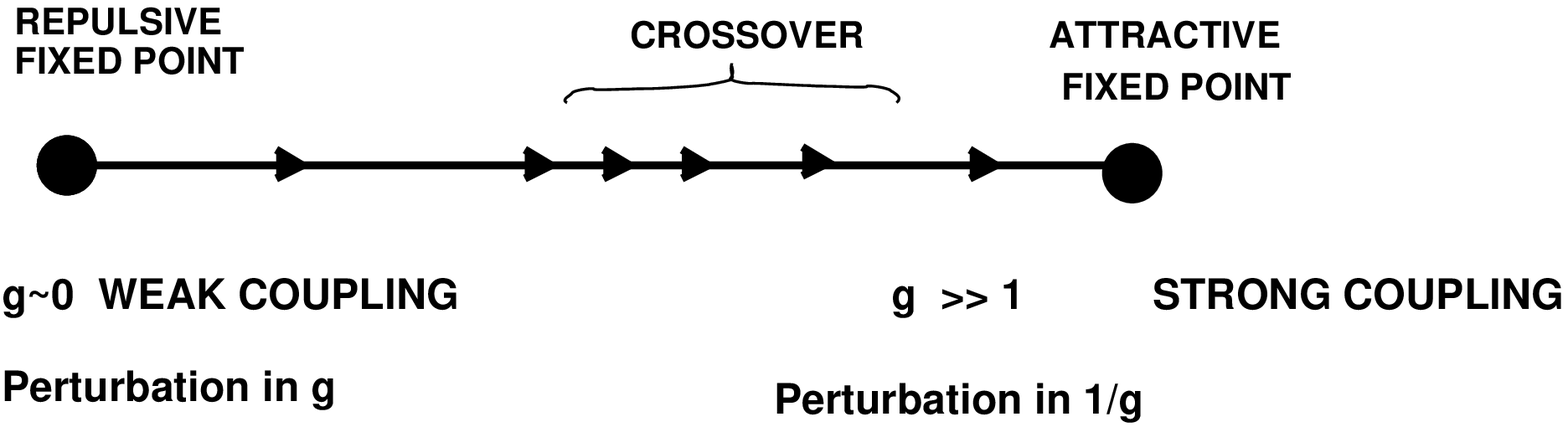}{Schematic illustration
of renormalization group flow from
a repulsive ``weak coupling'' fixed point, via a crossover to
an attractive ``strong coupling'' fixed point.
}{fig10}

Were we to take this equation literally, we would say that $g$
diverges at the scale $D'=T_{K}$. This  interpretation is too
literal, because the above scaling equation has only been calculated
to order $g^{2}$, nevertheless, this result does show us that
the Kondo interaction can only be treated perturbatively at energy
scales large compared with the Kondo temperature.  We also see that
once we have written the coupling constant in terms of the Kondo
temperature, all reference to the original cut-off energy scale vanishes
from the expression. This cut-off independence of the problem
is an indication that the physics of the Kondo problem does not depend
on the high energy details of the model: there is only one relevant
energy scale, the Kondo temperature. 

It is possible  to extend the above leading order renormalization calculation
to higher order in $g$. To do this requires a more systematic method
of calculating higher order scattering effects. One tool that is particularly
useful in this respect, is to use the Abrikosov pseudo-fermion representation
of the spin, writing
\begin{eqnarray}\label{}
\vec{ S}&=& d\dg _{\alpha }\left(\frac{\vec{\sigma }}{2} \right)_{\alpha \beta
}d_{\beta }\cr
n_{d}&=&1.
\end{eqnarray}
This has the advantage that the spin operator, which does not satisfy
Wick's theorem, is now factorized in terms of conventional fermions. 
Unfortunately, the second constraint  is required to enforce the condition that
$S^{2}=3/4$. This constraint proves very awkward for the development of
a Feynman diagram approach. One way around this problem, is to use the
Popov trick, whereby the d-electron is associated with a complex
chemical potential
\[
\mu = - i \pi \frac{T}{2}
\]
The partition function of the Hamiltonian is written as an unconstrained
trace over the conduction and pseudofermion Fock spaces, 
\begin{equation}\label{}
Z= \hbox{Tr}\left[e^{-\beta (H +i \pi \frac{T}{2} ( n_{d}-1))} \right]
\end{equation}
Now since the Hamiltonian conserves $n_{d}$, we can divide this
trace up into contributions from the $d^{0}$, $d^{1}$ and $d^{2}$
subspaces, as follows:
\[
Z= e^{i \pi /2} Z (d^{0})+ Z (d^{1})+ e^{-i\pi /2}
Z (d^{2})
\]
But since $S_{d}=0$ in the $d^{2}$and $d^{0}$ subspaces, $Z (d^{0})= Z
(d^{2})$
so that the contributions to the partition function from these two
unwanted subspaces exactly cancel. You can test this method
by applying it to a free spin in a magnetic field. (see exercise)

\fight=0.8\textwidth
\fg{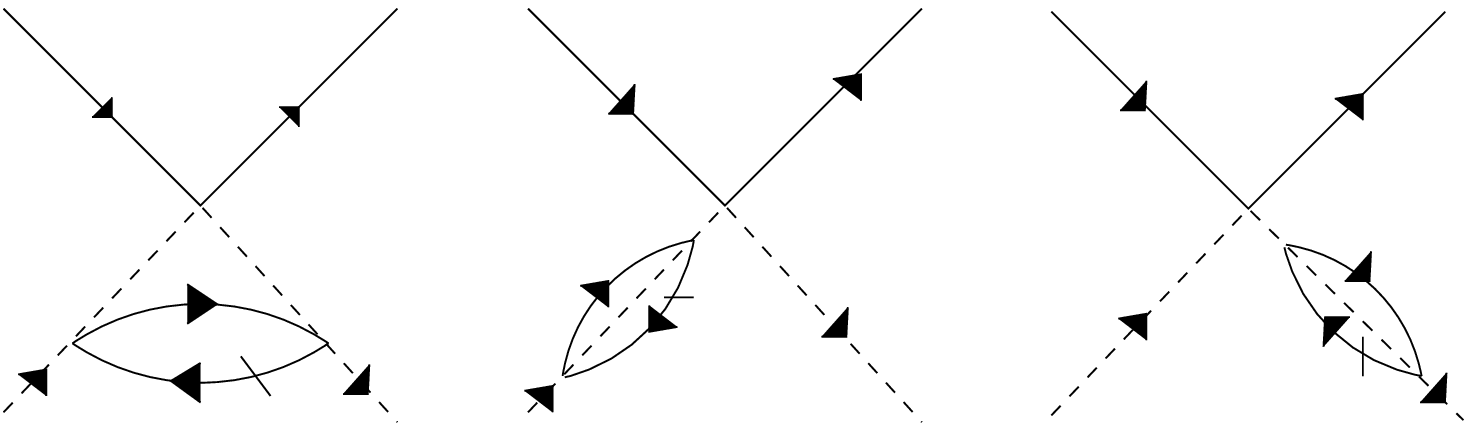}{Diagrams contributing to the third-order
term in the beta function. A ``crossed'' propagator line indicates
that the contribution from high-energy electrons with energies 
$\vert \epsilon _{k}\vert \in [D-\delta D,D]$ is taken from
this line.
}{figz}

By calculating the higher order diagrams shown in fig \ref{figz} , 
it is straightforward, though laborious to show that 
the beta-function to order $g^{3}$ is given by
\begin{equation}\label{}
\frac{\partial g}{\partial \ln  D} =\beta (g)
= - 2 g^{2} + 2 g^{3} +O (g^{4})
\end{equation}
One can integrate this equation to obtain
\[
\ln \left(\frac{D'}{D} \right)= \int_{g_{o}}^{g}\frac{dg'}{\beta (g')}
= -\frac{1}{2}\int_{g_{o}}^{g}dg\left[\frac{1}{g'^{2}} +\frac{1}{g'}+O (1)\right]
\]
A better estimate of the temperature $T_{K}$ where the system scales
to strong  coupling is obtained by setting $D'=T_{K}$ and $g=1$ in
this equation, which gives
\begin{eqnarray}\label{better}
\ln \left(\frac{T_{K}}{\tilde{D}} \right)= - \frac{1}{2g_{o}} +
\frac{1}{2}\ln  2 g_{o}+O (1) ,
\end{eqnarray}
where for convenience, we have absorbed a factor $\sqrt{\frac{e}{2}}$
into the cut-off, writing $\tilde{D}=D \sqrt{\frac{e}{2}}$.
Thus, 
\begin{equation}\label{}
T_{K}= \tilde{D} \sqrt{2g_{o}} e^{-\frac{1}{2g_{o}}}
\end{equation}
up to a constant factor.
The square-root pre-factor in $T_{K}$ is often dropped in qualitative
discussion, but it is important for more quantitative comparison. 

\subsection{Universality and the resistance minimum}

Provided the Kondo temperature is far smaller than the cut-off, then
at low energies it is the only scale governing the physics of the
Kondo effect.  For this reason, we expect all physical quantities
to  be expressed in terms of universal functions involving the ratio
of the temperature or field to the Kondo scale.  For example, the 
the susceptibility
\begin{equation}\label{}
\chi (T) = \frac{1}{4T}F (\frac{T}{T_{K}}),
\end{equation}
and the resistance
\begin{equation}\label{}
\frac{1}{\tau }(T)= \frac{1}{\tau _{o}} {\cal G} (\frac{T}{T_{K}})
\end{equation}
both display universal behavior. 

We can confirm the existence of universality by examining these
properties in the weak coupling limit, where $T>>T_{K}$. 
Here, we find 
\begin{eqnarray*}
\frac{1}{\tau } (T) &=& 2 \pi J^{2} \rho S (S+1)n_{i},\qquad (S=\frac{1}{2})\\
\chi (T) &=& \frac{n_{i}}{4T}\left[1 - 2J\rho \right]
\end{eqnarray*}
where $n_{i}$ is the density of impurities.
Scaling implies that at lower temperatures $J\rho \rightarrow J\rho +
2 (J\rho )^{2}\ln  \frac{D}{T}$, so that to next leading order we expect
\begin{eqnarray}\label{hello}
\frac{1}{\tau } (T) &=& n_{i}\frac{2\pi }{\rho } S (S+1)[J\rho + 2 (J\rho
)^{2 } \ln \frac{D}{T}]^{2},\\
\label{hello2}
\chi (T) &=& \frac{n_{i}}{4T}\left[
1 - 2J\rho 
- 4 (J\rho )^{2}\ln
\frac{D}{T} + O ((J\rho )^{3})\right ]
\end{eqnarray}
results that are confirmed from second-order perturbation theory.
The first result was obtained by Jun Kondo. Kondo was looking for a
consequence of the antiferromagnetic interaction predicted by the Anderson
model, so he computed the electron scattering rate to third order in
the magnetic coupling. The logarithm which appears in the electron scattering
rate means that as the temperature is lowered, the rate at which
electrons scatter off magnetic impurities rises. It is this phenomenon
that gives rise to the famous Kondo ``resistance minimum'' . 

Since we know the form of $T_{K}$, we can use this result to
deduce that the weak coupling limit of the scaling forms.
If we take  equation (\ref{better}),  and replace 
the cut-off by the temperature $D\rightarrow T$, and replace $g_{o}$
by the running coupling constant
$g_{o}\rightarrow g (T)$, we obtain
\begin{equation}\label{wkcoup}
g (T)= \frac{1}{2\ln \left(\frac{T}{T_{K}
} \right)+ \ln  2 g (T)
} 
\end{equation}
which we may iterate to obtain
\begin{equation}\label{wkcoup}
2g (T)= \frac{1}{\ln \left(\frac{T}{T_{K}} \right)
} +\frac{\ln (\ln (T/T_{K})) }{2\ln ^{2}\left(\frac{T}{T_{K}} \right)}.
\end{equation}
Using this expression to make  the replacement 
$J\rho \rightarrow g (T)$ in (\ref{hello}) 
and (\ref{hello2}), we obtain
\begin{eqnarray}\label{}
\chi (T) &= &\frac{n_{i}}{4T}\left[1 -\frac{1}{\ln (T/T_{K})}-
\frac{1}{2}\frac{\ln (\ln (T/T_{K}))}{\ln ^{2} (T/T_{K})} + \dots
\right]\\
\frac{1}{\tau } (T)&=& n_{i}\frac{ \pi S (S+1)}{2\rho }\left[ \frac{1}{\ln ^{2}(T/T_{K})}+
\frac{\ln (\ln (T/T_{K}))}{\ln ^{3} (T/T_{K})} + \dots\right]
\end{eqnarray}
From the second result, we see that the electron scattering rate
has the scale-invariant form
\begin{equation}\label{}
\frac{1}{\tau } (T) = \frac{n_{i}}{\rho } {\cal G} (T/T_{K}). 
\end{equation}
where ${\cal  G} (x)$ is a universal function.
The pre-factor in the electron scattering rate is essentially the
Fermi energy of the electron gas: it is the ``unitary scattering'' rate,
the maximum possible scattering rate that is obtained when an
electron experiences a resonant $\pi /2$ scattering phase shift.
From this result, we see that at absolute zero, the electron
scattering rate  will rise to the value 
$\frac{1}{\tau } (T) = \frac{n_{i}}{\rho }{\cal  G} (0)$, indicating
that at strong coupling, the scattering rate is of the same
order as the unitary scattering limit.  We shall now see how
this same result comes naturally out of a strong coupling analysis.

\subsection{Strong Coupling: Nozi{\`e}res Fermi Liquid Picture of the Kondo Ground-state}\label{}

The weak-coupling analysis tells us that at scales of order the Kondo
temperature, the Kondo coupling constant $g$ scales to a value of
order $O (1)$. Although perturbative renormalization group methods can  not
go past this point, Anderson and Yuval pointed out that it is not
unreasonable
to suppose that the Kondo coupling constant scales to a fixed point
where it is large compared to the conduction electron band-width $D$.
This assumption is the simplest possibility and if true, 
it means that the strong-coupling limit is an attractive fixed point,
being  stable under the renormalization group. Anderson and Yuval
conjectured that the Kondo singlet would be paramagnetic, with a temperature
independent magnetic susceptibility and a universal 
linear specific
heat  
given by $C_{V}= \gamma_{K}\frac{T}{T_{K}}
$
at low temperatures.

The first controlled treatment of this cross-over regime was
carried out by Wilson using a numerical renormalization group method.
Wilson's numerical renormalization method was able to confirm the
conjectured renormalization 
of the Kondo coupling constant to infinity. 
This limit is called the ``strong coupling'' limit of the Kondo
problem. Wilson carried out an analysis of the strong-coupling limit,
and
was able to show that the specific heat would be a linear
function of temperature, like a Fermi liquid.
Wilson showed that the linear specific heat could be written in a
universal form 
\begin{eqnarray}\label{}
C_{V}&=& \gamma T, \qquad\cr
\gamma &=&\frac{\pi^{2}}{3}\frac{0.4128 \pm 0.002}{8T_{K}}
\end{eqnarray}
Wilson also compared the 
ratio between the magnetic susceptibility and the linear
specific heat with the corresponding value in a 
non-interacting system, computing 
\begin{equation}\label{}
W = \frac{\chi /\chi ^{0}}{\gamma/\gamma^{0}}= \frac{\chi }{\gamma}
\left(\frac{\pi^{2}k_{B}^{2}}{3 (\mu _{B})^{2}}
 \right)
=2
\end{equation}
within the accuracy of the numerical calculation. 

Remarkably, the second result of Wilson's can be re-derived
using an exceptionally elegant set of  arguments due 
to Nozi{\`e}res\cite{Noz} that leads to an 
explicit form for the 
strong coupling fixed point Hamiltonian. 
Nozi{\`e}res began by considering an electron in a one-dimensional chain
as illustrated in Fig.~\ref{fig11}. The Hamiltonian for this situation is
\fg{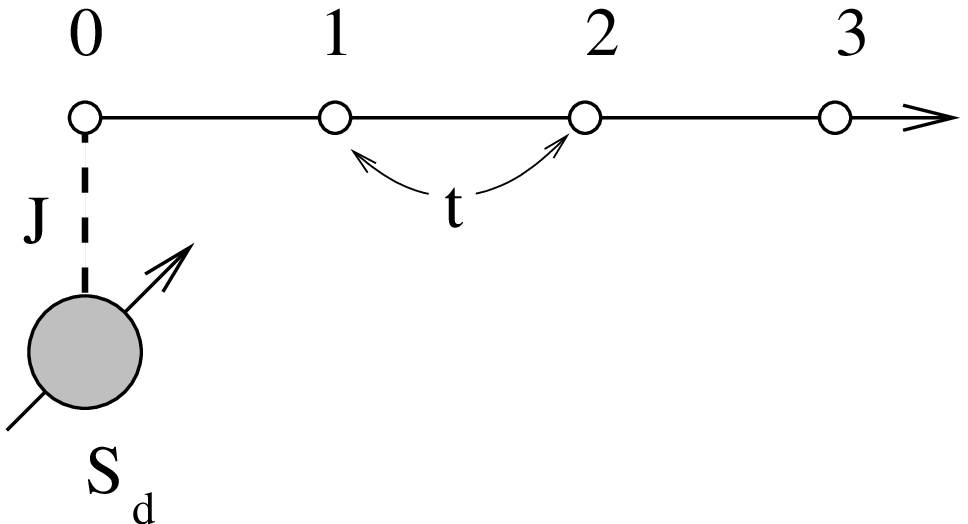}{Illustrating the strong-coupling limit of the Kondo model}{fig11}
\begin{equation}\label{}
H_{lattice} = - t \sum_{j=0,\infty }[c\dg_{\sigma } (j+1)c_{\sigma }
(j)+ \hbox{H.c}] + J c\dg _{\alpha } (0)\vec{\sigma }_{\alpha \beta
}c_{\beta } (0)
\cdot \vec{S_{d}}.
\end{equation}
Nozi{\`e}res argued that the strong coupling fixed point
will be described by the situation $J>>t$. In this limit, 
the kinetic energy of the electrons in the band can be treated
as  a perturbation to the Kondo singlet. 
The local moment couples to 
an electron at the origin, forming  a ``Kondo singlet'' 
denoted by
\begin{equation}\label{}
\vert GS\rangle = \frac{1}{\sqrt{2}}\left( \vert \Uparrow \downarrow
\rangle - \vert \Downarrow \uparrow \rangle 
\right)
\end{equation}
where the thick arrow refers to the spin state of the local moment
and the thin arrow refers to the spin state of the electron at
site $0$. 
Any electron which migrates from site $1$ to site $0$
will automatically break this singlet state, raising its energy
by $3J/4$. This will have the effect of \underline{excluding}
electrons (or holes) from the origin. The fixed point Hamiltonian must
then take the form 
\begin{equation}\label{}
H_{lattice} = - t \sum_{j=1,\infty }[c\dg_{\sigma } (j+1)c_{\sigma }
(j)+ \hbox{H.c}] + \hbox{weak interaction}
\end{equation}
where the second-term refers to the weak-interactions induced in
the conduction sea by virtual fluctuations onto site $0$. 
If the wavefunction of electrons far from the impurity
has the form $\psi (x)\sim \sin ( k_{F} x)$, where $k_{F}$ is the
Fermi momentum, then the exclusion
of electrons from site $1$ has the effect
of phase-shifting the electron wavefunctions  by one
the lattice spacing $a$, so that now $\psi (x)\sim \sin (k_{F}x-\delta  )$
where $\delta =k_{F}a$.  But if there is one electron per site,
then $2 ( 2k_{F}a/ (2\pi ))= 1$ by the Luttinger sum rule, so that
$k_{F}=\pi / (2a)$ and hence  the Kondo singlet acts as a
spinless, elastic scattering center with scattering phase shift
\begin{equation}\label{}
\delta = \pi /2.
\end{equation}
The appearance
of $\delta =\pi/2$  could also be deduced by appealing to
the Friedel sum rule, which states that the number of bound-electrons
at the magnetic impurity site is $\sum_{\sigma }
\frac{\delta _{\sigma=\pm 1 }}{\pi }= 2 \delta /\pi $, so that $\delta = \pi
/2$. By  considering
virtual fluctuations of electrons between site $1$ and $0$, 
Nozi{\`e}res argued that the induced interaction at site
$1$ must take the form
\begin{equation}\label{}
H_{int} \sim \frac{t^{4}}{J^{3}} n_{1\uparrow}n_{1\downarrow }
\end{equation}
because fourth order hopping processes lower the energy of the
singly occupied state, but they do not occur for the doubly
occupied state. This is a repulsive interaction amongst
the conduction electrons, and it is known to be a marginal
operator  under the renormalization group, leading to the conclusion
that the effective Hamiltonian describes a weakly interacting
``local'' Fermi  liquid. 

Nozi{\`e}res formulated this local Fermi liquid in the language
of an occupancy-dependent phase shift. Suppose the $k\sigma $ 
scattering state has occupancy $n_{k\sigma }$, then the 
the ground-state energy will be a functional of these occupancies
$E[\{n_{k\sigma } \}]$. The differential of this quantity with respect
to occupancies defines a {\sl phase shift} as follows
\begin{equation}\label{}
\frac{\delta E}{\delta n_{k\sigma }}= \epsilon _{k}- \frac{\Delta
\epsilon }{\pi }{\delta (\{ n_{k'\sigma '}\},\epsilon _{k})}
.
\end{equation}
The first term is just the energy of an unscattered conduction
electron, while $\delta (\{ n_{k'\sigma '}\},\epsilon _{k})$ is
the scattering phase shift of the Fermi liquid. This phase shift
can be expanded
\begin{equation}\label{}
{\delta (\{ n_{k'\sigma '}\},\epsilon _{k})}= \frac{\pi }{2} + 
\alpha (\epsilon_{k}-\mu ) + \Phi \sum_{k}\delta n_{k,-\sigma }
\end{equation}
where the term with coefficient $\Phi $ describes the interaction
between opposite spin states of the Fermi liquid.  
Nozi{\`e}res  argued that when the chemical potential
of the conduction sea is changed, the occupancy of the 
localized $d$ state will not change,   which implies that
the phase shift is invariant under changes in $\mu $. 
Now under a shift $\delta \mu $, the change in the 
occupancy $\sum_{k}
\delta n_{k\sigma } \rightarrow \delta \mu \rho $, so that changing
the chemical potential modifies the phase shift by an amount
\begin{equation}\label{}
\Delta \delta = (\alpha + \Phi \rho )\Delta \mu =0
\end{equation}
so that 
$\alpha = - \rho \Phi$. We are now in a position to calculate
the impurity contribution to the magnetic susceptibility and specific heat.
First note that the density of quasiparticle states is given by
\begin{equation}\label{}
\rho = \frac{dN}{dE}= \rho_{o} + \frac{1}{\pi}\frac{\partial \delta
}{\partial \epsilon }= \rho _{o} + \frac{\alpha }{\pi }
\end{equation}
so that the low temperature specific heat is given by  $C_{V}=
(\gamma_{bulk}+\gamma _{i})$ where
\begin{equation}\label{}
\gamma_{i} = 2 \left(\frac{\pi ^{2}k_{B}^{2}}{3} \right)\frac{\alpha
}{\pi }
\end{equation}
where the prefactor ``$2$'' is derived from the spin up and spin-down
bands.
Now in a magnetic field, the impurity magnetization is  given by
\begin{equation}\label{}
M= \frac{\delta _{\uparrow}}{\pi } - \frac{\delta _{\downarrow}}{\pi }  
\end{equation}
Since the Fermi energies of the up and down quasiparticles are shifted to 
$\epsilon _{F\sigma }\rightarrow \epsilon _{F}-\sigma B$, we have
$\sum_{k}\delta
n_{k\sigma }= \sigma \rho B$, 
so that the phase-shift at the Fermi
surface in the up and down scattering channels becomes
\begin{eqnarray}\label{}
\delta _{\sigma }&=& \frac{\pi }{2}+ \alpha \delta \epsilon _{F\sigma } + \Phi (\sum_{k}\delta
n_{k\sigma }\cr
&=& \frac{\pi }{2}+ \alpha \sigma B - \Phi \rho \sigma B\cr
&=& \frac{\pi }{2} + 2 \alpha \sigma B
\end{eqnarray}
so that the presence of the interaction term \underline{doubles} the size
of the change in the phase shift due to a magnetic field. The impurity
magnetization then becomes
\begin{equation}\label{}
M_{i} = \chi_{i} B = 2 \left(\frac{2\alpha }{\pi } \right)\mu _{B}^{2}
 B 
\end{equation}
where we have reinstated the magnetic moment of the electron. This is
twice the value expected for a ``rigid'' resonance, and it means that
the Wilson ratio is
\begin{equation}\label{}
W = 
\frac{\chi_{i} \pi^{2}k_{B}^{2}}{\gamma_{i} 3 (\mu _{B})^{2}}=2
\end{equation}

\subsection{Experimental observation of Kondo effect in real materials and
quantum dots }

Experimentally, there is now a wealth of observations that confirm our
understanding of the single impurity Kondo effect.  Here is a brief
itemization of some of the most important observations. (Fig. \ref{fig11b}.)
\fight=4truein
\fg{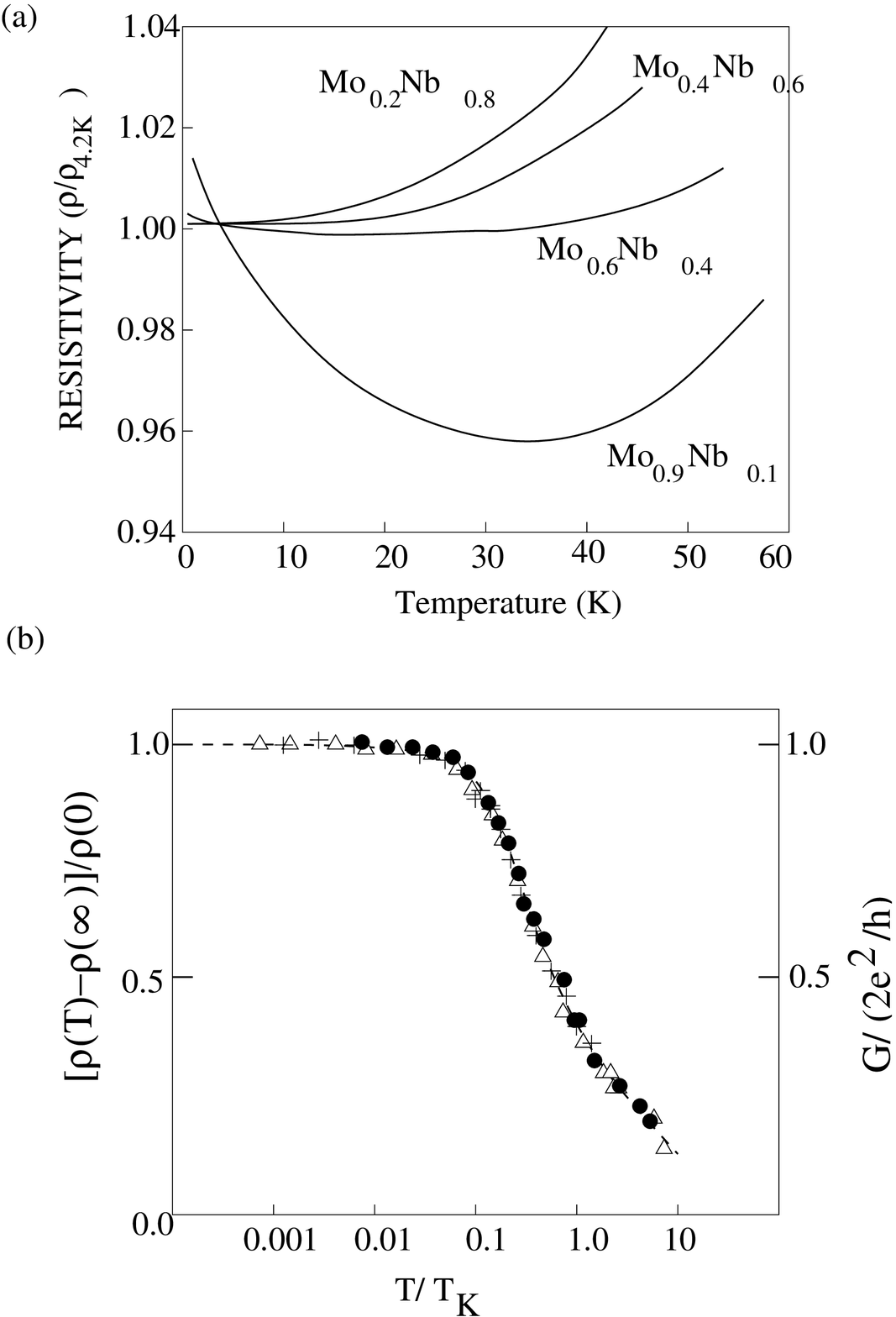}{(a) Sketch of resistance minimum in $Mo_{x}Nb_{1-x}$ (b)
Sketch of excess resistivity associated with scattering from an
impurity spin. Right hand-scale- differential conductivity of a
quantum dot. 
}{fig11b}
\begin{itemize}

\item A resistance minimum appears when local moments develop in a
material. For example, in $Nb_{1-x}Mo_{x}$ alloys, a local moment
develops for $x>0.4$, and the resistance is seen to develop a minimum beyond
this point.\cite{sarachik64,clogston62}

\item Universality seen in the specific heat $C_{V}= \frac{n_{i}}{T}F
(T/T_{K})$ of metals doped with dilute concentrations of impurities.  
Thus the specific heat of  $\underline{Cu}-Fe$ 
(iron impurities in copper)  can be superimposed on the specific heat
of $\underline{Cu}-Cr$, with a suitable rescaling of the temperature
scale. \cite{white79,triplett70}

\item Universality is observed 
in the differential conductance of quantum dots\cite{Cronenwett:1998,vanderWiel:2000} and
spin-fluctuation resistivity of metals  with a dilute concentration of
impurities.\cite{hedgcock63}  
Actually, both properties are dependent
on the same thermal average of the imaginary part of the scattering
T-matrix 
\begin{eqnarray}\label{}
\rho_{i} &=& n_{i}\frac{ne^{2}}{m} \int d\omega 
\left(- \frac{\partial f}{\partial
\omega } \right)
2 Im [T (\omega )] \cr
G&=& \frac{2e^{2}}{\hbar } \int d\omega \left(- \frac{\partial f}{\partial
\omega } \right)
 \pi \rho  Im [T (\omega )] .
\end{eqnarray}
Putting $\pi \rho \int d\omega \left(- \frac{\partial f}{\partial
\omega } \right)
\hbox{Im} T (\omega )= t (\omega /T_{K},
T/T_{K})$, we see that both properties have the form
\begin{eqnarray}\label{}
\rho_{i} &=& n_{i}\frac{2 n e^{2}}{\pi m \rho } t (T/T_{K})\cr
G &=& \frac{2 e^{2}}{\hbar } t (T/T_{K})
\end{eqnarray}
where $t (T/T_{K})$ is a universal function. 
This result is born out by experiment. 
\end{itemize}
\subsection{Exercises }

\begin{problems}
\item  Generalize the scaling equations to the anisotropic Kondo model
with an anisotropic interaction
\begin{equation}\label{kondomodel3}
H_{I}= \sum_{\vert \epsilon_{k}\vert,\vert \epsilon_{k'}, a= (x,y,z)}
J^{a} c\dg _{k\alpha }{\sigma}^{a}_{\alpha \beta }
c_{k'\beta }\cdot {S^{a}}_{d}
\end{equation}
and show that the scaling equations take the form
\[
\frac{\partial J_{a}}{\partial \ln  D} = - 2 J_{b}J_{c}\rho  + O (J^{3}),
\]
where and $(a,b,c) $ are a cyclic permutation of
$(x,y,z)$. Show that in the special case where
$J_{x}=J_{y}=J_{\perp}$, the scaling equations become
\begin{eqnarray}\label{}
\frac{\partial J_{\perp }}{\partial \ln  D} 
&=& - 2 J_{z}J_{\perp }\rho  + O (J^{3}),\cr
\frac{\partial J_{z }}{\partial \ln  D} &=& - 2 ( J_{z})^{2}\rho  + O (J^{3}),
\end{eqnarray}
so that $J_{z}^{2}- J_{\perp }^{2}=\hbox{constant}$. Draw the corresponding
scaling diagram.

\item Consider the symmetric Anderson model, with  a symmetric  band-structure
at half filling. In this model, the $d^{0}$ and $d^{2}$ states are
degenerate and there is the possibility of a ``charged Kondo effect''
when the interaction $U$ is negative. 
Show that
under the ``particle-hole'' transformation
\begin{eqnarray}\label{}
c_{k\uparrow}&\rightarrow& c_{k\uparrow}, \qquad
d_{\uparrow}\rightarrow d_{\uparrow}\cr
c_{k\downarrow}&\rightarrow&- c\dg _{k\downarrow}, \qquad
d_{\downarrow}\rightarrow -d\dg _{\downarrow}
\end{eqnarray}
the positive $U$ model is transformed to the negative $U$ model.
Show that the spin operators of the local moment are transformed into
Nambu ``isospin operators''  which describe the charge and pair
degrees of freedom of the d-state. 
Use this transformation 
to argue that when U is negative, a charged Kondo effect will occur
at exactly half-filling involving quantum fluctuations between the
degenerate $d^{0}$ and $d^{2}$ configurations.

\item 
What happens to the Schrieffer-Wolff transformation in the infinite
U limit? Rederive the Schrieffer-Wolff transformation for an N-fold
degenerate version of the infinite U Anderson model.  This is actually valid
for Ce and Yb ions.

\item 
Rederive the Nozi{\`e}res Fermi liquid picture for an SU (N) degenerate
Kondo model. Explain why this picture is relevant for magnetic rare earth ions
such as $Ce^{3+}$ or $Yb^{3+}$. 

\item 
Check the Popov trick works for a magnetic moment in an external
field. Derive the partition function for a spin in a magnetic field using
this method. 

\item 
Use the Popov trick to calculate the T-matrix diagrams for the
leading Kondo renormalization diagramatically. 

\end{problems}

\section{Heavy Fermions}

Although the single impurity 
Kondo problem was essentially solved by the early seventies,
it took a further decade before the physics community was ready to 
accept the notion that the same phenomenon could occur within a dense
lattice environment.    This resistance to change was rooted
in a number of popular misconceptions about the spin physics and the 
Kondo effect. 

At the beginning of the seventies, it
was well known that local magnetic moments
severely suppress superconductivity, so that typically, a few
percent is all that is required to destroy the superconductivity.  
Conventional
superconductivity is largely immune to the effects of non-magnetic
disorder \footnote{Anderson argued in his ``dirty superconductor
theorem''  that BCS superconductivity involves pairing of
electrons in states that are the time-reverse transform of one
another. Non-magnetic disorder does not break time reversal symmetry,
and so the one particle eigenstates of a dirty system can still be
grouped into time-reverse pairs from which s-wave pairs can be constructed.
For this reason, s-wave pairing is largely unaffected by non-magnetic
disorder. } but highly sensitive to magnetic impurities, which 
destroy the time-reversal symmetry necessary for s-wave pairing.
The arrival of a new class of superconducting material containing dense arrays
of local moments took the physics community completely by surprise. 
Indeed, the first observations of superconductivity in 
$UBe_{13}$, made in 1973 \cite{bucher} were dismissed as an artifact
and had to await a further ten years before they were revisited and acclaimed
as heavy fermion superconductivity. \cite{steglich,ott}

Normally, local moment systems develop  antiferromagnetic order at
low temperatures. When a magnetic moment is introduced into a metal
it 
induces Friedel oscillations in the 
spin density around the magnetic ion, given by 
\[
\langle \vec{M} (x)\rangle = -J\chi (\vec{x}-\vec{x}')
\langle \vec{S} (\vec{x}')\rangle 
\]
where $J$ is the strength of the Kondo coupling and 
\begin{eqnarray}\label{}
\chi (x)&=& \sum_{\vec{q}}\chi (\vec{q})e^{i \vec{q}\cdot \vec{x}}\cr
\chi (\vec{q})&=& 2 \sum_{\vec{k}}\frac{f (\epsilon _{\vec{k}})-f (\epsilon
_{\vec{k}+\vec{q}})
}{\epsilon _{\vec{k}+\vec{q}}-
\epsilon _{\vec{k}}}
\end{eqnarray}
is the the non-local susceptibility of the metal. 
\fight=0.8\textwidth
\fg{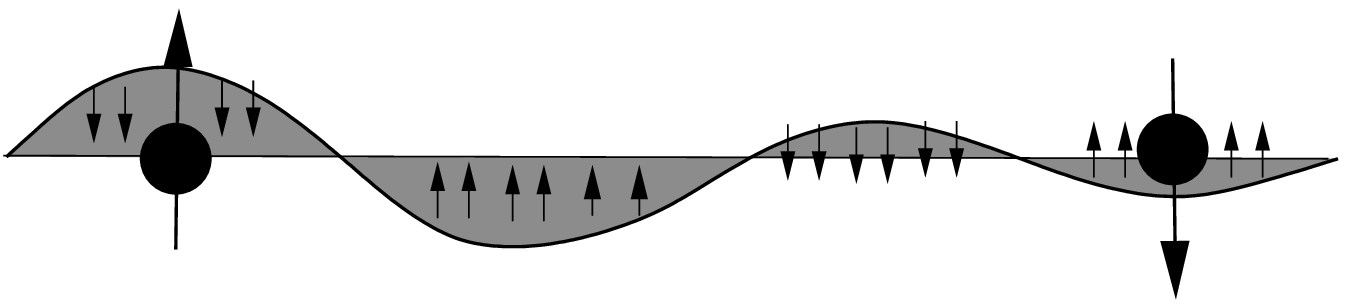}{Illustrating how the polarization of spin around a
magnetic
impurity gives rise to Friedel oscillations and induces an RKKY
interaction
between the spins
}{fig12}
If a second local moment is introduced at location $\vec{x}$, then 
it couples to $\langle M (\vec{x})\rangle $ giving rise to 
a long-range magnetic  interaction  called the ``RKKY''\cite{rkky}
interaction, 
\footnote{named after Ruderman, Kittel, Kasuya and Yosida
}
\begin{equation}\label{}
H_{RKKY} = \overbrace {- J^{2 }\chi (\vec{x}- \vec{x}')}^{{J_{RKKY}
(\vec{x}-\vec{x}')}} \vec{S} (x)\cdot
 \vec{S} (x').
\end{equation}
The sharp discontinuity in the occupancies at the Fermi surface 
produces slowly decaying 
Friedel oscillations in the RKKY interaction
given by
\begin{equation}\label{}
J_{RKKY} (r) \sim -J^{2}\rho  \frac{\cos 2 k_{F}r  }{\vert
k_{F}r\vert ^{3}
}
\end{equation} 
where $\rho $ is the conduction electron density of states and $r$ is
the distance from the impurity, so 
the RKKY interaction oscillates in sign, depending on the 
distance between impurities. The approximate size of the RKKY interaction
is given by 
$E_{RKKY}\sim J^{2}\rho $.

Normally, the oscillatory nature of this 
magnetic interaction favors the development of
antiferromagnetism. In alloys containing a dilute concentration of
magnetic transition metal ions, the RKKY interaction gives rise 
to a frustrated, glassy magnetic state known as a spin glass
in which the magnetic moments freeze into a fixed, but random
orientation.  In dense systems, the RKKY interaction typically 
gives rise to an ordered antiferromagnetic state with a N{\'e}el temperature
$T_{N}\sim J^{2}\rho $.

In 
1976 Andres, Ott and Graebner discovered the heavy fermion metal
$CeAl_{3}$. \cite{ott76}
This metal
has the following features:
\begin{itemize}
\item A Curie susceptibility $\chi ^{-1}\sim T$ at high
temperatures. 

\item  A paramagnetic spin susceptibility $\chi \sim constant$
at low temperatures. 

\item A linear specific heat capacity $C_{V}= \gamma T$, where 
$\gamma\sim 1600 mJ/mol/K^{2}$ is approximately $1600$ times larger
than in a conventional metal. 

\item A quadratic temperature dependence of the low temperature
resistivity $\rho = \rho _{o}+ A T^{2}$

\end{itemize}
Andres, Ott and Grabner pointed out that the low temperature properties
are those of a \underline{Fermi liquid}, but one in which the
effective masses of the quasiparticles are approximately $1000$ larger
than the bare electron mass.  The Fermi liquid expressions for the
magnetic susceptibility $\chi $ and the linear specific heat 
coefficient $\gamma $ are 
\begin{eqnarray}\label{}
\chi &=& (\mu_{B})^{2}\frac{N^* (0)}{1 + F_{o}^{a}}\cr
\gamma &=& \frac{\pi^{2}k_{B}^{2}}{3} N^{*} (0)
\end{eqnarray}
where $N^{*} (0)= \frac{m^{*}}{m}N (0)$ is the renormalized
density of states and $F_{0}^{a}$ is the spin-dependent part 
of the s-wave interaction between quasiparticles. 
What could be the origin of this huge mass renormalization?
Like other Cerium heavy fermion materials, the Cerium
atoms in this metal are in
a $Ce^{3+} (4f^{1})$ configuration, and because they are spin-orbit
coupled, they form huge local moments with a spin of $J=5/2$. In their
paper, Andres, Ott and Graebner suggested that a lattice version of
the Kondo effect might be responsible. 
\fight=\textwidth\fg{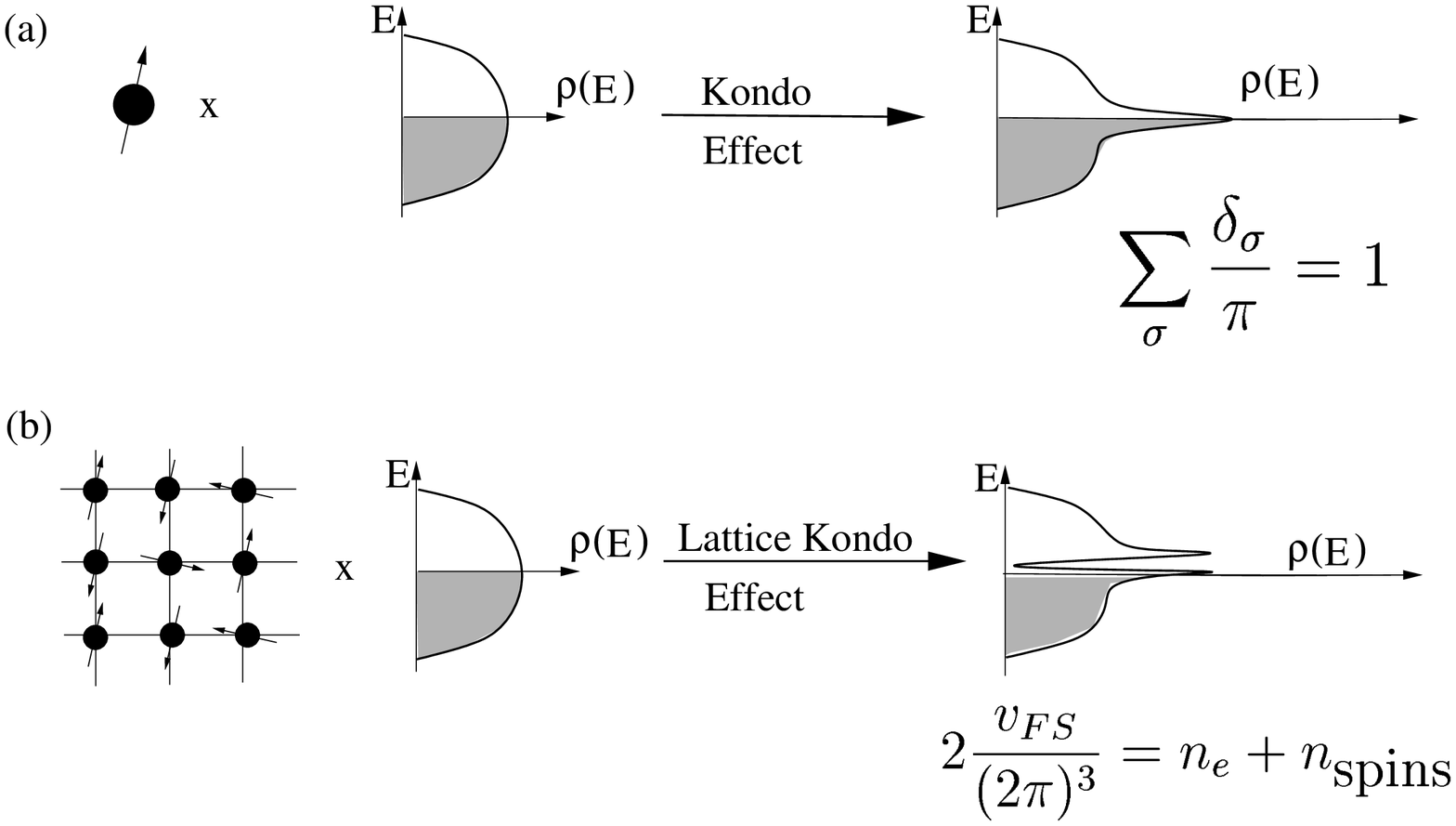}{(a) Single impurity Kondo effect
builds a single fermionic level into the conduction sea, which
gives rise to a resonance in the conduction electron density of states (b)
Lattice Kondo effect builds a fermionic resonance into the conduction
sea in each unit cell. The elastic scattering off this lattice of
resonances
leads to formation of a heavy electron band, of width $T_{K}$. 
}{fig14}
 
This discovery prompted Sebastian Doniach\cite{doniach78} to propose that the origin
of these heavy electrons derived from a dense version of the Kondo effect.
Doniach proposed that heavy electron systems should be modeled by the
``Kondo-lattice Hamiltonian'' 
where a dense array of local moments  interact with the conduction
sea. For a Kondo lattice with spin $1/2$ local moments, the Kondo
lattice Hamiltonian\cite{kasuya52} takes the form 
\begin{equation}\label{}
H=\sum_{\vec{k}\sigma }\epsilon_{\vec{k}}c\dg _{\vec{k}\si
}c_{\vec{k}\si}
+ J\sum_{j} \vec{S}_{j}\cdot c\dg _{\vec{k}\alpha
}\left(\frac{\vec{\sigma }}{2} \right)_{\alpha \beta }c_{\vec{k}'\beta
}e^{i (\vec{k}'- \vec{k})\cdot \vec{R }_{j}}
\end{equation}
Doniach argued that there are two scales in the Kondo lattice, the Kondo
temperature $T_{K} $ and $E_{RKKY}$, given by
\begin{eqnarray}\label{}
T_{K}&=& D e^{-1 /2 J \rho }\cr
E_{RKKY}&=& J^{2}\rho 
\end{eqnarray}
\fight=0.6\textwidth \fg{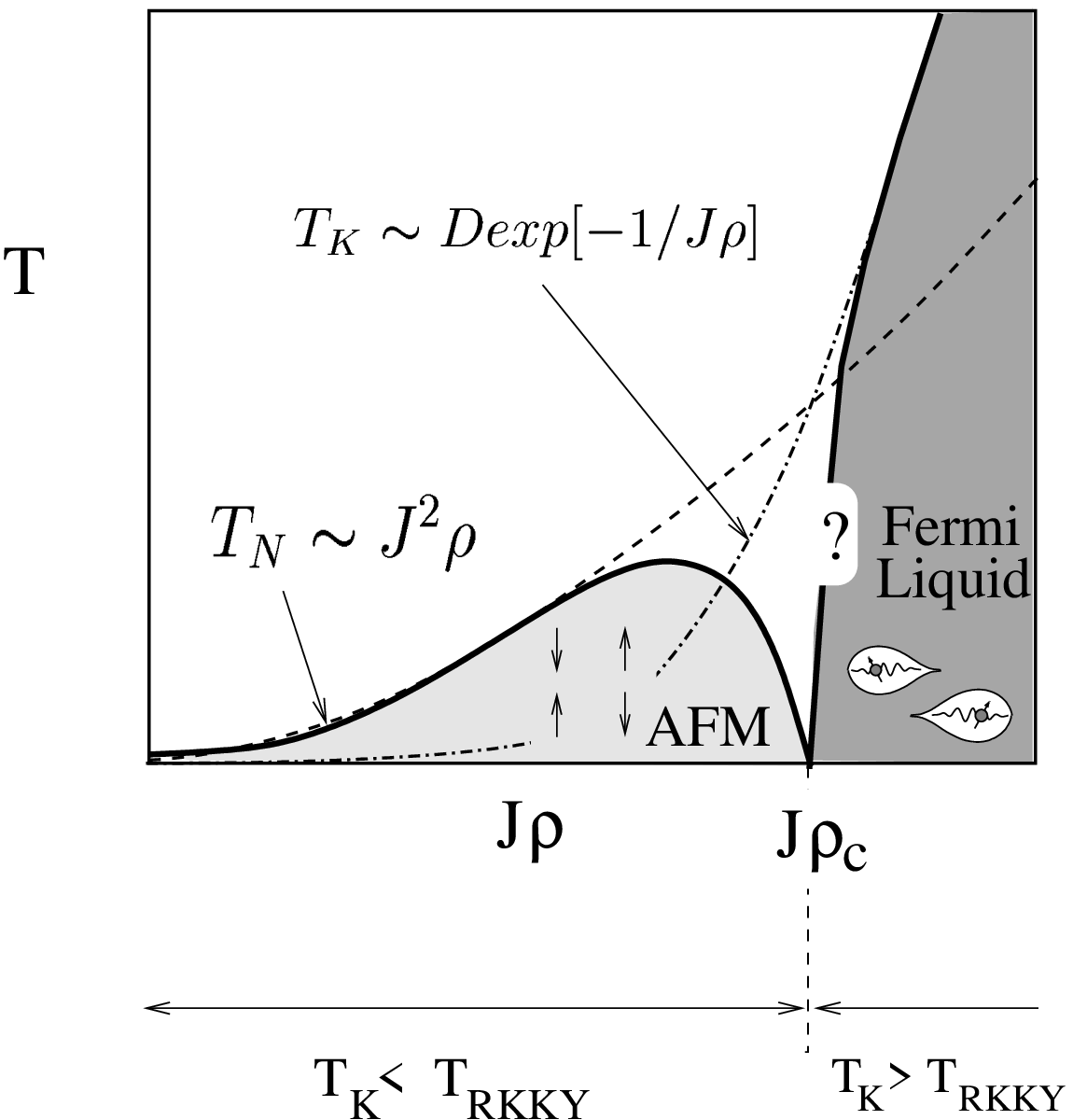}{Doniach diagram, illustrating the antiferromagnetic
regime,
where $T_{K}<T_{RKKY}$ and the heavy fermion regime, where $T_{K}>
T_{RKKY}$. Experiment has told us in recent times that the transition
between these two regimes is a quantum critical point.  The effective 
Fermi temperature of the heavy Fermi liquid is indicated as a solid
line. Circumstantial experimental evidence suggests that this scale
drops to zero at the antiferromagnetic quantum critical point, but 
this is still a matter of controversy. 
}{fig16}
When $J\rho  $ is small, then $E_{RKKY}>> T_{K}$, and an antiferromagnetic
state is formed, but when 
the Kondo temperature is larger than the RKKY interaction scale,
$T_{K}>>E_{RKKY}$, Doniach argued that 
a dense Kondo lattice ground-state is formed in which
each site resonantly scatters electrons. Bloch's theorem then insures
that the resonant elastic scattering at each site will form
a highly renormalized band, of width $\sim T_{K}$. 
By contrast to the single impurity Kondo effect, in the heavy electron
phase of the Kondo lattice the strong elastic scattering at each site
acts in a coherent fashion, and does not give rise to a resistance.
For this reason, as the heavy electron state forms, the resistance of
the system drops towards zero. 
\fight=0.7\textwidth
\fg{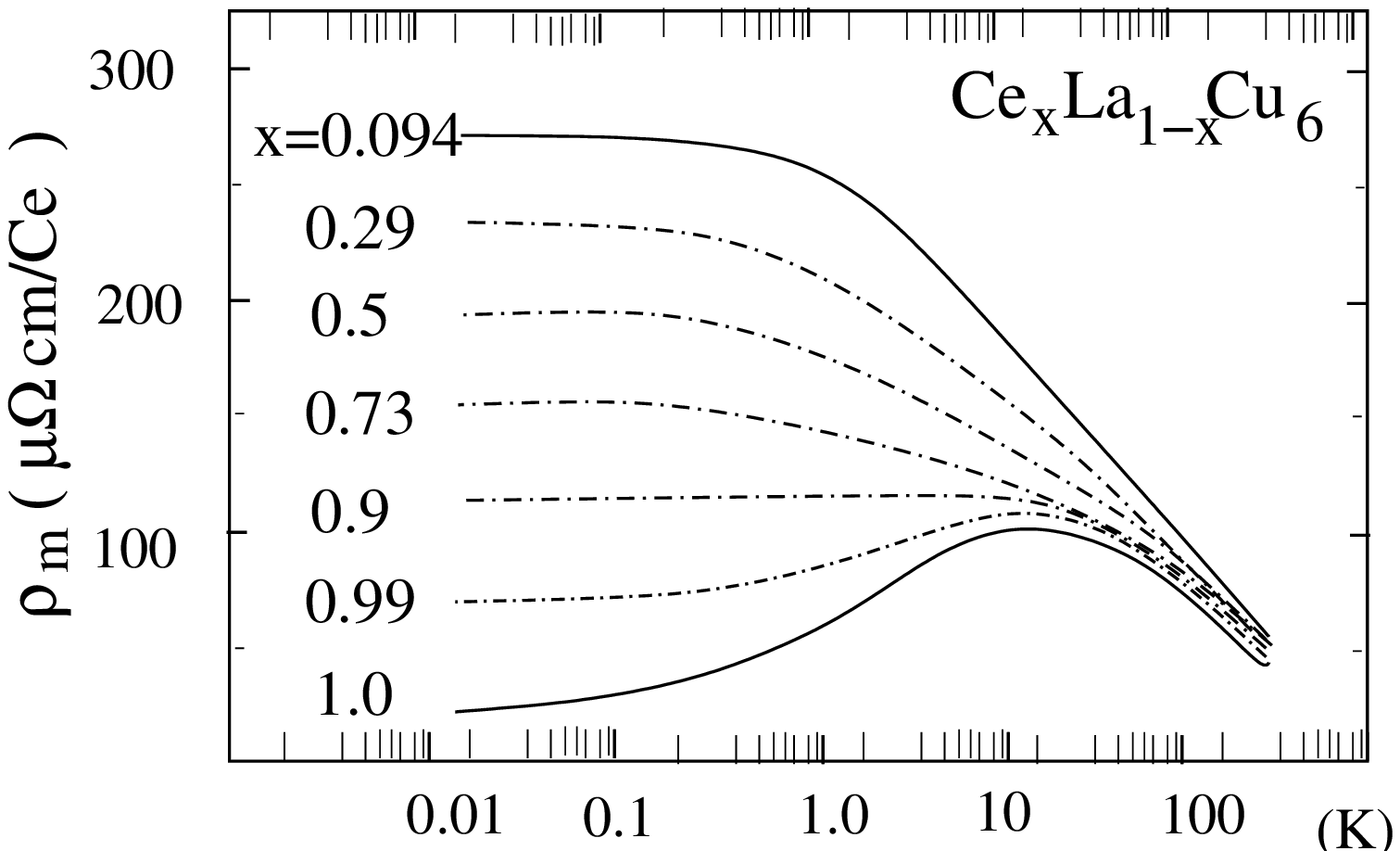}{Development of coherence in heavy fermion
systems. Resistance in $Ce_{1-x}La_{x}Cu_{6} $ after Onuki and
Komatsubara\cite{onuki87}}{fig15}
One of the fascinating aspects of the Kondo lattice concerns the Luttinger
sum rule.  This aspect was first discussed in detail by Martin\cite{martin82},
who pointed out that the Kondo model can be regarded as the result of
adiabatically increasing the interaction strength $U$ in the Anderson
model, whilst preserving the valence of the magnetic ion.   
During this process, one expects sum rules to be preserved. 
In the impurity, the scattering phase shift at the Fermi energy 
counts the number of localized electrons, according to the Friedel sum
rule
\[
\sum_{\sigma }\frac{\delta _{\sigma }}{\pi}= n_{f}=1
\]
This sum rule survives to large $U$, and reappears as the
constraint on the scattering phase shift created by the
Abrikosov Suhl resonance. 
In the lattice, the corresponding sum rule is the Luttinger sum
rule, which states that the Fermi surface volume counts the number of
electrons, which at small $U$ is just the number of localized (4f, 5f
or 3d) and conduction electrons. When $U$ becomes large, number of
localized electrons is now the number of spins, so that
\[
2 \frac{{\cal V}_{FS}}{(2 \pi )^{3}}= n_{e }+n_{\hbox{spins}}
\]
This sum rule is thought to hold for the Kondo lattice Hamiltonian,
independently of the origin of the localized moments. Such a sum
rule would work, for example, even if the spins in the model
were derived from nuclear spins, provided the Kondo temperature
were large enough to guarantee a paramagnetic state. 

Experimentally, there is a great deal of support for the above
picture. It is possible, for example, to examine the effect
of progressively increasing the concentration of $Ce$ in the non-magnetic
host $LaCu_{6}$.(\ref{fig15} )  At dilute concentrations, the resistivity rises to 
a maximum at low temperatures. At dense concentrations, the
resistivity  shows the same high temperature behavior, but at low temperatures
coherence between the sites leads to a dramatic drop in the
resistivity.   The thermodynamics of the dense and dilute system are
essentially identical, but the transport properties display the
effects of coherence. 

There are also indications that the Fermi surface of heavy 
electron systems does have the volume which counts both spins and
conduction electrons, derived from Fermi surface studies. 
\cite{lonzarich1,lonzarich2}

\subsection{Some difficulties to overcome. }\label{}

The Doniach scenario for heavy fermion development is purely a comparison
of energy scales: it does not tell us how the heavy fermion phase
evolves from the antiferromagnet. 
There were two early objections to Doniach's idea:

\begin{itemize}

\item Size of the Kondo temperature $T_{K}$. Simple estimates of the value
of $J\rho $ required for heavy electron behavior give a value $J\rho
\sim 1$. Yet in the Anderson model,  $J\rho \sim 1$  would imply a mixed
valent situation, with no local moment formation.

\item Exhaustion paradox. The naive picture of the Kondo model
imagines that the local moment is screened by conduction electrons
within an energy range $T_{K}$ of the Fermi energy. The
number of conduction electrons in this range is of order $T_{K}/D<<1$
per unit cell, where $D$ is the band-width of the conduction electrons,
suggesting  that there are not enough conduction electrons to screen the
local moments. 

\end{itemize}

The resolution of these two issues are quite intriguing.

\subsubsection{Enhancement of the Kondo temperature by spin
degeneracy}
\label{}The
resolution of the first issue has its origins in the large spin-orbit coupling 
of the rare earth or actinide ions in heavy electron systems. This 
protects 
the orbital angular momentum against quenching by the crystal fields.
Rare earth and actinide ions consequently display a large total
angular momentum 
degeneracy $N= 2j+1$, which has the effect of dramatically enhancing 
the Kondo temperature. 
Take for example the case of the Cerium ion, where the $4f^{1}$ electron
is spin-orbit coupled into a state with $j=5/2$, giving a spin
degeneracy of $N= 2 j +1 = 6$. Ytterbium heavy fermion materials
involve the $Yb:4f^{13}$ configuration, which has an angular momentum
$j=7/2$, or $N=8$. 

To take account of these large spin degeneracies, we need to generalize the
Kondo model.  This was done in the mid-sixties by Coqblin and
Schrieffer\cite{coqblin}. Coqblin and Schrieffer considered a degenerate
version of the infinite $U$ Anderson model in which the spin component
of the electrons runs from $-j$ to $j$,
\[
H = \sum_{{k}\sigma  }\epsilon_{{k}}c\dg _{{k}\sigma 
}c_{{k}\sigma } + E_{f}\sum_{\sigma }\vert f^{1}:\sigma \rangle \langle f^{1}:\sigma
\vert +\sum_{k,\sigma } V\left[c\dg _{{k}\sigma }\vert f^{0}\rangle\langle
f^{1}:\sigma \vert + \hbox{H.c.}
\right]. 
\]
Here the conduction electron states are also labeled by spin indices
that run from $-j$ to $j$. This is 
because the spin-orbit coupled $f$ states couple to partial wave 
states of the conduction electrons 
in which the orbital and spin angular
momentum are combined into a state of 
definite $j$.  Suppose $\vert \vec{k}\si\rangle  $ represents a plane
wave of momentum $\vec{k}$, then one can construct a state of definite
orbital angular momentum $l $ by integrating the plane wave with a
spherical harmonic, as follows:
\[
\vert klm \sigma \rangle = \int \frac{d\Omega}{4\pi }
\vert \vec k \sigma \rangle  Y^{*}_{lm }
(\hat k) 
\]
When spin orbit interactions are strong, one must work with a partial
wave of definite $j$, obtained by combining these states
in the following linear combinations. Thus for the case $j=l+1/2$
(relevant
for Ytterbium ions), we have
\[
\vert km \rangle =  \sum_{\sigma =\pm 1} \sqrt{\frac{l+ \sigma m+
\frac{1}{2}}{2l+1}}
\vert k l m-\frac{\sigma
}{2},\frac{\sigma }{2}\rangle .
\]
An electron creation operator is constructed in a similar way. 
This  construction 
is unfortunately, 
not simultaneously possible at more than one site.  

When $E_{f}<<0$, the valence of the ion approaches unity and
$n_{f}\rightarrow 1$. In this limit, one can integrate out the
virtual fluctuations $f^{1}\rightleftharpoons f^{0 }+ e^{-}$ via a 
Schrieffer Wolff
transformation. This leads to the Coqblin Schrieffer model
\[
H_{CS}=\sum_{{k}\sigma  }\epsilon_{{k}}c\dg _{{k}\sigma 
}c_{{k}\sigma }
+J \sum_{k ,k', \alpha \beta } 
 c\dg _{k\beta } c _{{k'}\alpha }\Gamma_{\alpha \beta }
, 
\qquad (\sigma ,\alpha
,\beta \in [-j,j]
).
\]
where $J= V^{2}/ |E_{f}| $ is the induced antiferromagnetic
interaction
strength. This interaction 
is 
understood as the result of  virtual charge fluctuations into the
$f^{0}$
state,  $f^{1}\rightleftharpoons f^{0}+e^{-}$.  The spin
indices run from $-j$ to $j$, and we have introduced the notation
\[
\Gamma_{\alpha \beta }\equiv f\dg _{\alpha }f_{\beta }= \vert
f^{1}:\alpha \rangle \langle f^{1}:\beta \vert 
\]
Notice that the charge $Q=n_{f}$ of the $f-$electron, normally taken to
be
unity, is conserved by the spin-exchange 
interaction in this Hamiltonian.

To get an idea of how the Kondo effect is modified
by the larger degeneracy, consider the renormalization of the
interaction, which is 
given by the diagram \\
\bxwidth=3in\upit = -0.1in
\begin{eqnarray}\label{}
J_{eff} (D') &=& \raiser{\frm{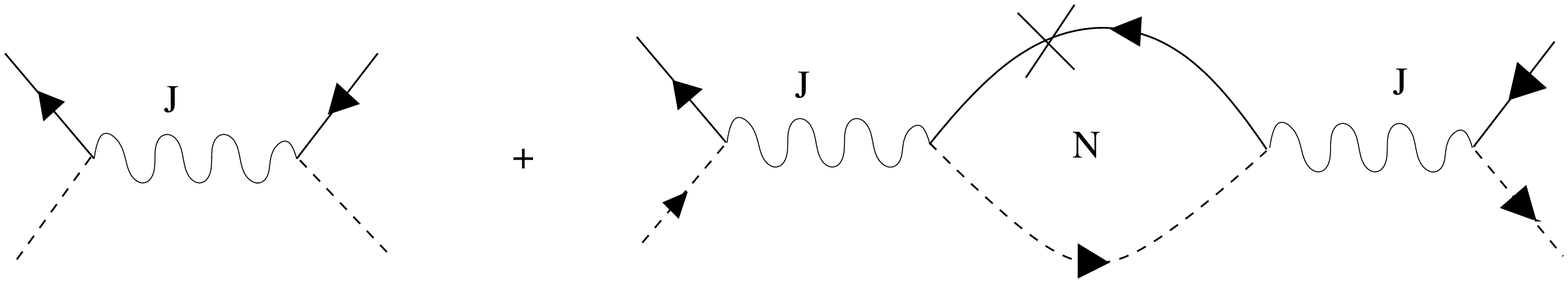}}\cr\cr
&=& J + N J^{2}\rho \ln \left(\frac{D}{D'} \right)
\end{eqnarray}
( where the cross on the intermediate conduction electron state
indicates that all states with energy $|\epsilon_{k}|\in [D',D]$ are
integrate over). 
From this result,  we see that $\beta (g)=  \partial g 
(D)/\partial \ln  D = - N g^{2}$, where $g=J \rho$
has an $N-$ fold enhancement, derived
from the $N$ intermediate hole states. 
A more extensive calculation shows that 
the beta function to third order takes the form
\begin{equation}\label{}
\beta (g)= -Ng^{2}+  Ng^{3}.
\end{equation} 
This then leads to the Kondo temperature
\[
T_{K } = D (NJ \rho )^{\frac{1}{N}}\exp \left[ - \frac{1}{N J \rho } \right]
\]
so that large degeneracy enhances the Kondo temperature in the
exponential factor. By contrast, the RKKY interaction strength is given
by $T_{RKKY}\sim J^{2}\rho $, and it does not involve any $N$ fold
enhancement factors, thus in systems with large spin degeneracy, the
enhancement of the Kondo temperature favors the formation of the heavy
fermion ground-state. 

In practice,  rare-earth ions are exposed to 
the crystal fields of their host, which splits the $N=2j+1$ fold
degeneracy into many multiplets. 
Even in this case, the
large degeneracy is helpful, because the crystal field splitting is
small compared with the band-width. At energies $D'$ large compared
with the crystal field splitting $T_{x}$,  $D'>>T_{x}$, the physics is that
of an $N$ fold degenerate ion, whereas at energies $D'$ small compared
with the crystal field splitting, the physics is typically that of a
Kramers doublet,  i.e.
\bxwidth=1.4in\upit = -0.4in
\begin{eqnarray}\label{}\cr\cr
\raiser{\frm{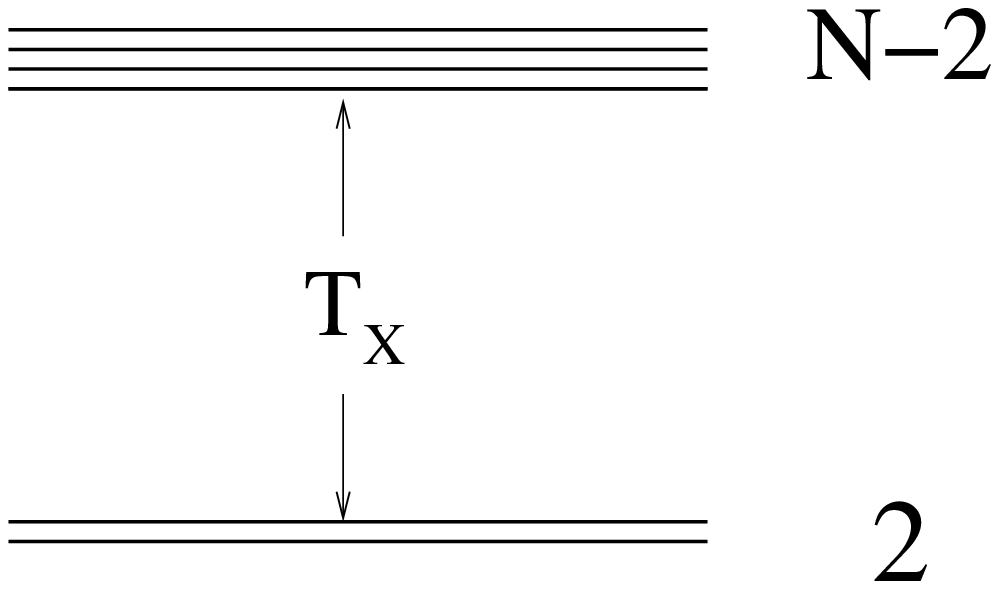}}\nonumber
\end{eqnarray}\vskip 0.2in
\begin{eqnarray}\label{}
\frac{\partial g}{\partial \ln  D} &=& \left\{\begin{array}{lr}
-N g^{2} & (D>> T_{x})\cr
-2 g^{2} & (D<< T_{x})
\end{array} \right.
\end{eqnarray}
from which we see that at low energy scales, the leading order
renormalization of $g$ is given by
\[
\frac{1}{g (D')}= \frac{1}{g_{o}} -N \ln \left(\frac{D}{T_{x}} \right)
-2\ln \left(\frac{T_{x}}{D'} \right)
\]
where the first logarithm describes the high energy screening
with spin degeneracy $N$, and the second logarithm describes the
low-energy screening, with spin degeneracy 2.
This expression is $\sim 0$ when $D'\sim T_{K}^{*}$, the Kondo
temperature, so that
\[
0=\frac{1}{g_{o}} -N \ln \left(\frac{D}{T_{x}} \right)
-2\ln \left(\frac{T_{x}}{T^{*}_{K}} \right)
\]
from which we deduce that 
the renormalized Kondo temperature has the
form\cite{sato}
\[
T_{K}^{*} = D \exp \left(- \frac{1}{2 J_{o}\rho } \right)
\left(\frac{D}{T_{x}} \right)^{\frac{N}{2}-1}.
\]
Here the first term is the expression for the Kondo temperature of a
spin $1/2$ Kondo model. The second term captures the enhancement 
of the Kondo temperature coming from the renormalization effects
at scales larger than the crystal field splitting.   Suppose 
$T_{x}\sim 100K$, and $D\sim 1000K $, and $N=6$, then the enhancement
factor is  order $100$.  This effect enhances the
Kondo temperature of rare earth heavy fermion systems to values that
are indeed, up to a hundred times bigger than those in transition
metal systems.  This is the simple reason why heavy fermion behavior
is rare in transition metal systems. \cite{takagi}
In short-  spin-orbit coupling, even in the presence of crystal fields,
substantially enhances the Kondo temperature. 

\subsubsection{The exhaustion problem}\label{}

At temperatures 
$T\ltappr T_{K}$, a local moment is ``screened'' by conduction
electrons. What does this actually mean?   The conventional view
of the Kondo effect interprets it 
in terms of the formation of a ``magnetic screening cloud''
around the local moment. 
According to the screening cloud picture, 
the electrons  which magnetically screen each local moment are confined
within an energy range of order $\delta \epsilon \sim T_{K}$ around
the Fermi surface, giving rise to a spatially extended screening cloud
of dimension ${l}=v_{F}/T_{K} \sim a \frac{\epsilon_{F}}{T_{K}}
$, where $a$ is a lattice constant and $\epsilon_{F}$ is the Fermi temperature.
In a typical heavy fermion system, this length  would extend
over hundreds of lattice constants. This leads to the following 
two dilemmas\begin{enumerate}
\item It suggests that when the density of magnetic ions is greater
than $\rho \sim 1/l^{3}$, the screening clouds will
interfere. Experimentally no such interference is observed, and
features of single ion  Kondo behavior are seen at much higher
densities. 

\item `` The exhaustion paradox'' 
The number of ``screening''electrons per unit cell 
within energy $T_{K}$ of the Fermi surface 
roughly $T_{K}/W$,  where $W$ is the
bandwidth, so there would never be enough
low energy electrons to screen a dense array of local moments.
\end{enumerate}

In this lecture I shall argue that the screening cloud picture
of the Kondo effect is conceptually incorrect. 
Although the Kondo effect does involve a binding of 
local moments to electrons, the binding process takes place between
the local moment and 
\underline{high
energy electrons },
spanning decades of energy from the Kondo temperature
up to the band-width. (Fig. \ref{fig17})
I shall argue that the key physics of the Kondo effect, both in the dilute
impurity and dense Kondo lattice, involves the formation of a 
\underline{composite heavy fermion} formed by binding electrons on
logarithmically large energy scales out to the band-width. 
\fight=0.7\textwidth
\fg{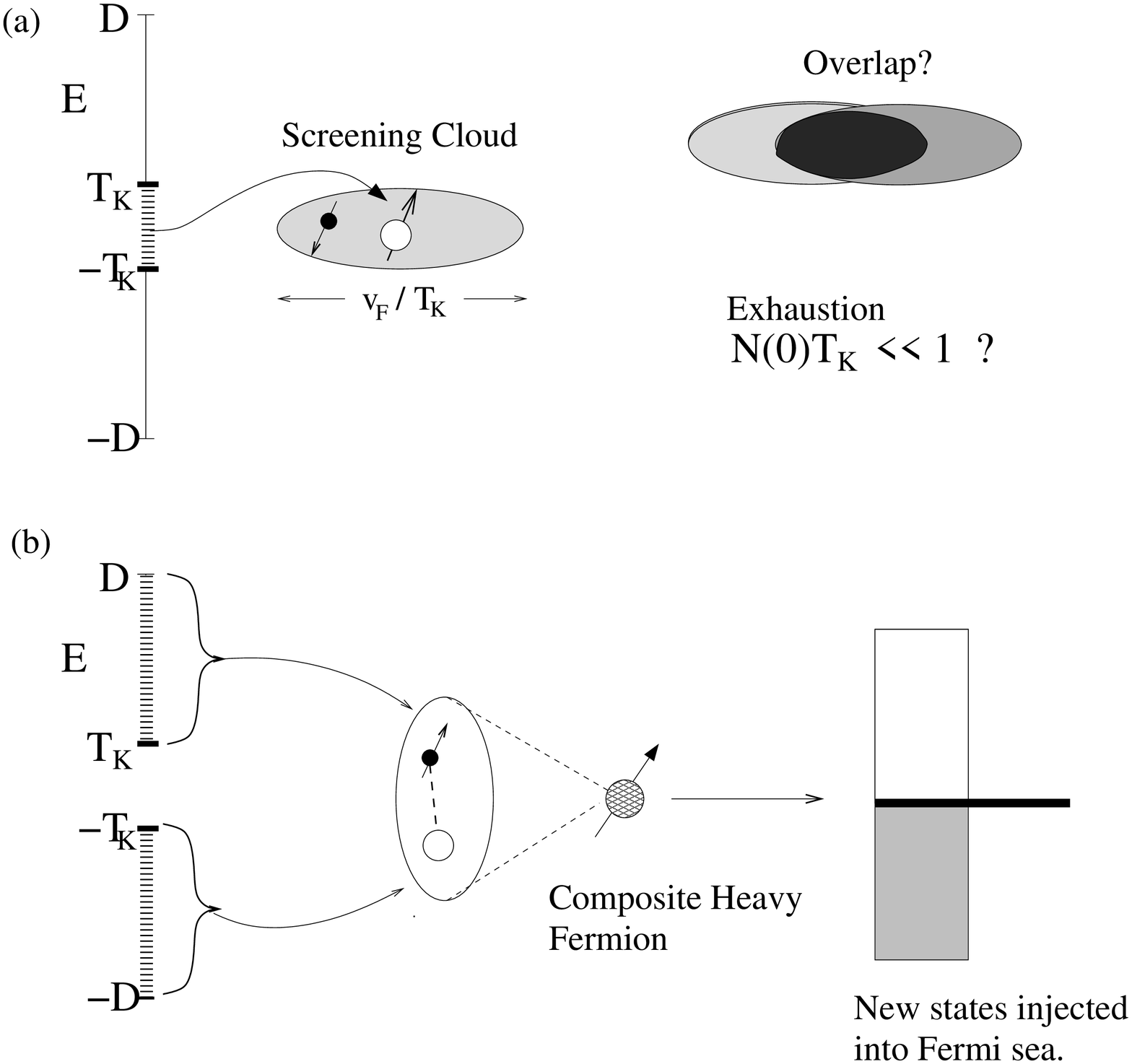}{Contrasting (a) the ``screening cloud'' picture of the
Kondo effect with (b) the composite fermion picture. In (a), 
low energy electrons form the Kondo singlet, leading to
the exhaustion problem. 
In (b) the composite heavy electron is a highly localized
bound-state between local moments and high energy electrons which
injects new electronic states into the conduction sea at the chemical
potential.  Hybridization of these states with conduction electrons 
produces a singlet ground-state, forming 
a Kondo resonance in the single impurity, and a coherent heavy
electron band in the Kondo lattice.
}{fig17}
\fight=3 truein
These 
\underline{new electronic states} are injected into the 
conduction electron sea near the Fermi energy. For a single impurity,
this leads to a single isolated resonance. 
In the lattice, the presence of a new multiplet of fermionic
states at each site leads to the formation of a coherent heavy electron
band with an expanded  Fermi surface. ( \ref{fig17})

\subsection{Large N Approach}

We shall now solve the Kondo model, both the single impurity and the
lattice, in the large $N$ limit. In the early
eighties, Anderson\cite{andersonlargeN} pointed out that the large
spin degeneracy $N=2j+1$ furnishes a small parameter $1/N$ 
which might be used to develop a controlled expansion about 
the limit $N\rightarrow \infty $.  Anderson's observation 
immediately provided a new tool for examining the heavy fermion problem:
the so called ``large $N$ expansion''.   
\cite{witten}. 

The basic idea behind the large $N$ expansion, is to take a limit
where every term in the Hamiltonian grows extensively with $N$. 
In this limit, quantum  fluctuations in intensive variables, such as the
electron density, become smaller and smaller, scaling as $1/N$, and in 
this sense, 
\[
\frac{1}{N} \sim \hbar_{eff}
\]
behaves as an effective Planck's constant for the theory. In this sense,
a large $N$ expansion is a semi-classical treatment of the quantum mechanics,
but instead of expanding around $\hbar =0$, one can obtain new, non trivial
results by expanding around the non trivial solvable limit $\frac{1}{N}=0$. 
For the Kondo model, we are lucky, because the important physics of
the Kondo effect is already captured by the large $N$ limit as we 
shall now see. 

Our model for a Kondo lattice or an ensemble of Kondo
impurities localized at sites $j$ is
\begin{equation}\label{}
H=\sum_{\vec{k}\sigma }\epsilon_{\vec{k}}c\dg _{\vec{k}\si
}c_{\vec{k}\si}
+ \sum_{j} H_{I} (j)
\end{equation}
where 
\[
H_{I} (j)= \frac{J}{N} \Gamma_{\alpha \beta } (j)
\psi \dg _{\beta } (j)\psi _{\alpha } (j)
\]
is the interaction Hamiltonian between the local moment and conduction
sea. Here, the spin of the local moment at site  $j$
is represented using pseudo-fermions
\[
\Gamma _{\alpha \beta } (j)=f\dg _{j\alpha } f_{j\beta },
\]
and
\[
\psi \dg _{\alpha } (j)= \sum_{\vec{k}}c\dg _{\vec{k}\alpha
}e^{-i\vec{k}\cdot \vec{R}_{j}}
\]
creates an electron localized at site $j$. 

There are a number of technical points about this model that need to 
be discussed:

\begin{itemize}

\item {\bf  The spherical cow approximation}. For simplicity, we assume that electrons have a spin degeneracy
$N=2j+1$.  This is a theorists' idealization- a ``spherical cow approximation''
which can only be strictly justified for a single impurity. Nevertheless,
the basic properties of this toy model allow us to 
understand how the Kondo effect works in a Kondo lattice.  With an $N$-fold
conduction electron degeneracy, it is clear that the Kinetic energy 
will grow as $O (N)$.  

\item {\bf Scaling the interaction.} 
Now the interaction part of the Hamiltonian $H_{I} (j)$ involves
two sums over the spin variables, giving rise to a contribution
that scales as $O (N^{2})$. To ensure that the interaction energy grows
extensively with $N$, we need to  scale the coupling constant
as $O (1/N)$. 

\item {\bf  Constraint $n_{f}=Q$}. Irreducible representations of  the rotation group SU (N) 
require that the number of $f-$electrons at a given site
is constrained to equal to $n_{f}= Q$. In the large $N$ limit, it
is sufficient to apply this constraint on the average
$\langle n_{f}\rangle = Q$, though at finite $N$ a time dependent Lagrange
multiplier coupled to the difference $n_{f}-Q$ 
is required to enforce the constraint dynamically. 
With $Q$  $f-$electrons, the spin operators $\Gamma_{ab}=f\dg
_{a}f_{b}$ provide an irreducible {\sl antisymmetric } representation
of $SU (N)$ that is described by column  Young Tableau with $Q$  boxes.
As $N$ is made large, we need to ensure that 
$q= Q/ N$ remains fixed, so that $Q\sim O (N)$ is an extensive
variable.  Thus, for instance, if we are interested in $N=2$, this
corresponds to $q= n_{f}/N = \frac{1}{2}$.  We may obtain insight into this
case by considering the large $N$ limit with $q=1/2$. 
\end{itemize}

The next step in the large $N$ limit is to carry out a ``Hubbard Stratonovich''
transformation on the interaction. We first write
\[
H_{I} (j) = -\frac{J}{N}\left(\psi \dg _{j\beta }f_{j\beta } \right)
\left(f\dg _{j\alpha } \psi_{j\alpha } \right)
,
\]
with a summation convention on the spin indices.
We now factorize this\cite{lacroix,read} as 
\[
H_{I} (j)\rightarrow H_{I}[V,j]=
\bar V_{j}\left(\psi \dg _{j\alpha  }f_{j\alpha  } \right)
+
\left( f\dg _{j\alpha } \psi_{j\alpha } \right)V_{j}
 +N\frac{\bar V_{j}V_{j}}{J}
\]
This is an exact transformation, provided the 
hybridization variables $V_{j} (\tau )$ are regarded as fluctuating
variables inside a path integral, so formally,
\begin{eqnarray}\label{}
Z= \int {\mathcal{D}}[V,\lambda ] 
\overbrace {{\rm  Tr}[T
\exp \left[{-\int_{0}^{\beta }H[V,\lambda ]} \right]]}^{Z[\lambda ,V]}
\end{eqnarray} where
\begin{eqnarray}\label{}
H[V,\lambda ]=\sum_{\vec{k}\sigma }\epsilon_{\vec{k}}c\dg _{\vec{k}\si
}c_{\vec{k}\si}
+ \sum_{j} \left(H_{I}[V_{j},j]+\lambda _{j} [n_{f} (j)-Q] \right),
\end{eqnarray}
is exact. In this expression, 
${\mathcal{D}}[V,\lambda] $ denotes a path integral over all
possible time-dependences of $V_{j}$ and $\lambda _{j} (\tau )$, and
$T$ denotes time ordering. 
The important point for our discussion here however, is that 
in the large $N$ limit, the Hamiltonian entering into this path integral
grows extensively with $N$, so that we may write the partition
function in the form
\begin{eqnarray}\label{}
Z&=& \int {\mathcal{D}}[V,\lambda ] {\rm  Tr}[T
\exp \left[{-N\int_{0}^{\beta }{\mathcal{H}}[V,\lambda ]} \right]
\end{eqnarray}
where ${\mathcal{H}}[V,\lambda ]= \frac{1}{N}H[V,\lambda]\sim O (1) $
is an intensive variable in $N$. 
The appearance of a large factor $N$ in the exponential means that
this path integral becomes dominated by its saddle points in the large
$N$ limit- i.e, if we choose 
\[
V_{j}=V_{o},\qquad \lambda_{j}=\lambda_{o}
\]
where  the saddle point values $V_{o}$ and $\lambda _{o}$ are chosen
so that 
\[
\left. 
\frac{\partial \ln  Z[V,\lambda ]}{\partial
V}
\right|_{V_{j}=V_{o},\lambda _{j}=\lambda _{o}}
= 
\left. 
\frac{\partial \ln  Z[V,\lambda ]}{\partial \lambda }\right|_{V_{j}=V_{o},\lambda _{j}=\lambda _{o}}
= 0
\]
then in the large $N$ limit, 
\[
Z= {\rm  Tr}e^{-\beta H[V_{o},\lambda _{o}]}
\]
In this way, we have converted the problem to a mean-field theory,
in which the fluctuating variables $V_{j} (\tau )$ and $\lambda _{j}
(\tau )$ are
replaced by their saddle-point values. Our mean-field Hamiltonian
is then
\[
H_{MFT}=\sum_{\vec{k}\sigma }\epsilon_{\vec{k}}c\dg _{\vec{k}\si
}c_{\vec{k}\si}
+ \sum_{j,\alpha }
\left(f\dg _{j\alpha } \psi_{j\alpha }
V_{o}+
\bar V_{o}\psi \dg _{j\beta }f_{j\beta } + \lambda _{o}f\dg _{j\alpha
}f_{j\alpha }\right)
 +Nn\left(\frac{\bar V_{o}V_{o}}{J}- \lambda _{o}q
 \right),
\]
where n is the number of sites in the lattice.
We shall now illustrate the use of this mean-field theory in two cases-
the Kondo impurity, and the Kondo lattice.  In the former, there is just
one site; in the latter, translational invariance permits us to set
$V_{j}=V_{o}$ at every site, and for convenience we shall choose this value
to be real.  

\subsection{Mean-field theory of the Kondo impurity }\label{}

\subsubsection{Diagonalization of MF Hamiltonian}

The Kondo effect is at  heart, the formation of a many body resonance.
To understand this phenomenon at its conceptually simplest, we begin
with the impurity model. 
We shall begin by writing down the mean-field Hamiltonian for a single Kondo
ion 
\begin{equation}\label{humdy}H= \sum_{k\sigma }\epsilon_{k}c\dg _{k\sigma
}c_{k\sigma }+ \sum_{k\sigma }V[c\dg _{k\sigma }f_{\sigma }+f\dg
_{\sigma }c_{k\sigma }]+\lambda \sum_{\sigma }n_{f\sigma  }-
\lambda Q + \frac{NV^{2}}{J}
\end{equation}
By making a mean-field approximation, we have reduced the problem to
one of a self-consistently determined resonant level model. 
Now, suppose we 
diagonalize this Hamiltonian, writing it in the form
\begin{equation}\label{}H = \sum_{\gamma \sigma }E_{\gamma }a\dg _{\gamma
\sigma }a_{\gamma\sigma  }+ \frac{NV^{2}}{J}- \lambda Q
\end{equation}
where the ``quasiparticle operators'' $\alpha _{\gamma }$ are related
via a unitary transformation to the original operators
\begin{equation}\label{expansion}
a\dg _{\gamma \sigma }= \sum_{k}\alpha _{k}c\dg _{k\sigma }+\beta f\dg
_{\sigma }.
\end{equation}
commuting $a\dg _{\gamma \sigma }$ with $H$, we obtain
\begin{equation}\label{junko}
[H, \ a\dg _{\gamma \sigma }] = E_{\gamma }a\dg_{} 
\end{equation}
Expanding the right and left-hand side of (\ref{junko})
in terms of (\ref{expansion}) and (\ref{humdy}), 
we obtain, 
\begin{eqnarray}\label{}
( E_{\gamma }-\epsilon _{k})\alpha _{k}- V \beta &=&0\cr
-V\sum_{k}\alpha _{k }+ (E_{\gamma }-\lambda )\beta &=&0
\end{eqnarray}
Solving for $\alpha _{k}$ using the first equation, and substituting
into the second equation, we obtain
\begin{equation}\label{pole1}
E_{\gamma }-\lambda - \sum_{k}\frac{V^{2}}{E_{\gamma }-\epsilon _{k}}=0
\end{equation}
We could have equally well obtained these eigenvalue equations
by noting the electron eigenvalues $E_{\gamma }$ 
must correspond
to the poles of the f-Green function, 
$G_{f} (E_{\gamma })^{-1}=0$, where from an earlier subsection, 
\begin{equation}\label{pole2}
G_{f}^{-1} (\omega )= \left[ \omega -\lambda - \sum_{k}\frac{V^{2}}{\omega -\epsilon _{k}}\right] 
\end{equation}
Either way, the one-particle excitation energies $E_{\gamma }$ must satisfy
\begin{equation}\label{}
E_{\gamma }= \lambda +\sum_{k}\frac{V_{o}^{2}}{E_{\gamma }-
\epsilon _{k}}
\end{equation}
The solutions of this eigenvalue equation are illustrated graphically in
Fig. (\ref{fig18}). 
\fight=0.8 \textwidth
\fg{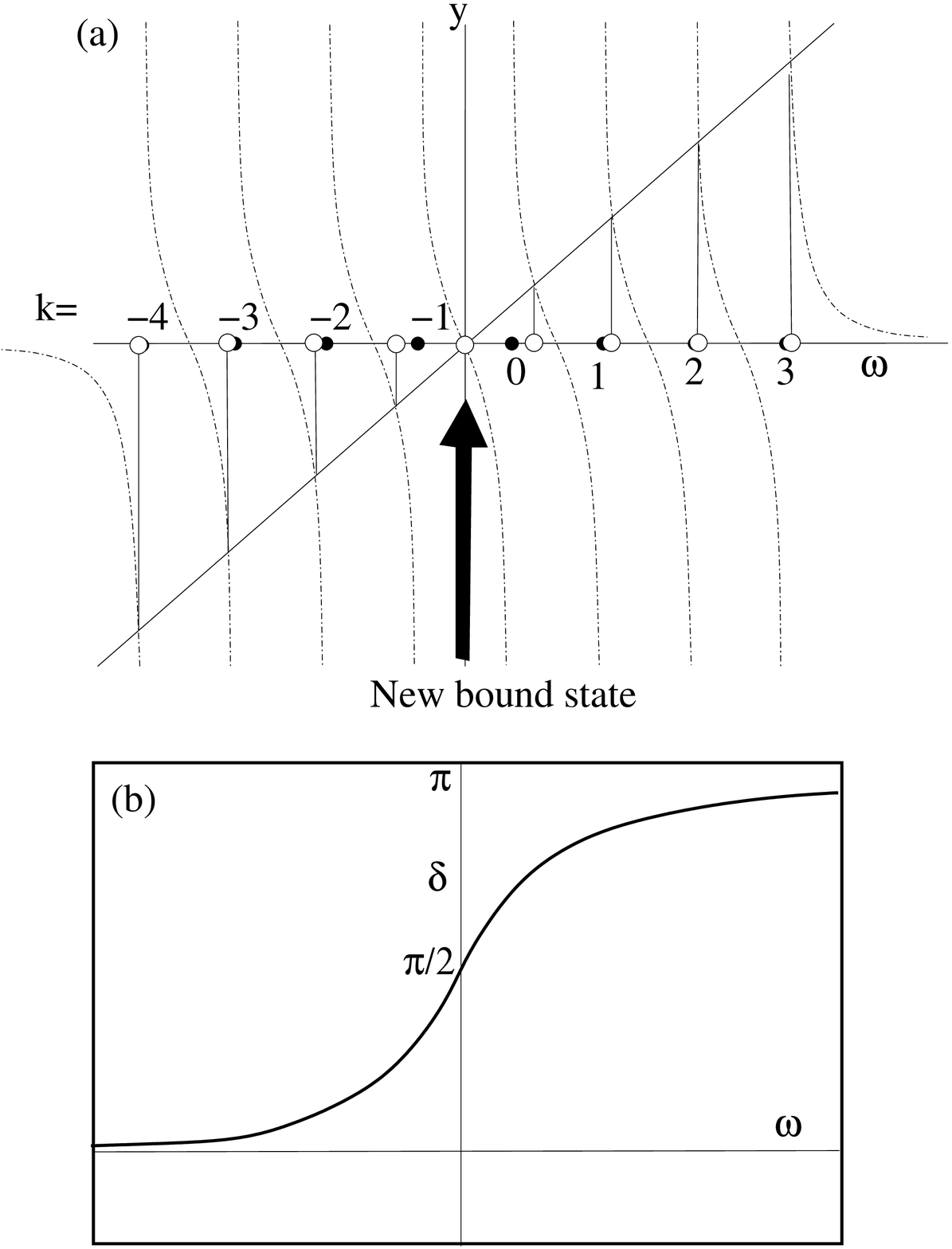}{(a)Graphical solution of the equation $y=-
\sum_{k}\frac{V^{2}}{y-\epsilon _{k}}$, for eight equally spaced
conduction electron energies. Notice how the introduction of a new
bound-state at $y=0$ displaces electron band-states both up and down
in energy. In this way, the Kondo effect injects new bound-state
fermion states into the conduction sea. (b) Energy dependence of the
scattering phase shift. 
}{fig18}
Suppose the energies of the conduction sea
are given by the $2M$ discrete values
\[
\epsilon _{k }= (k +\frac{1}{2})\Delta \epsilon , \qquad k 
\in \{-M,\dots ,M-1\}
\]
Suppose we restrict our attention to the particle-hole case when the 
f-state
is exactly half filled, i.e. when $Q=N/2$. In this situation, $\lambda
=0$. We see that one solution to the eigenvalue equation  corresponds to
$E_{\gamma }=0$.  The original band-electron energies are now displaced
to both lower and higher energies, forming a band of $2M+1$
eigenvalues. Clearly, the effect of the hybridization is to inject
one new fermionic eigenstate into the band. Notice however, that
the electron states  are displaced symmetrically either-side of the
new bound-state at $E_{\gamma }=0$. 

Each new eigenvalue is shifted relative to the original conduction
electron energy by an amount of order $\Delta \epsilon $. Let us write
\[
E_{\gamma  }
= \epsilon _{\gamma } - \Delta \epsilon \frac{\delta _{\gamma  }}{\pi
}
\]
where $\delta \in [0,\pi]$ is called the ``phase shift''. Substituting 
this into the eigenvalue equation, we obtain
\[
E_{\gamma }= \lambda + \sum_{n=\gamma +1-M}^{{\gamma +M}}\frac{V_{o}^{2}}{\Delta
\epsilon (n-\frac{\delta }{\pi })}
\]
Now if $M$ is large, we can replace the sum over states in the above
equation by an unbounded sum
\[
E_{\gamma }= \lambda + \frac{V_{o}^{2}}{\Delta \epsilon }
\sum_{n=-\infty }^{\infty }\frac{1}{ (n-\frac{\delta }{\pi })}
\]
Using contour integration methods, one can readily show that
\[
\sum_{n=-\infty }^{{\infty }}\frac{1}{ (n-\frac{\delta }{\pi })}
= - \pi \cot \delta 
\]
so that the phase shift is given by $\delta _{\gamma }= \delta
(E_{\gamma })$, where
\[
\tan \delta [\epsilon ]= \frac{\pi \rho V_{o }^{2}}{\lambda -\epsilon
}
\] where we have replaced $\rho =\frac{1}{\Delta \epsilon }$ as the
density of conduction electron states. This can also be written
\begin{equation}\label{phaseshift}
\delta (\epsilon ) = \tan ^{-1}\left[ \frac{\Delta }{\lambda -\epsilon
} 
\right]
= {\rm  Im}\ln [\lambda +i\Delta -\epsilon ]
\end{equation}
where $\Delta = \pi \rho
V_{o}^{2}$ is the width of the  resonant level induced by the Kondo
effect. Notice that for $\lambda =0$, $\delta = \pi /2$ at the Fermi energy.
\begin{itemize}

\item The phase shift varies from $\delta =0$ at 
$E_{\gamma }=-\infty $ 
to $\delta =\pi$ at $E_{\gamma }=\infty $ , passing through $\delta
=\pi/2$ at the Fermi energy.

\item An extra state has been \underline{inserted} into the
band, squeezing the original electron states both down and up in
energy
to accommodate the additional state:
states beneath the Fermi sea
are pushed downwards, whereas states above the Fermi energy are pushed upwards.
From the relation 
\[
E_{\gamma }= \epsilon_{\gamma }- \frac{\Delta \epsilon }{\pi }\delta
(E_{\gamma } )
\]
we deduce that 
\begin{eqnarray}\label{}
\frac{d\epsilon }{dE}&=& 1 + \frac{\Delta \epsilon }{\pi }
\frac{d \delta (E )}{d E}
\cr
&=&1 + \frac{1 }{\pi \rho }
\frac{d \delta (E )}{d E}
\end{eqnarray}
where $\rho =1/\Delta \epsilon $ is the density of states in the continuum.
The new density of states $\rho ^{*} (E)$is given by $\rho ^{*} (E)dE
= \rho d\epsilon $, so that 
\begin{equation}\label{newdens}
\rho ^{*} (E) = \rho  (0)\frac{d\epsilon }{dE} = 
\rho + \rho _{i} (E)
\end{equation}
where 
\begin{equation}\label{dostates}
\rho _{i} (E) = \frac{1}{\pi }\frac{d\delta (E)}{dE}= \frac{1}{\pi
}\frac{\Delta}{(E-\lambda )^{2}+ \Delta ^{2}}
\end{equation}
corresponds to the enhancement of the conduction electron density of
states due to injection of resonant bound-state.

\end{itemize}

\subsubsection{Minimization of Free energy}

With these results, let us now calculate the Free energy and minimize
it to self-consistently evaluate $\lambda $ and $\Delta $. The Free
energy is given by
\begin{equation}\label{freebie}
F = - N T \sum_{\gamma }\ln [1+e^{-\beta E_{\gamma }}] 
- \lambda Q +\frac{NV_{o}^{2}}{J}.
\end{equation}
In the continuum limit, where $\epsilon \rightarrow 0$, we can use 
the relation $E_{\gamma }=\epsilon _{\gamma }
- \Delta
\epsilon \frac{\delta }{\pi }
$ to write 
\begin{eqnarray}\label{shifty}
-  T \ln [1+e^{-\beta E_{\gamma }}]&=& 
-T\ln [1+e^{-\beta (
\epsilon _{\gamma }- \Delta
\epsilon \frac{\delta }{\pi })
}] \cr
&=&
\overbrace {-T\ln [1+e^{-\beta 
\epsilon _{\gamma }
}]}^{{\rightarrow F_{0}}}
 - \frac{\Delta \epsilon }{\pi }
{\delta (\epsilon _{\gamma })}
f (\epsilon _{\gamma })
\end{eqnarray}
where $f (x)= 1/ (e^{\beta x}+1)$ is the Fermi function.
The first term in (\ref{shifty}) is the Free energy associated with a
state in the continuum. The second term results from the displacement
of continuum states due to the injection of
a resonance into the continuum. Inserting this result into
(\ref{freebie}), we obtain
\begin{eqnarray}\label{next}
F&=&F_{0} - N\sum_{\gamma }
\frac{\Delta \epsilon }{\pi }
{\delta (\epsilon _{\gamma })}
f (\epsilon _{\gamma })
- \lambda Q +\frac{NV_{o}^{2}}{J}\cr
&=& F_{0} - N\int_{-\infty }^{{\infty }}
\frac{d\epsilon }{\pi }f (\epsilon )
\delta (\epsilon)
- \lambda Q +\frac{NV_{o}^{2}}{J}
\end{eqnarray}
The shift in the Free energy due to the Kondo effect is then 
\begin{equation}\label{}
\Delta F = -N\int_{-\infty }^{{\infty }}
\frac{d\epsilon }{\pi }f (\epsilon )
{\rm  Im}\ln  [ \zeta -\epsilon ]
- \lambda Q +\frac{N\Delta }{\pi J\rho }
\end{equation}
where we have introduced $\zeta = \lambda + i\Delta $.
This integral can be done at finite temperature, but 
for simplicity, let us carry it out at $T=0$, when the Fermi function
is just at step function, $f (x)=\theta (-x)$. This 
gives 
\begin{eqnarray}\label{}
\Delta E &=&  \frac{N}{\pi }{\rm  Im}\biggl[(\zeta -\epsilon )\ln \left[
\frac{\zeta -\epsilon }{e}
\right]
 \biggr]_{-D}^{0}
- \lambda Q +\frac{N\Delta }{\pi J\rho }\cr
&=&  \frac{N}{\pi }{\rm  Im}
\biggl[
\zeta  \ln \left[
\frac{\zeta }{eD}\right]-
D \ln \left[
\frac{D}{e}\right]
 \biggr]
- \lambda Q +\frac{N\Delta }{\pi J\rho }
\end{eqnarray}
where we have expanded $(\zeta +D)\ln \left[
\frac{D+\zeta }{e}\right]\rightarrow D\ln \left[
\frac{D}{e}\right] + \zeta \ln D$ to obtain the second line. 
We can further simplify this expression by noting that
\begin{equation}\label{}
- \lambda Q +\frac{N\Delta }{\pi J\rho }= - \frac{N}{\pi }
{\rm  Im } \left[ 
\zeta  \ln \left[ e^{-\frac{1}{\rho J}+ i\pi q}
\right]\right]
\end{equation}
where $q=Q/N$. With this simplification, the shift in the ground-state
energy due to the Kondo effect is 
\begin{equation}\label{}
\Delta E=\frac{N}{\pi }{\rm  Im}
\biggl[
\zeta  \ln \left[
\frac{\zeta }{e T_{K}e^{i\pi q} }\right]
 \biggr]
\end{equation}
where we have dropped the constant term and introduced the Kondo temperature
$T_{K}=De^{-\frac{1}{J\rho }}$.  The stationary point
$\partial E/\partial \zeta =0$ is given by
\[
\zeta =\lambda +i\Delta = T_{K}e^{i\pi q}\qquad \qquad 
\left\{
\begin{array}{rcl}
T_{K}&=&\sqrt{\lambda ^{2}+\Delta ^{2}}\cr
\tan (\pi q)&=&\frac{\Delta }{\lambda }
\end{array} \right.
\]
Notice that 
\begin{itemize}
\item The phase shift $\delta = \pi q$ is the same in each spin scattering
channel, reflecting the singlet nature of the ground state. 
The relationship between the filling  of the resonance and the phase shift
$Q =\sum_{\sigma } 
\frac{\delta _{\sigma }}{\pi } = N \frac{\delta }{\pi }$ is nothing more
than Friedel's sum rule.

\item The energy is stationary with respect to small variations in
$\lambda $ and $\Delta $.  
It is only a local minimum once the
condition $\partial E/\partial \lambda $, corresponding to the
constraint
$\langle \hat n _{f }\rangle =Q $, or $\lambda = \Delta \cot (\pi q) $
is imposed. It is instructive to study the energy for the special case
$q=\frac{1}{2}$, $\lambda =0$ 
which is physically closest to the $S=1/2$, $N=2$
case. In this case, the energy takes the simplified form
\begin{equation}\label{}
\Delta E=\frac{N}{\pi }
\biggl[
\Delta  \ln \left[
\frac{\Delta }{e T_{K}}\right]
 \biggr]
\end{equation}
Plotted as a function of $V$, this is the classic ``Mexican Hat'' potential,
with a minimum where $\partial E/\partial V =0$ at  $\Delta= \pi \rho |V|^{2}
=T_{K}$. (Fig. \ref{fig19})  

\item According to (\ref{newdens}), the enhancement of the density of
states at the Fermi energy is 
\begin{eqnarray}\label{}
\rho ^{*} (0) &=& \rho + \frac{\Delta }{\pi (\Delta ^{2}+\lambda ^{2})}\cr
&=& \rho + \frac{\sin (\pi q)}{\pi T_{K}}
\end{eqnarray}
per spin channel.
When the temperature is changed or a
magnetic field introduced, one can neglect
changes in $\Delta $ and $\lambda $, since the Free energy
is  stationary.  This implies that in the large $N$ limit, the
susceptibility and linear specific heat are those of a
non-interacting resonance of width $\Delta $. The change in
linear specific heat $\Delta C_{V}=\Delta \gamma T$
and the change in the paramagnetic susceptibility $\Delta \chi $
are given by
\begin{eqnarray}\label{sheatetc}
\Delta \gamma&=& \left[\frac{N \pi^{2}k_{B}^{2}}{3}
 \right]\rho _{i} (0)
= \left[\frac{N \pi^{2}k_{B}^{2}}{3}
 \right]
\frac{\sin  (\pi q)
}{\pi T_{K}}\cr
\Delta \chi &=&\left[ N\frac{j (j+1) (g\mu _{B})^{2}}{3} \right]
\rho _{i} (0)
=\left[ N\frac{j (j+1) (g\mu _{B})^{2}}{3} \right]
\frac{\sin (\pi q)}{\pi T_{K}}
\end{eqnarray}
Notice how it is the Kondo temperature that determines the size
of these two quantities. The dimensionless ``Wilson'' 
ratio of these two quantities is
\[
W = \left[ \frac{(\pi k_{B})^{2}}{(g\mu _{B})^{2}j (j+1)}
\right] \frac{\Delta \chi }{\Delta \gamma }= 1
\]
At finite $N$, fluctuations in the mean-field theory can no longer be
ignored.  These fluctuations induce \underline{interactions} amongst
the quasiparticles, and the Wilson ratio becomes
\[
W= \frac{1}{1-\frac{1}{N}}.
\]
The dimensionless Wilson ratio of a large variety of heavy electron materials
lies remarkably close to this value. 

\end{itemize}

\subsection{Gauge invariance and the composite nature of the $f-$electron}

We now discuss the nature of the $f-$electron. 
In particular, we shall discuss how
\begin{itemize}
\item the $f-$electron is actually a composite
object, formed from the binding of high-energy conduction electrons
to the local moment.  

\item although the broken symmetry associated with the large $N$
mean-field  theory  does not persist to finite $N$, the phase
stiffness associated with the mean-field theory continues to finite
$N$. This phase stiffness is responsible for the charge of the
composite $f$ electron. 
\end{itemize}

\subsubsection{Composite nature of the heavy $f-$electron}

Let us begin by discussing the composite structure of the $f-$electron.  
In real materials, the Kondo effect we have described involves spins
formed from localized f- or d-electrons. Though it is tempting to 
associate 
the composite $f-$electron in the Kondo effect with the 
the $f-$electron locked inside the local moment, we 
should also bear in mind that 
the Kondo effect could have occurred equally well with a
\underline{nuclear} spin!
Nuclear spins do couple antiferromagnetically with a conduction
electron, but the coupling is far too small for 
an observable nuclear Kondo effect. 
Nevertheless, we could conduct a thought experiment where 
a nuclear spin is coupled to conduction electrons via a strong 
antiferromagnetic coupling.  In this case, a resonant bound-state
would also form from the nuclear spin.  
The composite bound-state formed in
the Kondo effect clearly does not depend on the origin of the spin partaking
in the Kondo effect. 

There are some useful analogies between
the formation of the composite $f-$electron in the Kondo problem and 
the formation of Cooper pairs in superconductivity, which we shall 
try to draw upon. 
One of the best examples of a composite bound-state is the Cooper pair.
Inside a superconductor, pairs of electrons behave as composite bosonic
particles. One of the signatures of pair formation, is the fact that
Cooper 
pairs of electron operators  behave as a single composite at low energies,
\[
\psi _{\uparrow} (x)\psi _{\downarrow} (x') \equiv  F (x-x') 
\]
The Cooper pair operator is a boson, and it 
behaves as a c-number because the Cooper pairs condense. 
The Cooper pair wavefunction is extremely extended in space, extending out
to distances of order $\xi\sim v_{F}/T_{c}$. 
A similar phenomenon takes place in the Kondo effect, but here the
bound-state is a {\sl fermion} and it does not condense For the Kondo
effect the fermionic composite $(\vec{\sigma }\cdot \vec{ S} (x))_{\alpha
\beta }\psi _{\beta } (x)$ behaves as a single charged electron
operator.  
The analogy between superconductivity and the Kondo effect
involves the temporal correlation between spin-flips of the conduction
sea and spin-flips of the local moment, so that at low energies
\[
[\vec{\sigma} _{\alpha \beta }\cdot \vec{S} (t)]\psi _{\beta } (t')
\sim \Delta (t-t')f_{\alpha } (t').
\]
The function $\Delta (t-t')$ is the analog of the Cooper pair
wavefunction, and it extends out to times $\tau_{K}\sim
\hbar /T_{K}$. 

To see this in a more detailed fashion, 
consider how the interaction term behaves.
In the path integral we 
factorize the interaction as follows
\[
H_{I} = \frac{J}{N} 
\psi \dg _{\beta } \Gamma_{\alpha \beta }  \psi _{\alpha } 
\longrightarrow 
\bar V\left(\psi \dg _{\beta   }f_{\beta   } \right)
+
\left( f\dg _{\beta  } \psi_{\beta  } \right)V
 +N\frac{\bar VV}{J}
\]
By comparing these two terms, we see that the composite operator
$\Gamma_{\alpha \beta } (j) \psi _{\alpha } (j)$ behaves as a single
fermi field:\\

\boxit{\[
\frac{1}{N}\Gamma_{\alpha \beta } (t)  \psi _{\alpha } (t)
\longrightarrow \left(\frac{\bar V }{J} \right)f_{\beta } (t)
\]
{\sl Evidently, a localized conduction electron is bound to a spin-flip
of the local moment at the same site, creating a new 
\underline{independent} fermionic excitation.
The correlated action of adding a conduction electron
with a simultaneous spin flip of the local moment at the same site 
creates a \underline{composite $f-$electron}.
}}
\\
\vskip 0.1truein
\noindent It is worth noting that this fermionic object only hybridizes
with conduction electrons at a single point: it is thus \underline{local} in space.

Let us now try to decompose the composite fermion 
in terms of the electrons that contribute to the bound-state amplitude.
We start by
writing the local moment in the fermionic representation,
\footnote{Important and subtle point: The emergence of a composite
fermion does not depend on a fermionic representation of the spin. 
The fermionic representation for the spin is simply the most convenient
because it naturally furnishes us with an operator in the theory
that represents the composite bound-state. 
}
\[
\frac{1}{N}\Gamma_{\alpha \beta } \psi _{\alpha } = 
-\frac{1}{N}
f\dg _{\alpha }\psi _{\alpha } f_{\beta 
}\longrightarrow 
-\frac{1}{N}\langle f\dg _{\alpha}\psi _{\alpha}
\rangle  
f_{\beta }
\]
where we have replaced the bilinear product between the conduction and
$f-$electron by its expectation value. 
We can evaluate this ``bound-state amplitude'' from the
corresponding Green-function 
\begin{eqnarray}\label{bstate}
-\frac{V_{0}}{J}=\frac{1}{N}\langle f\dg _{\beta}\psi _{\beta }
\rangle  &=& \int \frac{d\omega }{\pi }  f (\omega ){\rm  Im} G_{\psi f}
(\omega -i\delta )\cr
&=& V_{o}\int f (\omega ) \frac{d\omega }{\pi } {\rm  Im} \left[\sum_{k} 
\frac{1}{\omega -\epsilon _{k}-i\delta }\frac{1}{\omega -i\Delta }\right]
\end{eqnarray}
where we have chosen the half-filled
case $Q/N=1/2$, $\lambda =0$. 
In the large band-width limit, the main contribution to this integral
is obtained by neglecting the principal
part of the conduction electron propagator
$1/ (\omega -\epsilon _{k}-i\delta
)\rightarrow i\pi \delta (\omega -\epsilon _{k})$, so that
\begin{eqnarray}\label{}
\frac{1}{N}\langle f\dg _{\beta}\psi _{\beta }
\rangle  
&=& \sum_{k} f (\epsilon _{k})
\left(\frac{\epsilon _{k}}{\epsilon _{k}^{2} +\Delta^{2} }\right)
\end{eqnarray}
From this expression, we can see that the contribution of a given $k$
state in the Fermi sea
to the bound-state amplitude is given by
\[
\frac{1}{N}\langle f\dg _{\beta}c _{k\beta }
\rangle  =f (\epsilon _{k}) 
\left( 
\frac{\epsilon _{k}}{\epsilon _{k}^{2} +\Delta^{2}}
\right)  
\]
This function decays with the inverse of the energy, 
right out to the band-width. Indeed, if we
break-down the contribution to the overall bound-state amplitude, we see
that each decade of energy counts equally. Let us take $T=0$ and divide
the band on a logarithmic scale into $n$ equal parts, where the ratio
of the lower and upper energies is $s>1$, then 
\begin{eqnarray}\label{}
\frac{V_{o}}{J} &=& \rho V_{o}\int _{-D}^{0}d\epsilon \frac{-\epsilon
}{\epsilon ^{2}+\Delta ^{2}}
\sim \rho V_{o}\int _{\Delta }^{D}d\epsilon \frac{1
}{\epsilon }\cr
&=& \rho V_{o} \left\{\int_{D/s}^{D} + \int_{D/s^{2}}^{D/s} +\dots
\int_{D/s^{n}}^{D/s^{n-1}} + \int _{\Delta }^{{D/s^{n}}}
\right\}\frac{d\epsilon }{\epsilon }\cr
&=& \rho V_{0}\left\{ \ln s + \ln s +\dots \ln s + \ln
\frac{Ds^{-n}}{\Delta } \right\}
\end{eqnarray}
This demonstrates that the composite bound-state involves
electrons spread out over decades of energy out to the band-width.
If we complete the integral, we find that
\[
\frac{V_{o}}{J}= \rho V_{o}\ln \frac{D}{\Delta }\Rightarrow \Delta =
De^{-\frac{1}{J\rho }}= T_{K}
\]
as expected from the minimization of the energy.
Another way of presenting this discussion, is to write the composite
bound-state in the time-domain, as 
\begin{eqnarray}\label{}
\frac{1}{N}\Gamma_{\alpha \beta } (t')  \psi _{\alpha } (t)
\longrightarrow \Delta (t-t')f_{\alpha } (t')
\end{eqnarray}
where now
\[
\Delta (t-t') = \frac{1}{N}\langle f_{\beta }\dg (t)\psi _{\beta }
(t')\rangle 
\]
This is the direct analog of Cooper pair bound-state wavefunction, 
except that the
relevant variable is time, rather than space.
If one evaluates the function $\Delta (t)$ at a finite $t$, 
we find that 
\[
\Delta (t-t') = \sum_{k} f (\epsilon _{k})
\left( \frac{\epsilon _{k}}{\epsilon _{k}^{2} +\Delta^{2} }\right)
e^{-i\epsilon _{k} (t-t')}
\]
Heuristically, 
the finite time cuts off the energy integral over the Fermi surface
at an energy of order $\hbar /t$, so that
\[
\Delta (t)\sim \left\{ 
\begin{array}{lr}
\rho V_{o}\ln \left(\frac{Dt}{\hbar } \right) 
&(t<<\hbar /T_{K})
\cr
\rho V_{o}\ln \left(\frac{D}{T_{K}} \right) 
&(t>>\hbar /T_{K})
\end{array}
\right.
\]
emphasizing the  fact that the Kondo effect involves a correlation
between the spin-flips of the conduction sea and the local moment
over decades of time scales from the the inverse band-width up to
the Kondo time $\hbar /T_{K}$. 

From these discussions, we see that the Kondo effect is
\begin{itemize}

\item entirely localized  in space. 

\item extremely non-local in time and energy.

\end{itemize}
This picture of the Kondo effect as a temporal, rather than a spatial
bound-state is vital if we are to understand the extension of the
Kondo effect from the single impurity to the lattice. 

\subsubsection{Gauge invariance and the charge of the $f-$electron}

One of the interesting points to emerge from the mean-field
theory is that the energy of mean-field theory does 
not depend on
the phase of the bound-state amplitude $V=|V|e^{i\theta } $.
This is analogous to the gauge invariance in superconductivity, 
which derives from the
conservation of the total electronic charge. Here, gauge invariance
arises 
because there are no charge fluctuations at the 
site of the local moment, a fact encoded by the
conservation of the total f-charge $Q$. 
Let us look at the full Lagrangian for the $f-$electron and interaction
term
\begin{eqnarray}\label{}
{\cal L}_{I}&=& f_{\sigma }\dg (i\partial_{t}-\lambda )f_{\sigma }  - H_{I}\cr
H_{I}&=&\bar V\left(\psi \dg _{\alpha  }f_{\alpha  } \right)
+
\left( f\dg _{\alpha } \psi_{\alpha } \right)V
 +N\frac{\bar VV}{J}
\end{eqnarray}
This is invariant under the ``Read-Newns''\cite{read} transformation
\begin{eqnarray}\label{readnewns}
f&\rightarrow& fe^{i\phi },\cr
V&\rightarrow&
Ve^{i\phi },\qquad \qquad (\theta \rightarrow \theta +\phi ),\cr
\lambda &\rightarrow& \lambda + \frac{\partial \phi }{\partial t} .
\end{eqnarray}
where the last relation arises from a consideration of the gauge invariance
of the dynamic part $f\dg (i\partial_{t}- \lambda )f$ of the
Lagrangian.  Now if $V (t)= \vert V(t)\vert e^{i\theta (t)}$, where $V (t)$ is
real, Read and Newns observed that by making 
the gauge choice  $\phi
(t)=-\theta (t)$, the resulting $V=\vert V \vert e^{i (\theta +\phi)
}=\vert V\vert   $ is real. 
In this way,  once the Kondo effect takes
place the phase of 
$V= |V|e^{i\theta  }$ is dynamically absorbed into the constraint field
$\lambda $ : effectively $\lambda \equiv \partial_{t}\phi $ represents
the phase  precession rate  of the hybridization field. 
The absorption of the phase of an order parameter into 
a dynamical gauge field is called the ``Anderson Higgs ''
mechanism.\cite{andersonhiggs}
By this mechanism, once the Kondo effect takes place, 
$V$ behaves as a real, and hence 
neutral object under gauge transformations, this in turn implies
that the composite $f-$electron has to transform under real
electromagnetic
gauge transformations, in other words the Anderson Higgs effect
in the Kondo problem endows the composite $f-$electron with charge. 

\fight=0.6\textwidth
\fg{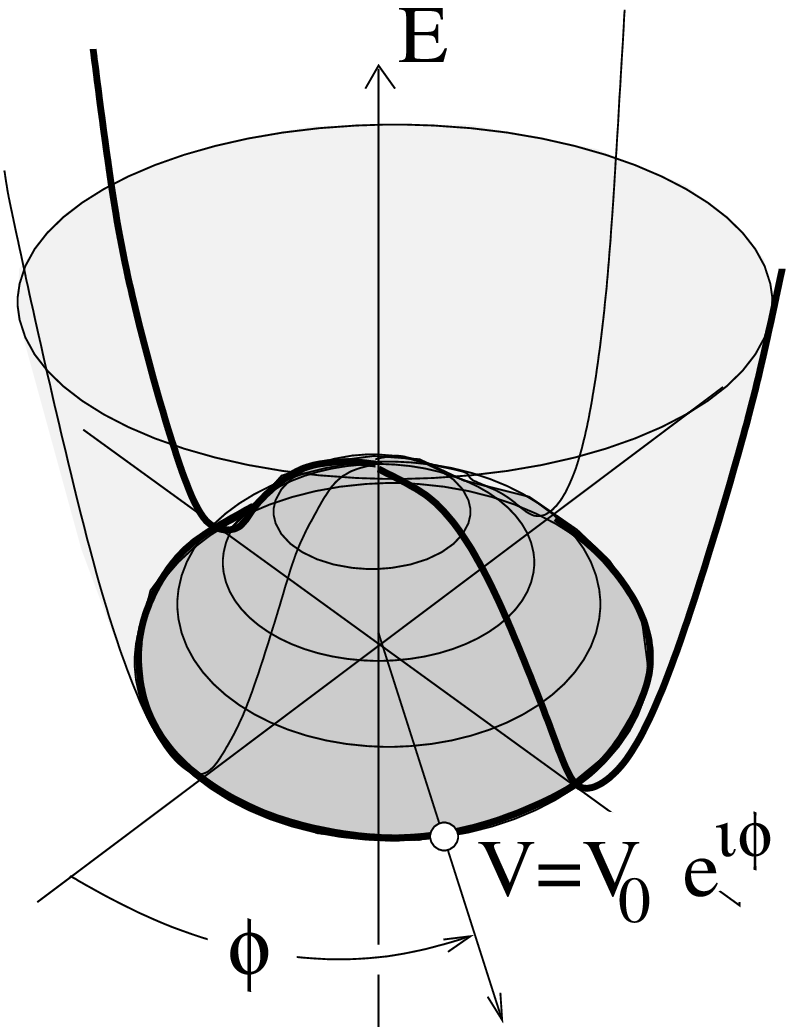}{``Mexican Hat Potential'' which determines minimum of
Free energy, and self-consistently determines the width of the Kondo
resonance. The Free energy displays this form provided the constraint
$\partial F/\partial \lambda = \langle n_{f} \rangle -Q=0$ is imposed.
}{fig19}

\input{parallels.table}
There is a paradox  here, for in the Kondo effect, there can
actually be no true broken symmetry, since we are
dealing with a system where the number of local degrees of freedom is finite.
\underline{Nevertheless}, the phase  $\phi $ 
does develop a stiffness- a stiffness against variation in time, and
the  order parameter 
consequently develops 
infinite range correlations in time.
There is a direct analogy between the spatial 
phase stiffness of a superconductor and the temporal phase stiffness
in the Kondo effect. 
In superconductivity, the energy depends
on spatial derivatives of the phase 
\[
E\propto \frac{\rho _{s}}{2} (\nabla \phi -{2e}{
}\vec{A})^{2}\Rightarrow \frac{1}{\lambda _{L}}^{2}\propto \rho _{s}
\]
( where we have set $\hbar=1$.)
Gauge invariance links this stiffness to the mass of the photon field,
which generates the Meissner effect; the inverse squared penetration
depth is directly proportional to the phase stiffness.
In an analogous fashion, 
in the Kondo effect, the energy depends on temporal derivatives of the
phase and the phase stiffness is \footnote{Note that
because $\lambda \sim \partial_{t}\phi $, the phase stiffness is
given by $\rho _{\phi }= \partial^{2}F/\partial \lambda ^{2}$
}
\[
E\propto \frac{\rho _{\phi }}{2} (\partial_{t}\phi )^{2}
\]


For a Kondo lattice, there is one independent Kondo phase for each spin site,
and the independent conservation of $Q$ at each site  guarantees
that there is no  spatial phase stiffness associated with $\phi $.
The temporal phase stiffness leads to a slow logarithmic growth in 
the phase -phase correlation functions, which in turn
leads to power-law
temporal correlations in  the order parameter $V (\tau )$:
\[
\langle \delta \phi (\tau )\delta \phi (\tau ')\rangle
 \sim \frac{1}{N}\ln  (\tau
-\tau '), \qquad \qquad \langle \bar V (\tau )V (\tau ')\rangle \sim
e^{- \langle \delta \phi (\tau )\delta \phi (\tau ')\rangle
}\sim 
(\tau -\tau ')^{-\frac{1}{N}}. \]   
In this respect, the Kondo ground-state resembles a two dimensional
superconductor, or a one dimensional metal: it is critical but has no
true long-range order.
As in the superconductor, 
the development of phase stiffness involves real physics.
When we make a gauge transformation of the electromagnetic field, 
\begin{eqnarray}\label{}
e\Phi  (x,t)&\rightarrow& e\Phi (x,t) + \partial_{t} \alpha  (x,t), \cr
e \vec{A} (x,t) &\rightarrow & e \vec{A} (x,t) +\nabla
 \alpha ,
\cr
\psi (x)& \rightarrow &\psi (x)e^{-i \alpha  (x,t)}
\end{eqnarray}
Because of the Anderson - Higgs effect, the hybridization is real and the
only way to keep $L_{I}$ invariant under the above transformation, is 
by gauge transforming the $f-$electron and the constraint field
\begin{eqnarray}\label{}
f_{\sigma } (j) &\rightarrow &f_{\sigma } (j) e^{-i\alpha  (x_{j},t)}\cr
\lambda  &\rightarrow & \lambda + \partial_{t} \alpha 
\end{eqnarray}
( Notice how $ \lambda $ transforms in exactly the same way as the
potential $e\Phi $.)\\

The non-trivial transformation of the $f-$electron under electromagnetic
gauge transformations confirm that it has acquired a charge.
Rigidity of the Kondo phase 
is thus intimately related to the formation of a composite charged
fermion. 
The  gauge invariant
form for the energy dependence of the Kondo effect on the Kondo phase 
$\phi $ must then be
\[
E\propto \frac{\rho _{\phi }}{2} (\partial_{t}\phi -{e}{
} \Phi )^{2}
\]
From the coefficient of $\Phi ^{2}$, we see that the Kondo cloud 
has an intrinsic capacitance $C= e^{2}\rho _{\phi }$ ($E\sim C\Phi
^{2}/2$). 
But since
the energy can also be written $ (e n_{f})^{2}/2C \sim U^{*}n_{f}^{2}/2$
we see that the stiffness of the Kondo phase 
can also be associated with an interaction between the $f-$electrons of strength
$U^{*}$, where 
\[
\frac{1}{U^{*}} = C/e^{2} = {\rho _{\phi }}
\]

\subsection{Mean-field theory of the Kondo Lattice}\label{}

\subsubsection{Diagonalization of the Hamiltonian}

We can now make the bold jump from the single impurity problem, to the lattice.
Most of the methods described in the last subsection generalize
very naturally from the impurity to the lattice: the main difficulty
is to understand the underlying physics. 
The mean-field Hamiltonian for the lattice\cite{auerbach,coleman87} takes the form
\[
H_{MFT}=\sum_{\vec{k}\sigma }\epsilon_{\vec{k}}c\dg _{\vec{k}\si
}c_{\vec{k}\si}
+ \sum_{j,\alpha }
\left(f\dg _{j\alpha } \psi_{j\alpha }
V_{o}+
\bar V_{o}\psi \dg _{j\beta }f_{j\beta } + \lambda _{o}f\dg _{j\alpha
}f_{j\alpha }\right)
 + {\cal  N}N\left(\frac{\bar V_{o}V_{o}}{J}- \lambda _{o}q
 \right),
\]
where $\cal N$ is the number of sites in the lattice.  Notice, before we begin,
that the composite  f-state at  each site of the lattice is entirely local,
in that hybridization occurs at one site only.  Were the composite f-state
to be in any way non-local, we would expect that the hybridization
of one f-state would involve conduction electrons at different sites. 
We begin by rewriting the  mean field Hamiltonian in momentum space, 
as follows 
\[
H_{MFT}= \sum_{\vec{k}\sigma }\left(c\dg _{\vec{k}\sigma },f\dg
_{\vec{k}\sigma } \right)
\pmatrix{\epsilon _{\vec{k}}& \bar V_{o}\cr
V_{o}&\lambda _{o}}
\pmatrix{c_{\vec{k}\sigma }\cr f_{\vec{k}\sigma }}
+{\cal N} N\left(\frac{\bar V_{o}V_{o}}{J}- \lambda _{o}q\right)\]
where
\[
f\dg _{\vec{ k}\sigma }= \frac{1}{\sqrt{\cal N}}\sum_{j}f\dg _{j\sigma }
e^{i \vec{ k}\cdot \vec{ R}_{j}}
\]
is the Fourier transform of the $f-$electron field. The absence of
$k-$ dependence in the hybridization is evident that each composite
$f-$electron is spatially local. 
This Hamiltonian can be diagonalized in the form
\[
H_{MFT}= \sum_{\vec{k}\sigma }\left(a\dg _{\vec{k}\sigma },b\dg
_{\vec{k}\sigma } \right)\pmatrix{E_{\vec{k}+}&0\cr
0&E_{\vec{k}-}}
\pmatrix{a_{\vec{k}\sigma }\cr b_{\vec{k}\sigma }}
+Nn\left(\frac{\bar V_{o}V_{o}}{J}- \lambda _{o}q\right)\]
where
$a\dg _{\vec{k}\sigma }
$ and $b\dg _{\vec{k}\sigma }$
are linear combinations of $c\dg _{\vec{k}\sigma }$ and $f\dg _{\vec{
k}\sigma} $, playing the role of 
``quasiparticle operators'' of the theory and
the momentum state eigenvalues $E_{\vec{k\pm}}$ of this Hamiltonian are determined by
the condition
\[
{\rm  Det}
\left[E_{\vec{k\pm}}
\underline{1}-\pmatrix{\epsilon _{\vec{k}}& \bar V_{o}\cr
V_{o}&\lambda _{o}}
 \right]=0,
\]
which gives
\begin{equation}\label{}
E_{\vec{k}\pm } = \frac{\epsilon _{\vec{k}}+\lambda_{o}}{2}\pm
\left[\left(\frac{\epsilon _{\vec{k}}-\lambda _{o}}{2}
\right)^{2}+ \vert V_{o }\vert ^{2} \right]^{\frac{1}{2}}
\end{equation}
are the energies of the upper and lower bands. The dispersion
described by these energies is shown in Fig. \ref{fig20} .
\fight=4 truein
\fg{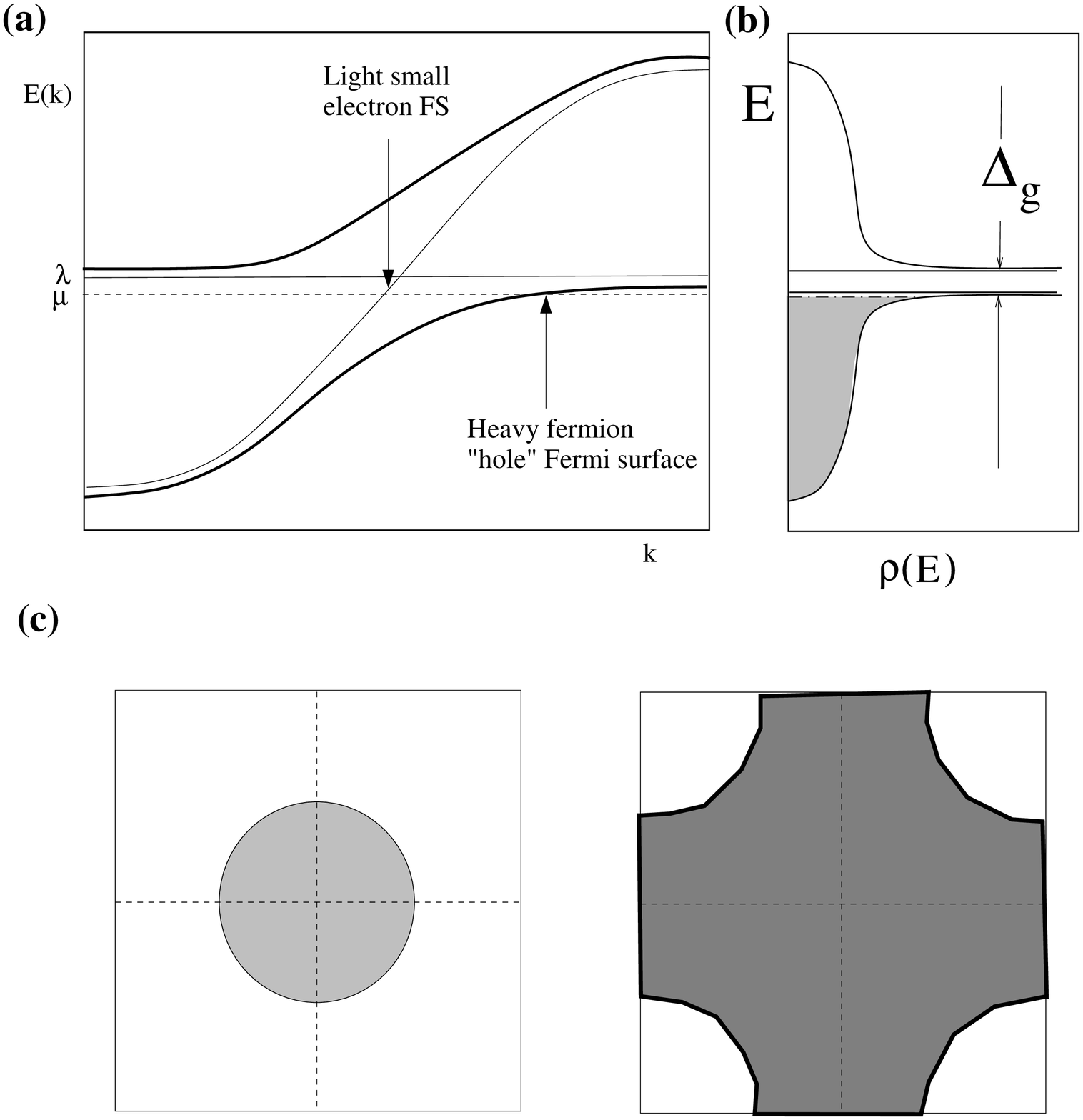}{(a) Dispersion produced by the injection of a
composite fermion into the conduction sea. 
(b) Renormalized density of states, showing ``hybridization gap''
($\Delta _{g}$).
(c) Transformation of the
Fermi surface from a light electron Fermi surface into a heavy
``hole''-like Fermi surface.
}{fig20}
A number of points can be made
about this dispersion:

\begin{itemize}

\item 
We see that
the Kondo effect \underline{injects} new fermionic states into the original
conduction band.  Hybridization between the heavy electron states and
the conduction electrons builds an upper and lower Fermi band
separated by a ``hybridization gap'' of width 
$\Delta _{g}=E_{g} (+)-E_{g} (-)$, such that energies in the range
\begin{eqnarray}\label{}
E_{g} (-)
<&E&<\lambda _{o}+E_{g} (+)\cr
E_{g} (\pm)&=&\lambda _{o}\pm\frac{V_{0}^{2}}{D_{\mp}}
\end{eqnarray}
are forbidden.  Here $\pm D_{\pm}$ are the top and bottom of the
conduction band. In the special case where $\lambda _{o}=0$, corresponding
to half filling, a \underline{Kondo} insulator is formed. 

\item The effective mass of the Fermi surface has the
\underline{opposite sign} to the original conduction sea from which it
is built, so naively, the Hall constant should change sign when coherence
develops. 

\item The Fermi surface volume 
\underline{expands} in response to the presence of the new
heavy electron bands.  The new Fermi surface volume now counts the total
number of particles. To see this note that 
\[
N_{tot}= \langle \sum_{k\lambda \sigma }n_{k\lambda \sigma }
\rangle = \langle n_{f}+ n_{c}\rangle 
\]
where $n_{k\lambda \sigma }=a\dg _{k\lambda \sigma
}a_{k\lambda \sigma }
$ is the number operator for the quasiparticles and 
$n_{c}$ is the total number of conduction electrons. This means  
\[
N_{tot}= N \frac{V_{FS}}{(2\pi )^{3}}= Q + n_{c}.
\]
This expansion of the Fermi surface is a direct manifestation of the
creation of new states by the Kondo effect. It is perhaps worth
stressing that these new states would form, 
\underline{even if the local moments were nuclear in origin}. 
In other words, 
it is only the 
rotational degrees of freedom of the local moments that 
are needed to form heavy electron bound-states with the conduction electrons.
\end{itemize}

The Free energy of this system is then
\[
\frac{F}{N} = -  T \sum_{\vec{ k},\pm}\ln \biggl[
1 + e^{-\beta E_{\vec{k}\pm}} \biggr]
+ {\cal N}\left(\frac{\bar V V}{J}- \lambda
q\right)
\]
Let us discuss the ground-state energy, $E_{o}$  -the $T\rightarrow 0$ limit
of this expression. We can write this in the
form 
\[
\frac{E_{o}}{{\cal N} N}= \int_{-\infty }^{0} dE\rho ^{*} (E)
E + \left(\frac{\bar V V}{J}- \lambda
q\right)
\]
where we have introduced the density of heavy electron 
states $\rho ^{*} (E)= \sum_{\vec{k},\pm}\delta
(E-E^{(\pm)}_{\vec{k}})$.
Now the relationship between the energy of the heavy electrons ($E$) and the
energy of the conduction electrons ($\epsilon $) is given by
\[
E= \epsilon + \frac{\bar V V
}{E-\lambda }
\]
so that the density of heavy electron states related to the conduction
electron density of states $\rho $ by
\begin{equation}\label{dens2}
\rho^{*} (E)= \rho \frac{d\epsilon }{dE} = \rho \left(1 +
\frac{\bar V V}{(E-\lambda )^{2}} \right)
\end{equation}

The originally flat conduction electron density of states
is now replaced by a ``hybridization gap'', flanked by two sharp peaks
of width approximately $\pi \rho V^{2}\sim T_{K}$.
With this information, we can carry out the integral over the energies,
to obtain
\begin{eqnarray}\label{latticeenergy}
\frac{E_{o}}{{\cal N}
N }&=& \frac{D^{2}\rho }{2}+ \int _{-D}^{0}dE \rho
{\bar VV}\frac{E}{(E-\lambda )^{2}}+ \left(\frac{\bar V V}{J}- \lambda
q\right)\cr
&=& \frac{D^{2}\rho }{2}- \frac{\Delta}{\pi}\ln\left(\frac{D}{\lambda e}\right)
+ \left(\frac{\bar V V}{J}- \lambda
q\right)\cr
&=& \frac{D^{2}\rho }{2}- \frac{\Delta}{\pi}\ln\left(\frac{T_K}{\lambda e}\right)
 - \lambda q
\end{eqnarray}
where we have assumed that the upper band is empty, and the lower band
is partially filled, and set $T_{K}=De^{-\frac{1}{J\rho }}$ as before.
If we impose the constraint 
\begin{equation}\label{conny}
\frac{\partial
F}{\partial \lambda }= \langle
n_{f}\rangle -Q=0
\end{equation}
we obtain
\[
\frac{\Delta }{\pi \lambda} - q =0
\]
so that the ground-state energy can be written
\begin{equation}\label{}
\frac{E_{o}}{N n_s}= \frac{\Delta }{\pi }\ln \left(
\frac{\Delta e}{\pi q  T_{K}} 
\right).
\end{equation}

Let us pause for a moment to consider this energy functional
qualitatively. The Free energy surface
has  the form of the 
``Mexican Hat'' at low temperatures.   The minimum of this functional will
then determine a family of saddle point values $V= V_{o}e^{i\theta
}$, where $\theta $ can have any value. 
If we differentiate the ground-state energy with respect to $V^{2}$,
we obtain
\begin{equation}\label{}
0=\frac{1}{\pi }\ln \left(
\frac{\Delta e^{2}}{\pi q  T_{K}} 
 \right)
\end{equation}
or 
\begin{equation}\label{jummy}
\Delta = \frac{\pi q}{e^{2}} T_{K}
\end{equation}
confirming that $\Delta \sim T_{K}$.

\subsubsection{Composite Nature of the heavy quasiparticle in the Kondo lattice.}

We now turn to discuss the nature of the heavy quasiparticles in the
Kondo lattice.  
Clearly, at an operational level, the composite 
$f-$electrons are formed in the same way as in the impurity model, 
but at each site, i.e
\[
\frac{1}{N}\Gamma_{\alpha \beta } (j,t)  \psi _{j\alpha } (t)
\longrightarrow \left(\frac{\bar V }{J} \right)f_{j\beta } (t)
\]
This composite object admixes with conduction electrons at a single
site- site j.  The bound-state amplitude in this expression can be
written 
\begin{eqnarray}\label{}
-\frac{\bar V_{o}}{J}=\frac{1}{N}\langle f\dg _{\beta}\psi _{\beta }
\rangle  
\end{eqnarray}
To evaluate the contributions to this sum, it is useful to notice that 
the condition $\partial E/\partial V=0$ can be written
\begin{eqnarray}\label{}
\frac{1}{N}\frac{\partial E}{\partial V_{o}}
&=&0= \frac{ \bar  V_{o}}{J}
+\frac{1}{N}\langle f\dg
_{\beta}\psi _{\beta }\rangle \cr
&=& \frac{   \bar V_{o}}{J}+V_{o}\int_{-D}^{0}dE \rho \frac{E}{(E-\lambda )^{2}}
\end{eqnarray}
where we have used (\ref{latticeenergy}) to evaluate the derivative.
From this we see that we can write
\begin{eqnarray}\label{}
\frac{\bar V_{o}}{J}&=& - \bar V_{o}\int_{-D}^{0}dE \rho 
\left(\frac{1}{E-\lambda }+\frac{\lambda }{(E-\lambda )^{2}} \right)\cr
&=& -V_{o}\rho \ln \left[\frac{\lambda e}{D} \right]
\end{eqnarray}
It is clear that as in the impurity, 
the composite $f-$electrons in
the Kondo lattice are
formed from {\sl high energy} electron states all the way out to
the bandwidth. In a similar fashion to the impurity, \underline{each decade}
of energy between $T_{K}$ and $D$ contributes \underline{equally to the overall
bound-state amplitude}. 
The above expression only differs from the corresponding  impurity
expression (\ref{bstate})
at low energies,
showing that low energy 
electrons play a comparatively unimportant role in forming the
composite heavy electron. It is this feature that permits a dense 
array of composite  fermions to co-exist throughout the crystal lattice.

These composite $f-$electrons admix with the conduction electrons to
produce a heavy electron band with a density of states given by
(\ref{dens2}), 
\[
\rho ^{*} (E)= \rho \frac{d\epsilon }{dE} = \rho \left(1 +
\frac{V_{0}^{2}}{(E-\lambda )^{2}} \right)
\]
which,  setting $E=0$ and using (\ref{conny}) and (\ref{jummy}),  becomes
\[
\rho ^{*} (0) = \rho + \frac{q}{\lambda}=\rho + \frac{qe^{2}}{T_{K}}
\]
at the Fermi energy. The mass enhancement of the heavy
electrons is then
\[
\frac{m^{*}}{m} = 1 + \frac{q e^{2}}{\rho T_{K}}\sim \frac{q D}{T_{K}}
\]
This large factor  in the effective mass enhancement can be as much
as $1000$ in the most severely renormalized heavy electron systems.

\subsubsection{Consequences of mass renormalization}
The effective mass enhancement
of heavy electrons can be directly observed 
in a wide range of experimental quantities
including
\begin{itemize}

\item The large renormalization of the linear specific heat
coefficient $\gamma ^{*}\sim \frac{m^{*}}{m}\gamma$ 
and Pauli susceptibility $\chi ^{*}\sim \frac{m^{*}}{m}\chi $.

\item The quadratic temperature (`` A'' ) 
coefficient of the resistivity. At low temperatures the resistivity of
a Fermi liquid has a quadratic temperature dependence, $\rho \sim \rho
_{o}+A T^{2}$, where $A\sim \left( \frac{1}{T_{F}}\right)^{2}\sim
\left(\frac{m^{*}}{m} \right)^{2}\sim \gamma^{2}
$ is related to the density of three-particle excitations.  
The approximate constancy of the ratio $A/\gamma ^{2}$ in heavy fermion
systems is known as the ``Kadowaki-Woods'' relation.\cite{kadowaki}

\item  The renormalization of the effective mass as measured by
dHvA measurements of heavy electron Fermi surfaces.\cite{onuki,lonzarich1,lonzarich2}

\item  The appearance of a heavy quasiparticle Drude
feature in the frequency dependent optical conductivity $\sigma (\omega )$.
(See discussion below).

\end{itemize}
\fg{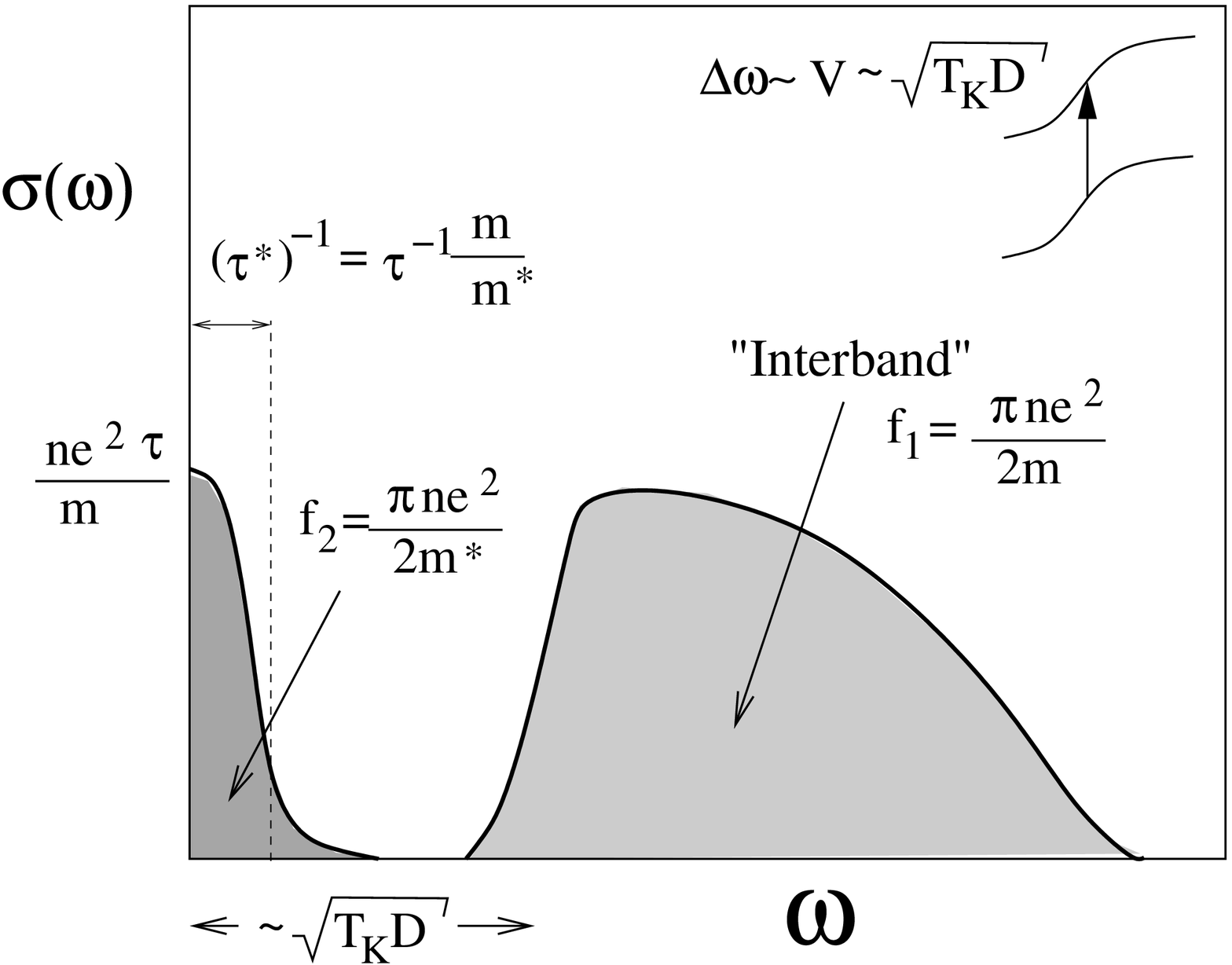}{Separation of the optical sum rule
in a heavy fermion system into a high energy ``inter-band'' component
of weight $f_{2}\sim ne^{2}/m$
and a low energy Drude peak of weight $f_{1}\sim ne^{2}/m^{*}$.
}{fig21}

The optical conductivity of heavy fermion metals deserves 
special discussion. 
According to the f-sum rule, the total integrated optical conductivity
is determined by the plasma frequency
\[
\int_{0}^{\infty } \frac{d\omega }{\pi }\sigma (\omega ) =
f_{1}=\frac{\pi }{2}\left( \frac{ne^{2}}{m}\right)
\]
where $n$ is the density of electrons. \footnote{The f-sum rule is a
statement about the instantaneous, or short-time diamagnetic response of the metal. At
short times $ dj/dt = ( ne^{2 }/m) E$, so the high frequency
limit of the conductivity is $\sigma (\omega )=
\frac{ne^{2}}{m}\frac{1}{\delta -i\omega }$. But using the Kramers
Kr{\"o}nig relation 
\[
\sigma (\omega )= \int \frac{dx}{i\pi }\frac{\sigma
(x)}{x-\omega-i\delta }
\] 
at large frequencies, 
\[
\sigma (\omega )= \frac{1}{\delta -i\omega }\int \frac{dx}{\pi }\sigma
(x)
\] 
so that the short-time diamagnetic response implies the f-sum rule.
}
In the absence of local moments, this is the total spectral weight
inside the Drude peak of the optical conductivity. 

What happens to this spectral
weight when the heavy electron fluid forms? 
Whilst  we expect this sum rule to be preserved, we also expect 
a new ``quasiparticle'' Drude peak to form in which 
\[
\int {d\omega }\sigma (\omega ) = f_{2}
\frac{\pi }{2}\frac{ne^{2}}{m^{*}}= f_{1}\frac{m}{m^{*}}
\]
In other words, we 
expect 
the total spectral weight to divide up into a tiny ``heavy fermion''
Drude peak, of total weight $f_{2}$, 
where 
\[
\sigma (\omega )= \frac{ne^{2}}{m^{*}}\frac{1}{(\tau
^{*})^{-1}-i\omega }
\]
is split off by an energy of order $V\sim \sqrt{T_{K}D}$
from an ``inter-band'' component associated with
excitations between the lower and upper Kondo
bands.\cite{millis,anders} 
This second term 
carries
the bulk $\sim f_{1}$ of the spectral weight. (Fig. \ref{fig21}  ).

Simple calculations, based on the Kubo formula confirm
this basic expectation,\cite{millis,anders} showing that the relationship between the
original relaxation rate of the conduction sea and the heavy electron
relaxation rate $\tau ^{*}$ is 
\begin{equation}\label{}
(\tau ^{*})^{-1} = \frac{m}{m^{*}}(\tau )^{-1} .
\end{equation}
Notice that this means that the residual resistivity
\[
\rho _{o}= \frac{m^{*}}{ne^{2}\tau ^{*}}= \frac{m}{ne^{2}\tau }
\]
is unaffected by the effects of mass renormalization. 
This can be understood by observing that the 
heavy electron Fermi velocity is also renormalized by the effective mass,
$v_{F}^{*}= \frac{m}{m^{*}}$, so that the mean-free path of the
heavy electron quasiparticles is unaffected by the Kondo effect.
\[
l^{*}= v_{F}^{*}\tau ^{*}= v_{F}\tau 
\]
This is yet one more reminder that the Kondo effect is local in space,
yet non-local in time.

These basic features- the formation of a narrow Drude peak, and the
presence of a hybridization gap, have been seen in optical measurements
on heavy electron systems\cite{schlesinger,gruner88,gruner97}

\subsection{Summary}

In this lecture 
we have presented Doniach's argument that the enhancement of the Kondo temperature over and
above the characteristic RKKY magnetic interaction energy between
spins leads to the formation of a heavy electron ground-state.
This enhancement is thought to be generated by the large spin
degeneracies of rare earth, or actinide ions.
A simple mean-field theory of the Kondo model and Kondo lattice,
which ignores the RKKY interactions, 
provides
a unified picture of heavy electrons and the Kondo effect.
The essential physics involves composite quasiparticle formation between
high energy conduction band electrons  and local moments. 
This basic physical effect is 
local in space, but non-local in time. Certain analogies
can be struck between Cooper pair formation, and the formation of the
heavy electron bound-state, in particular, the  charge on the $f-$electron
can be seen as a direct consequence of the temporal phase stiffness of
the  Kondo bound-state. 
This bound-state hybridizes with conduction electrons- producing a single
isolated resonance in a Kondo impurity, and an entire renormalized
Fermi surface in the Kondo lattice.

\subsection{Exercises}

\begin{problems}

\item  
\begin{enumerate}
\item Directly confirm the Read-Newns gauge transformation (\ref{readnewns}).

\item Directly calculate the ``phase stiffness'' $\rho _{\phi }= -
\frac{d^{2}F}{d\lambda ^{2}}$ of the large $N$
Kondo model and show that at $T=0$.
\[
\rho _{\phi }= \frac{N}{\pi }\left(\frac{\sin (\pi q)}{T_{K}} \right).
\]

\end{enumerate}

\item  
\begin{enumerate}
\item Introduce a simple relaxation time into the conduction electron
propagator, writing
\begin{equation}\label{impurity}
G (\vec{k},i\omega_{n} )^{-1}= 
i\omega_{n }+ i{\rm sgn} (\omega
_{n})/2\tau 
+ \frac{V^{2}}{i\omega _{n}-\lambda }
\end{equation}
Show that the poles of this Greens function occur at
\[
\omega = E_{k}\pm \frac{i}{2\tau ^{*}}
\]
where 
\[
\tau ^{*}= \frac{m^{*}}{m}\tau 
\]
is the renormalized elastic scattering time.

\item The Kubo formula for the optical conductivity of an isotropic
 one-band system is
\[
\sigma (\nu)= -\frac{Ne^{2}}{3}\sum_{k} v_{k}^2 \frac{\Pi (\nu )}{i\nu }
\]
where we have used the $N$ fold spin degeneracy, and 
$\Pi (\nu )$  is the analytic extension of 
\[
\Pi (i\nu _{n})= T\sum_{m} G (\vec{k},i\omega _{m})
\left[
G (\vec{k},i\omega _{m}+i\nu _{n})- G (\vec{k},i\omega _{m})
 \right]
\]
where in our case, $G (\vec{k},i\omega _{n})$ is the conduction
electron propagator.
Using (\ref{impurity}), and approximating the momentum sum  by 
an integral over energy, show that the  low frequency conductivity
of the large $N$ Kondo lattice is given by
\[
\sigma (\nu  )= \frac{ne^{2}}{m^{*}}\frac{1}{(\tau
^{*})^{-1}-i\nu  }.
\]
\end{enumerate}

\end{problems}

\section{Quantum Criticality in Heavy Electron Systems}

\subsection{Introduction}

This section provides a brief introduction to the unsolved 
problem of quantum criticality in
heavy fermion materials.  Many of the ideas summarized here are the result
of collaborations, and much of the material in this section
is published in review form. \cite{questions,susy,susy2}
Heavy electron materials lie on the verge of magnetic instability.
In the discussion of the last section, we ignored magnetism and
focussed on the dense Kondo effect.   What
happens when they are pushed to the very edge of magnetic
instability? Such a question was first posed in the context of itinerant
magnetic order in a pioneering work by John Hertz, almost thirty years
ago.  Hertz concluded that a metallic system at the edge of
magnetic instability would develop a new kind of critical behavior-
quantum critical behavior.

A quantum critical point (QCP) is a zero-temperature 
instability between two phases of matter
where quantum fluctuations develop long range
correlations in both space and time\cite{sachdevbook}.  
At a finite temperature critical point, the critical 
long-wavelength fluctuations of the order parameter
do not involve quantum mechanics. This is because 
thermal fluctuations
destroy the coherence of quantum fluctuations on time-scales longer
than
\begin{equation}\label{}
\tau \sim \frac{\hbar }{k_{B}T},
\end{equation}
The great revolution in our understanding  of critical phenomena which
occured in the 1970s  involved many tools borrowed from relativistic
field theory, but the physics was entirely classical.

\fg{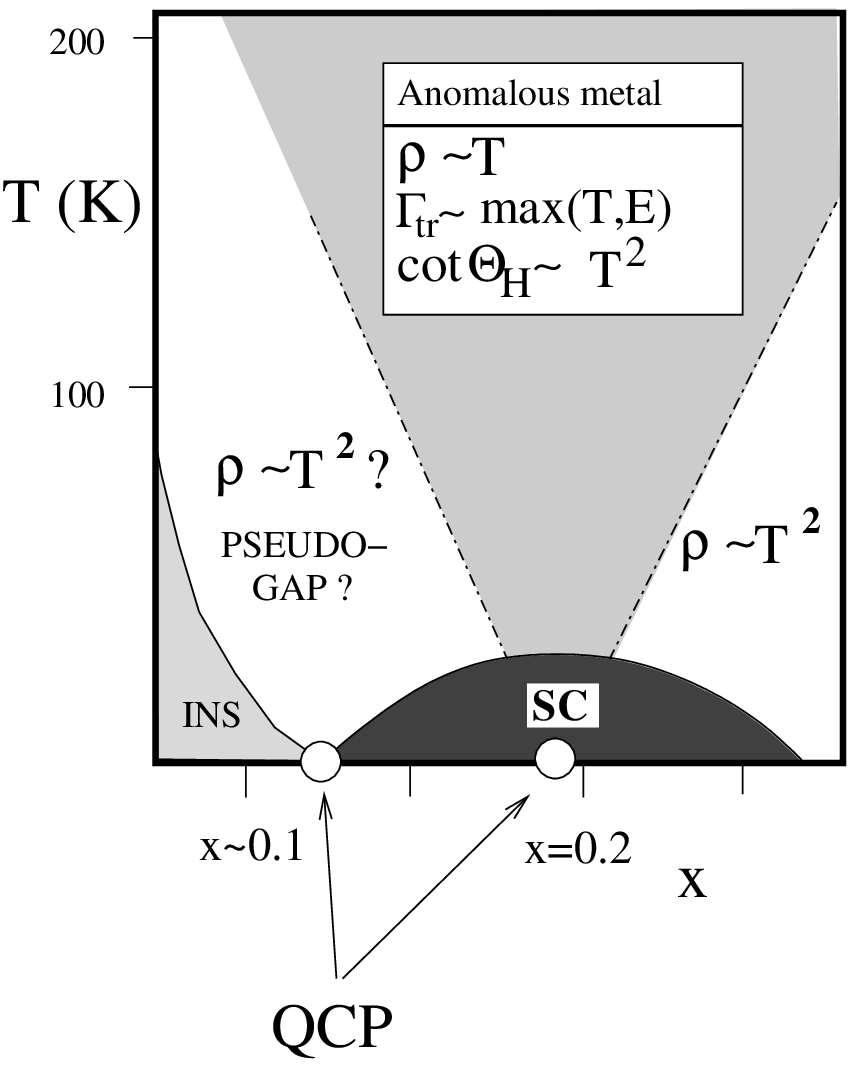}{Schematic phase diagram for cuprate superconductors showing
location of possible quantum critical points.  
One of these QCP may be responsible for the anomalous normal state
which develops above the pseudogap scale. 
}{invfig1x}

Experimental developments of the past decade
have brought a new awareness of
the importance of quantum critical points in condensed matter
physics. 
These special points exert
a profound influence on the finite temperature properties 
of a material. 
Materials close 
to quantum criticality develop a new excitation structure, 
they display novel thermodynamic, transport and magnetic
behavior. They
also 
have marked 
a predeliction towards the
development of new kinds of order, such as anisotropic
superconductivity.
A dramatic 
example is provided by the cuprate superconductors. By doping with holes, 
these materials pass through one or more quantum phase transitions: 
from an  insulator to a 
metal with a spin gap at low doping, and
at
higher doping 
a second QPT 
appears to occur when the spin gap closes 
\cite{loram} (Fig.~\ref{invfig1x})
The singular interactions induced by quantum criticality 
are thought to 
the driving force for both the high temperature
superconductivity and 
the anomalous metallic state 
above the spin gap temperature $T^{*}$.\cite{castellani}

\fight=0.7\textwidth
\fg{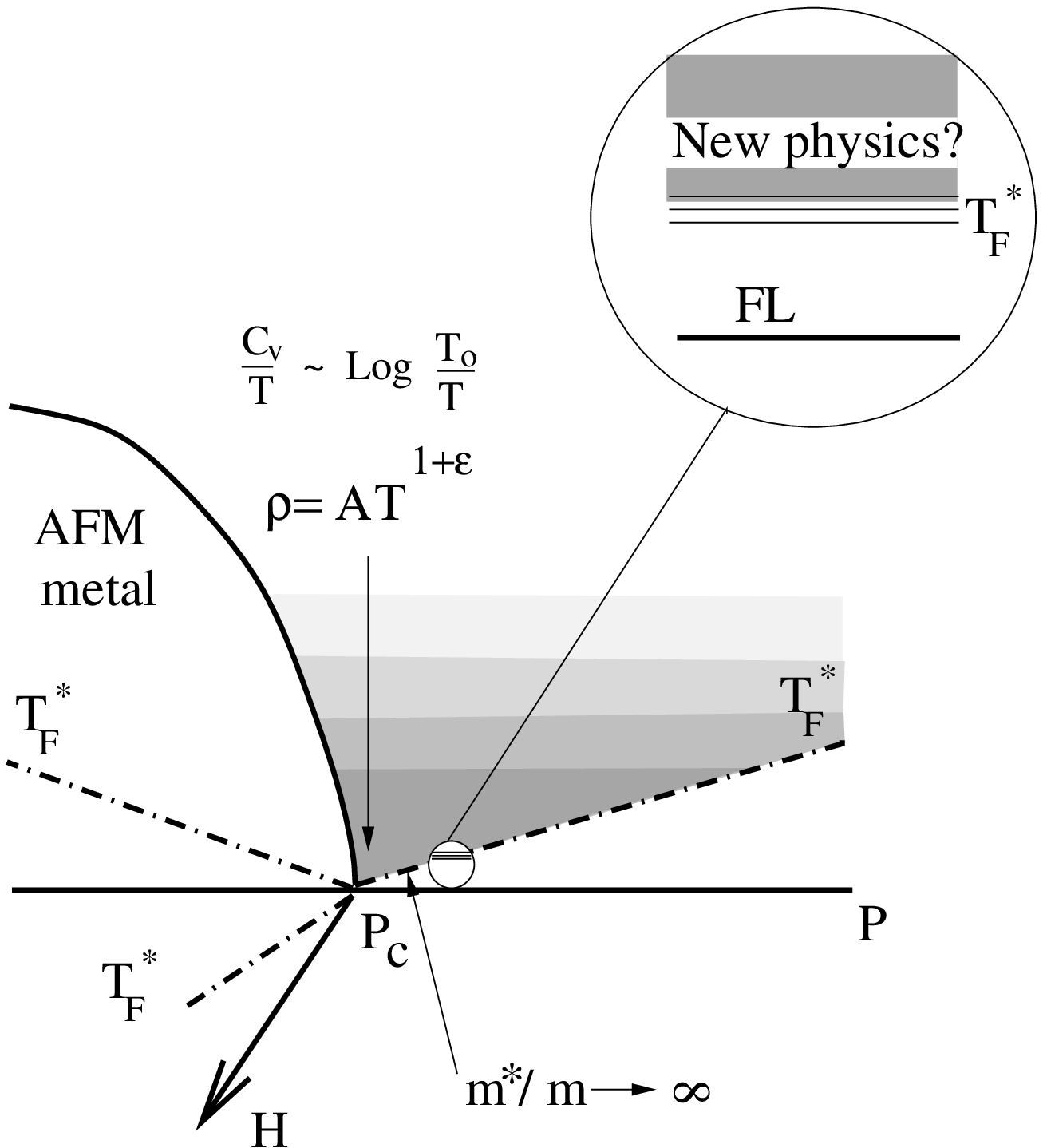}{Illustration of quantum critical physics
in heavy fermion metals. 
As criticality is 
approached from either side of the transition, the temperature 
scale $T_{F}^{*}$ on which Fermi liquid behavior breaks down goes to zero.
A key challenge is to characterize the 
new class of universal excitations which develops above 
$T_{F}^{*}$.
}{invfig1b}

Heavy Fermion materials
offer a unique opportunity to study quantum
criticality under controlled conditions.  By the application of
pressure,  doping and most recently, magnetic field, 
these materials can be tuned
through a quantum critical point from a metallic
antiferromagnet into a paramagnet (Fig.~\ref{invfig1b}). 
Unlike the cuprate metals, here 
the paramagnetic phase is a well characterized Fermi
liquid,\cite{landau,early,early2}   with heavy Landau
quasiparticles, or ``heavy electrons''.
A central property of these quasiparticles, is the existence of a
finite overlap ``$Z$'' between a
single quasiparticle state, denoted by $\vert \hbox{qp}^{-}\ra$ and
the state formed by adding a single electron to the ground-state,
denoted by $\vert e^{- }\ra = c\dg _{{\bf k }\sigma}\vert 0\ra$.
This quantity is closely related
to the ratio $m/m^{*}$ of the electron to quasiparticle mass,  
\begin{equation}\label{}
Z= \vert \la e^{- }\vert \hbox{qp}^{-}\ra \vert ^{2}\sim \frac{m}{m^{*}}.
\end{equation} 
A wide body of evidence suggests that $m^{*}/m$
diverges at a heavy fermion QCP, indicating
that 
\[
Z\rarrow 0 \qquad \qquad  (\hbox{$P\rarrow P_{c}$}).
\]
The state which forms above the QCP 
is referred to as a ``non-Fermi'' or ``singular Fermi
liquid''. \cite{continentino,varma2001,questions,questions2,stewart2001}
By what mechanism does this 
break-down in the Landau quasiparticle occur? 

\vskip 0.2truein
\noindent 
\vspace{0.1 truein}
\input{how.table}

\vspace{0.1 truein}
\noindent{\small $^{*}$ New  data\cite{newgegen,gegenwart} show a stronger
divergence at lower temperatures. 
}\\


\subsection{Properties of the Heavy Fermion Quantum Critical Point}

There is a  growing list of heavy fermion systems that
have been tuned to an antiferromagnetic QCP
by the application of pressure or by doping (Table 1.). 
These materials display many common properties

\begin{itemize}

\item {\bf Fermi liquid behavior in the paramagnet}, as indicated by
the emergence of a quadratic temperature dependence in the resistivity
in the approach to the QPT $\rho =\rho_{o}+A T^{2}$
\cite{devisser2,flouquet} at ever lower temperatures.

\item {\bf  Divergent A coefficient in resistivity} at the QCP.
In a typical Fermi
liquid the $A$ coefficient in the resistivity is proportional
to $\left( \frac{1}{1/T_{F}^{*}}\right)^{2}\sim \left(\frac{m^{*}}{m} \right)^{2}
$, where $T_{F}^{*}$ is the Fermi temperature. 
Support for the divergence of the effective mass 
is provided by the observation that
the quadratic coefficient $A$ of  the resistivity grows, and
apparantly diverges at the quantum critical point\cite{devisser}.

\item 
{\bf  Divergent  specific heat} 
at the QCP, with 
an asymptotic  logarithmic temperature dependence, 
\begin{eqnarray}\label{}
\gamma (T) = \frac{C_{v} (T)}{R T} = \frac{ Q}{T_{o}} \log
\left[\frac{T_{o}}{T} \right]+\gamma_{n},
\end{eqnarray}
where $R$ is the gas constant, and experimentally $
{ Q} \approx 0.4$, 
suggesting that the Fermi temperature vanishes
and the quasiparticle effective masses diverge 
\begin{equation}\label{}
T_{F}^{*}\rarrow 0, \qquad \frac{m^{*}}{m}\rarrow \infty 
\end{equation}
at the QCP.  The above expression has been written in a form
where the characteristic energy $T_{o}$ enters both inside the
logarithm and in the prefactor. There are a number of materials
where this one-parameter form holds, with $\gamma_{n}=0$, suggesting a new
kind of universality where no normal component to the Fermi surface
survives at the QCP. \cite{sereni}

\item {\bf  Quasi-linear resistivity}
\begin{equation}\label{}
\rho \propto T^{1+\epsilon},
\end{equation}
at the QCP with
$\epsilon$ in the range $ 0-0.6$.
In critical $YbRh_{2}Si_{2-x}Ge_{x}$, $\rho \propto T$
over three decades\cite{steglich00}.

\item {\bf  Non-Curie spin susceptibilities}
\begin{equation}\label{}
\chi ^{-1} (T)= \chi _{0}^{-1}
+ c T^{a} 
\end{equation}
with $a<1$ observed in critical 
$Ce Cu_{6-x}Au_{x}$ (x=0.1), $YbRh_{2} Si_{2-x}Ge_{x}$ (x=0.1)
and $CeNi_{2}Ge_{2}$.

\item {\bf $E/T$ and $H/T$ Scaling.} 
In critical $Ce Cu_{6-x}Au_{x}$ and $YbRh_{2} Si_{2-x}Ge_{x}$ the differential magnetic
susceptibility
$dM/dH$
exhibits $H/T$ scaling, 
\begin{equation}\label{}
(dM/dH)^{-1} = \chi _{0}^{-1}
+ c T^{a} g[H/T], 
\end{equation}
where $a\approx 0.75$.
Neutron measurements\cite{schroeder}
show $E/T$ scaling\cite{varma1989,aronson} in 
the dynamical spin susceptiblity of critical $Ce Cu_{6-x}Au_{x}$, 
throughout the Brillouin zone, parameterized in the form
\begin{equation}\label{lab1}
\chi^{-1} ({\bf q},\omega ) =  T^{a}f (E/T)+\chi _{0}^{-1} ({\bf q})
\end{equation}
$F[x]\propto (1-ix)^{a}$. 
Scaling behavior
with a single anomalous exponent in 
the momentum-independent
component of the dynamical spin susceptibility 
suggests an emergence of {\sl local} magnetic moments  which 
are {\sl critically correlated in time} at the quantum critical
point\cite{schroeder}. 
\end{itemize}
 
\subsection{Universality}

Usually, the physics of a metal above its
Fermi temperature depends on the 
detailed chemistry and
band-structure of the material: it is non-universal. 
However, if the renormalized Fermi temperature $T^{*}_{F}
(P)$ can be tuned to become arbitrarily small compared with the 
characteristic scales of the material as one approaches a QCP, 
we expect that 
the ``high energy'' physics
{\sl above} the Fermi temperature $T^{*}_{F}$ is itself, 
\underline{universal}. 

Quantum critical behavior implies a divergence of
the long distance and long-time correlations in the material. 
Finite temperatures introduce 
the cutoff timescale
\begin{equation}\label{}
\tau_{T} = \frac{\hbar }{k_{B}T}
\end{equation}
beyond which coherent quantum processes  are dephased by thermal
fluctuations. Renormalization group principles\cite{hertz}
imply that the quantum critical physics has an 
upper-critical dimension $d_{u}$. For $d<d_{u}$, $\tau _{T}$
becomes \underline{the }
correlation time $\tau $ of the system\cite{zinnjustin}, so frequency dependent
correlation functions and response functions take the form 
\begin{equation}\label{}
F (\omega,T)= \frac{1}{\omega^{\alpha }} f (\omega \tau_{T} )= 
\frac{1}{\omega^{\alpha }} f (\hbar \omega /k_{B}T).
\end{equation}
leading to 
$E/T $
scaling\cite{sachdev}.
By contrast, for $d>d_{u}$ the correlation
time is sensitive to the details of the short-distance interactions
between the critical modes, and in general  $\tau ^{-1}\propto T^{1+b}$, ($b>0$).
Thus $E/T$ scaling with a non-trivial exponent strongly suggests that 
the underlying physics of the heavy fermion quantum critical point is 
governed by universal physics with $d_{u}> 3$.

%
%
%
%
%

\subsection{Failure of the Spin Density Wave picture }

The standard model of the heavy fermion QCP
assumes the non-Fermi liquid behavior derives from Bragg diffraction of
the electrons off a quantum-critical spin density wave
(QSDW)\cite{hertz,paladium,moriya,millis}. 
The virtual emission  of these soft fluctuations,
\begin{equation}\label{}
e^{-} \rightleftharpoons e^{-}+ \hbox{spin fluctuation}
\end{equation}
generates 
a retarded interaction 
\begin{equation}\label{}
V_{eff} ({\bf q},\omega )= g^{2}\overbrace {\left[ \frac{\chi _{0}}{({\bf q}-{\bf
Q})^{2}+\xi^{-2}-\frac{i\omega }{\Gamma_{\bf Q}}}\right]}^{{\chi ({\bf
q},\omega)} }
\end{equation}
between the electrons, 
where $\chi ({\bf q},\omega )$
is the dynamical 
spin susceptibility of the collective modes. 
The damping term $-i\omega/\Gamma_{{\bf  Q}}$
of the magnetic fluctuations is derived from
the linear density of particle-hole states in the Fermi sea.
$\xi^{-1}
$ and 
$\tau ^{-1}= \Gamma
_{{\bf Q}}\xi^{-2}$ are the inverse spin correlation length and
correlation times respectively. 
In real space, 
\begin{equation}\label{}
V_{eff} (r,\omega =0)\propto \frac{e^{-r/\xi}}{r}e^{i {\bf Q}\cdot {\bf r}}
\end{equation}
is a  ``modulated ''
Yukawa potential whose range $\xi\sim
(P-P_{c})^{-\frac{1}{2}}
\rightarrow\infty  $ at the QCP. 
Unlike a ferromagnetic QCP, 
the modulated potential only  affects electron quasiparticles along 
``hot lines'' on the Fermi surface, 
that are separated by the wave-vector $\bf Q$ and satisfy
$\epsilon_{\bf k}= \epsilon_{{\bf  k}+{\bf  Q} }$. At a finite
temperature, electrons within a momentum range  $\sim \sqrt{T}$
are affected by this critical scattering (Fig.~\ref{invfig4}.). This limits
the ability of this singular potential to generate non Fermi liquid
behavior. 
\fight=0.6\textwidth
\fg{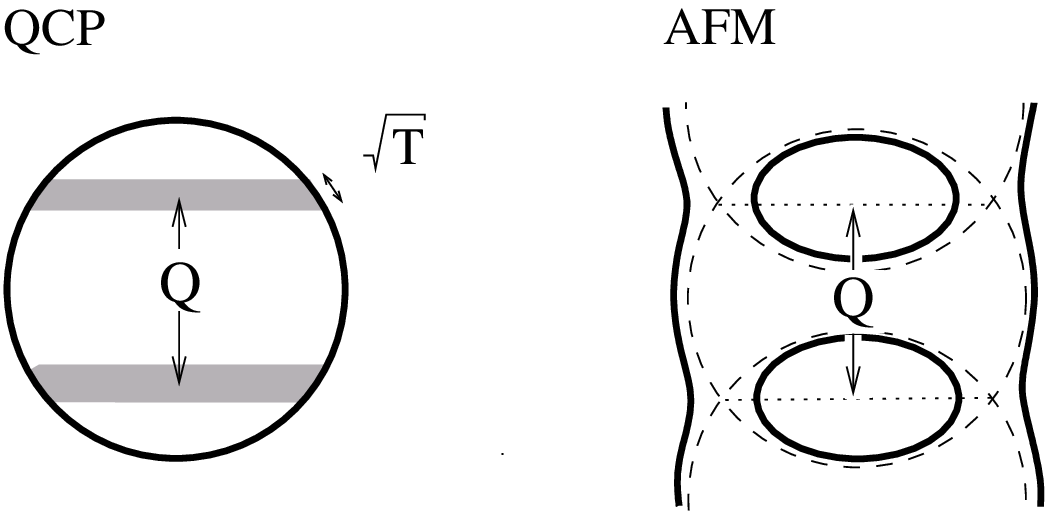}{
Quantum spin density wave
scenario, where the Fermi surface ``folds'' 
along lines separated by
the
magnetic 
$Q$ vector, pinching off into two separate Fermi surface sheets.
}{invfig4}
\noindent There are then two major difficulties with the (three dimensional) QSDW scenario 
for the heavy fermion QCP: 
\begin{enumerate}

\item {\bf  No breakdown of the Fermi liquid} Away from the hot lines, the Fermi
surface and Landau quasiparticles remain intact at the
QCP.  Thus the specific heat and typical quasiparticle mass do not diverge
but exhibit a weaker singularity, $C_{V}/T= \gamma_{o}
-A\sqrt{T}$ in the QSDW picture\cite{millis}.

\item {\bf  No $E/T$ scaling } The quantum critical behavior predicted by this model has been extensively
studied\cite{hertz,millis}.
In the 
interaction $V_{eff} ({\bf q},\omega)$
the momentum dependence enters
with twice the power of the frequency, so
\[
\tau\sim\xi^z,\qquad (z=2).
\]
In the renormalization group (RG) treatment\cite{hertz} 
time counts as $z$ space dimensions so 
the effective dimensionality is
$D_{eff}=d + z = d+2$. The upper critical dimension is set by
$D_{eff}=4$, or $d_{u}=2$\cite{millis}, so 3D
quantum spin fluctuations will not lead to $E/T$ scaling. 
In three dimensions, QSDW theory predicts that the scale
entering into the energy dependent response functions should scale
as $T^{3/2}$, with a non-universal prefactor\cite{sachdevbook}. 
\end{enumerate}


\subsection{Towards a new understanding. }\label{}

In this last lecture, I would like to give you a sense of the 
seriousness of the failure of the spin density wave scenario and 
share with you some of the new ideas that are circulating.
Some have argued that it may be possible to explain the 
of $E/T$ scaling and the logarithmically divergent 
specific heat\cite{rosch} by supposing  that the spin fluctuations 
form a \underline{quasi-two-dimensional} 
spin fluid\cite{mathur,rosch}, lying at the critical dimension. 
Inelastic neutron scattering experiments on $CeCu_{6-x}Au_{x}$,
(x=0.1) 
support a kind of reduced
dimensionality in which the 
critical scattering is concentrated
along linear, rather than at point-like regions in reciprocal
space\cite{schroeder,rosch}. More recent data\cite{fak}
may support quasi-2D spin fluctuations at intermediate scales in
$CeGe_{2}Ni_{2}$.

The assumption that the spin fluid is two dimensional is hard 
to reconcile with the fact that the developing order
is fully three dimensional, and with the fact that these systems
exhibit very little dimensional anisotropy.  
Even if we accept these
problems, there other difficulties. 
First- quasi-two dimensionality can furnish $E/T$ scaling, but 
it does not drop the theory below its critical dimension, and hence
has no way of accounting for the anomalous exponents in the $E/T$
scaling. 

Finally, there is 
another more serious
difficulty. It has recently become possible to examine 
the approach to the heavy electron quantum critical point 
through the use of field tuning\cite{stewart,stewartupt3,gegen2002}. 
The material 
$YbRh_{2}Si_{2-x}Ge_{x}$ with $x=0.1$ lies precisely at a quantum
critical point. By applying a small magnetic field, this system is
driven back into the Fermi liquid. As the field is reduced and the system
is tuned back towards the quantum critical point, the $A$ coefficient of the
resistivity is observed to diverge as
\[
A\propto \frac{1}{B}
\]
Such behavior can be obtained in a two dimensional spin fluid
model in which the inverse squared correlation length is assumed
to be proportional to $B$, $\xi^{-2}\propto B$.  The same model
predicts a weak dependence of the linear specific heat on magnetic
field
\[
\gamma_{th}\propto Log (1/B)
\]
so that the ratio
\[
\frac{A_{th}}{\gamma_{th}^{2}}\sim \frac{1}{B Log (1/B)}.
\]
The same 
experiments also show
that the  linear specific heat diverges much more rapidly with $B$, as 
$\gamma \propto
\frac{1}{\sqrt{B}}$, so that the Kadowaki Woods ratio
\begin{equation}\label{}
A/\gamma ^{2}=constant.
\end{equation}

It is difficult to understate the importance of this new result.
The constancy of the Kadowaki Woods ratio over more than a decade
in $\gamma$ indicates that the momentum dependence of the 
scattering amplitudes in the Fermi liquid are \underline{not}
radically affected by the magnetic field, as they would be
if the chief mechanism for the mass renormalization were derived from
the exchange of soft magnetic fluctuations in  a 2D spin fluid. 
These new results
can only be understood if, 	
in the approach to the quantum critical point, the Fermi liquid 
scattering amplitudes remain local, depending only on
the size of the renormalized Fermi temperature $T_{F}^{*}$\cite{coleman87}.
\bxwidth=1.2in\upit = -0.4in
\begin{eqnarray}\label{}
\cr
\raiser{\frm{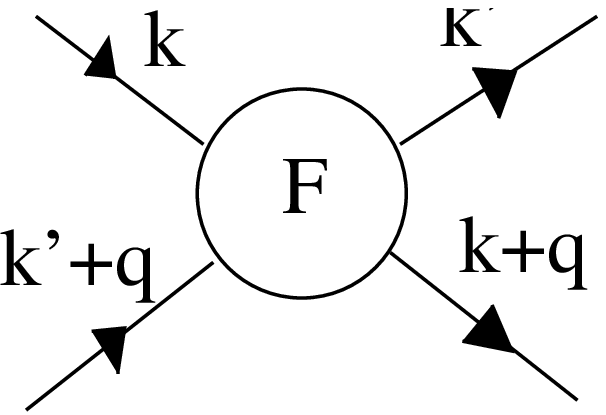}}= F_{k,k',k-q}=T_{F}^{*} {\cal F} (
\frac{E_{\vec{k}}}{T_{F}^{*}},
\frac{E_{\vec{k}'}}{T_{F}^{*}},
\frac{E_{\vec{k}-\vec{q}}}{T_{F}^{*}};\vec{k},\vec{k}',\vec{q}
)\cr
\end{eqnarray}
In the previous chapter, I argued that the effective Fermi temperature
of the Kondo lattice measures the ``phase stiffness'' associated with
the amplitude to form of a composite heavy electron, so that 
\[
\rho _{\phi }\sim {T_{F}^{*}}.
\]
The constancy of the Kadowaki Woods ratio, the lock-step divergence
of both $A$ and $\gamma^{2}$, and the appearance  of local features in the
spin correlations at the quantum critical point are all in keeping with
the idea that the Kondo bound-state phase stiffness
\underline{is going to zero} on the paramagnetic side of the heavy electron
quantum critical point, just as the spin-wave stiffness $\rho _{M}$
goes to zero on the magnetic side of the same point. In other words
\begin{eqnarray}\label{}
\left. 
\begin{array}{rcl}
\rho _{\phi }&\longrightarrow& 0, \cr
\rho _{M}&\longrightarrow&0
\end{array}
\right\}\hbox{\underline{quantum bicriticality?} }
\end{eqnarray}
In other words, 
the Kondo composite bound-state appears to die at exactly the same time
magnetic order develops. This strongly suggests to me, that perhaps
the heavy electron quantum critical point might be better understood
as a quantum \underline{bicritical} point, where two order parameters
go to zero at a point. 

Traditionally, theories of phase transition are built apon an underlying
mean-field theory.  The spin density wave scenario is a consequence of
examining fluctuations about the Stoner and Slater view of itinerant
magnetism.  If this approach fails, then perhaps it is a sign that we
should search for a new kind of mean-field theory to describe the
quantum phase transition between antiferromagnetism and the heavy
electron fluid.  
There are two kinds of suggestion that have 
been considered recently :
\begin{itemize}

\item Local quantum criticality.  The apparent momentum independence
of the localized critical correlations at the quantum critical
point\cite{schroeder} has led to the suggestion that the the correct
mean-field theory, is one that is local, yet fully
dynamical.\cite{xsach,sengupta,siandsmith} Such ``dynamical
mean-field theories'',\cite{krauth} are thought to asymptotically
exact in infinite dimensions.  In
this philosophy, the local physics remains strongly interacting even
in infinite dimensions, but the local character of the interactions is
supposed to be stable against finite dimensionality. This idea forms
the basis of a recent theory by Si et al.

\item Traditional RG approach on a new Lagrangian. 
Rather than abandon the  traditional RG approach first suggested by
Hertz,\cite{hertz}  we should continue to
embrace the notion that a Wilsonian approach, where 
interactions become weak in high enough dimensions does work
for quantum critical points. This approach argues that what is needed,
is a new description of magnetism, and the way it couples  to the Fermi
sea. One idea here, is that at the quantum critical point, the heavy
electron breaks up into its spin and charge components.\cite{questions2}   

\end{itemize}

We now discuss these ideas in more depth. 
\subsection{Local Quantum Criticality}

The momentum-independent scaling term in the inverse dynamic
susceptibility (7)
suggests that the critical behavior associated with the heavy fermion QCP
contains some kind of {\sl local} critical
excitation\cite{schroeder}.  
One possibility, 
is that this 
critical excitation is the spin itself, which would then presumably
develop 
a slow power-law decay\cite{xsach,sengupta,siandsmith}
\begin{equation}\label{}
\langle S (\tau )S (\tau ')\rangle =\frac{1}{(\tau -\tau ')^{2-\epsilon
}},
\end{equation}
where $\epsilon \ne 0$ signals non- Fermi liquid behavior. 

\fgb{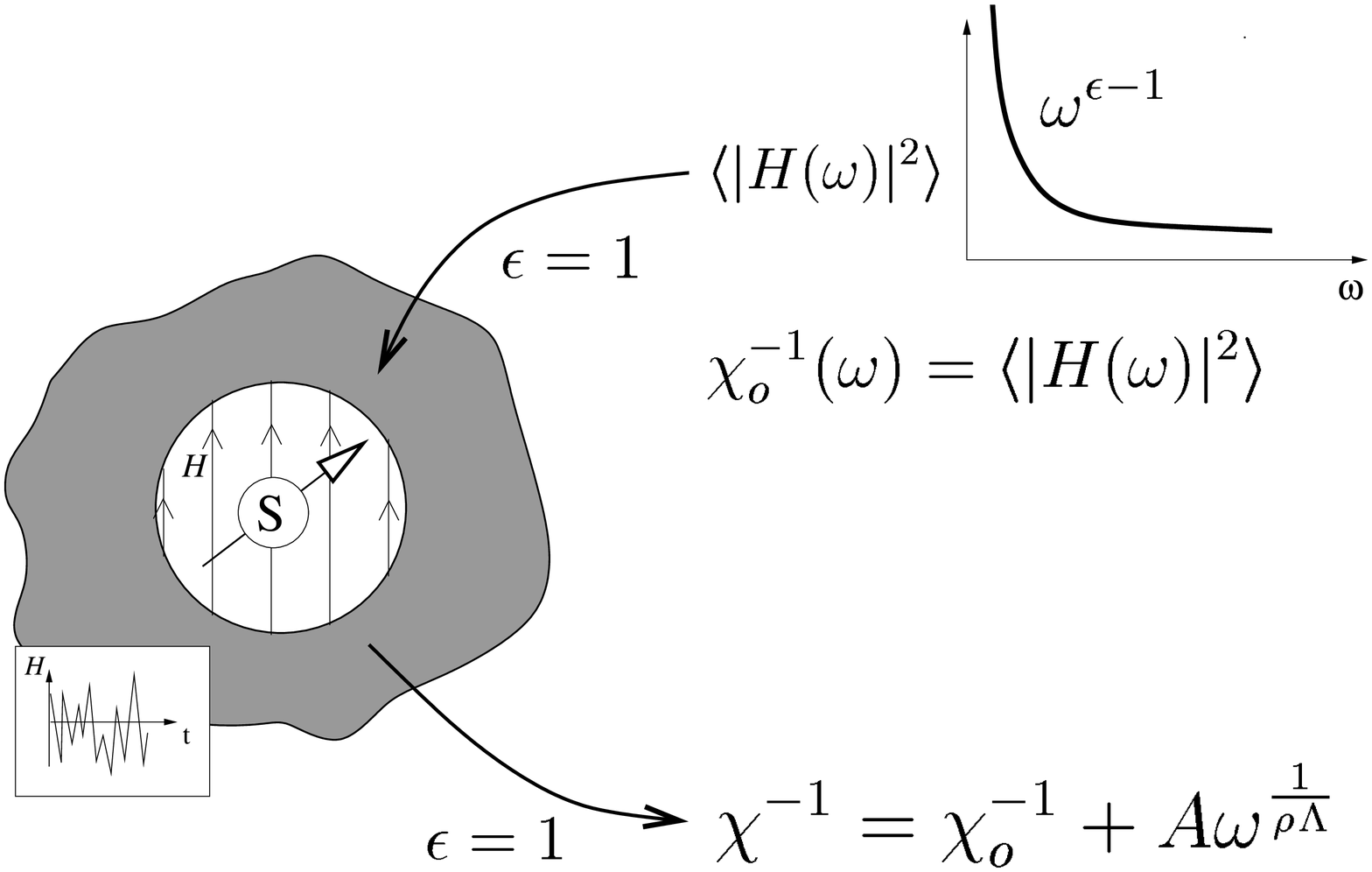}
{In the local quantum
critical theory, each spin behaves as a local moment in a fluctuating
Weiss field. In the theory of Si et al [102], a self-consistent solution
can be obtained for $\epsilon=1$ in which the local susceptibility
develops a self-energy with a non-universal exponent. 
$ M (\omega)\propto \omega ^{\frac{1}{\rho \Lambda}}$.}{invfig5}

Si\cite{si} et al. have extensively developed this idea, proposing 
that the
{\it local } spin susceptibility
$\chi_{loc}=\sum_{\vec{q}}\chi (\vec{q},\omega)\vert_{ \omega=0}$
diverges at a heavy fermion QCP. 
From (\ref{lab1}), 
\begin{equation}\label{}
\chi _{loc} (T)\sim \int d^{d }q \frac{1}{({\bf q}-{\bf  Q})^{2} +
T ^{\alpha
}}\sim T^{(d-2)\alpha /2}
\end{equation}
so a divergent local spin susceptibilty requires 
a spin fluid with $d\leq 2$.  Si et al are thus motivated 
to propose that the 
non-trivial physics of the heavy fermion QCP is driven by
the formation of a two-dimensional spin fluid.  
Si et al consider an impurity spin within an effective medium
in which the local Weiss field $H$ has 
a critical power-spectrum (Fig. \ref{invfig5}.)
\begin{equation}\label{}
\langle \vert H (\omega)\vert ^{2}\rangle \equiv \chi _{0}^{-1}
(\omega)= \omega^{\gamma}
\end{equation}
where $\epsilon$ is self-consistently evaluated using
a dynamical mean-field theory, where  $q-$
dependence of self-energies is dropped. In principle, the method
solves the dynamical spin
susceptibility of the impurity 
$\chi^{-1} 
(\omega)=\chi _{o}^{-1} (\omega)+M (\omega)$. This, in turn
furnishes a ``spin
self-energy'' $M (\omega)$ used to determine 
the spin susceptibility of the medium 
$\chi^{-1} (\vec{q},\omega)= J ( \vec{q})+ M(\omega)$.

Si et al find that a self-consistent solution is obtained for 
$\epsilon=1$,  {\sl if} the spin-self energy 
contains a separate power-law
dependence $M(\omega)\sim \omega^{\alpha }$ 
with an exponent $\alpha = 1/\rho \Lambda$ 
which is determined by the density of states $\rho$
and band-width $\Lambda$ of the bond-strengths in the
two-dimensional spin fluid.
Although self-consistency requires a new power-law in the spin 
susceptibility, 
independent solutions
of the impurity model have not yet shown 
that this feature is indeed  generated by a critical Weiss field. 
This theory nevertheless 
raises many interesting questions:

\begin{enumerate}
\item Is the requirement of a two dimensional spin fluid 
consistent with the ultimate emergence of three dimensional
magnetic order.  For example- does the
the cubic (and hence manifestly three dimensional) quantum critical
material, $CeIn_{3}$ display a divergent specific heat?

\item If the spin-fluids are quasi-two dimensional, do we expect an
ultimate cross-over to a
three-dimensional QSDW scenario? 

\item If $\alpha $ is non-universal, why are the 
critical exponents in $CeCu_{6-x}Au_{x}$
and $YbRh_{2}Si_{2-x}Ge_{x}$ so similar?

\item 
What stabilizes the local quantum criticality against intersite couplings?

\end{enumerate}

\subsection{Ideas of spin charge separation and supersymmetry.}


An alternative possibility, is that  the 
heavy fermion QCP is a truly three-dimensional phenomenon. In this  
case a different approach is needed- we need to search for a new class of
critical Lagrangian with $d_{u}>3$\cite{susy}. 
On general grounds, the existence of a Fermi liquid  in the paramagnetic phase
suggests that the new class of critical Lagrangians must 
find expression in terms of 
the quasiparticle fields $\psi $ in the Fermi liquid- but how do we
couple
these degrees of freedom to the magnetism, and how do we account
for the simultaneous loss of the resonant bound-state stiffness
at the same time that magnetism develops? The simplest possibility is
to write
\begin{equation}\label{lag}
L = L_{F}[\psi ] + L_{F-M}[\psi,M] + L_{M}[M].
\end{equation}
where $L_{F}$ describes the 
heavy  Fermi liquid,  far from the magnetic
instability, 
$L_{M}$ describes the magnetic excitations that 
emerge above the energy scale $T_{F}^{*} (P)$. 

$L_{F-M}$ describes
the way that the quasiparticles couple to and decay into critical 
magnetic modes; it also 
determines the type of transformation which takes place in the 
Fermi surface which occurs
at the QCP. This last point follows because 
away from the QCP, magnetic fluctuations can be ignored in the
ground-state, so that 
$L_{M}\rarrow 0$. In the 
paramagnetic phase, $\langle M \rangle =0$ so $L_{FM}\rightarrow 0$, but in the
antiferromagnetic phase $\langle M \rangle \neq 0$, i.e. 
\[
{{L}}_{\hbox{eff}}= \left\{ 
\begin{array}{cl}
{L} _{F}^{*}[\psi ]&\ \hbox{paramagnet}\cr \cr
{L}_{F}^{*}[\psi ]+{L}_{FM}[\psi , \langle M\rangle ]
&\  \hbox{a.f.m.}\end{array} \right.
\]
where the asterisk denotes the finite renormalizations 
derived from zero-point fluctuations in the magnetization.  

If the staggered magnetization
is the fundamental critical field, then we 
are forced
to 
couple the magnetic modes
directly to the spin density 
of the 
Fermi liquid 
\begin{eqnarray}\label{weak}
L_{F-M}^{(1)}&=& g \sum_{{\bf k},{\bf q} } 
\psi \dg_{\vec{k}-\vec{q}} \vec{\sigma}\psi _{\vec{k}}\cdot \vec{M}_{{\bf q}}. 
\end{eqnarray}
But once the staggered magnetization condenses, this leads 
directly back to a static spin density wave (Fig \ref{invfig4}).

An alternative 
possibility is suggested by the observation that the magnetism
develops spinorial character in the heavy Fermi liquid. 
The
Luttinger sum rule\cite{luttinger} governing the Fermi surface volume $V_{FS}$
``
counts'' both the electron density $n_{e}$
\underline{and} the number of 
local moments per unit cell
$n_{{spins}}$\cite{martin82,oshikawa00}
: \begin{equation}\label{}
2\frac{{\cal V}_{FS}}{(2\pi)^{3}}= n_{e} + n_{spins}.
\end{equation}
The appearance of the spin density in the Luttinger sum rule
reflects the composite nature of the heavy quasiparticles, formed
from 
bound-states between local moments and high energy electron
states. 
Suppose the spinorial character of the magnetic  degrees of freedom seen in the
paramagnet {\sl also } manifests itself
in the decay modes of the heavy quasiparticles.  This would imply that
at the QCP, the staggered magnetization factorizes 
into a spinorial degree of freedom
$\vec{M} (x) = z\dg(x)\vec{\sigma }z (x) $, where $z$ is a 
two-component spin $1/2$ Bose field.  
``Spinorial magnetism''
affords a direct coupling between the magnetic spinor $z$
and the heavy electron quasi-particles via an inner product, 
over the spin indices
\begin{equation}\label{poss2}
L_{F-M}^{(2)} = g \sum _{{\bf k}, {\bf q}}[ 
\chi _{\bf q} \dg \left(z\dg _{{\bf k}-{\bf
q}\sigma}
\psi _{{\bf
k}\sigma }
 \right)
  +\hbox{H.c}], 
\end{equation}
where conservation of exchange statistics obliges us to 
introduce
a spinless charge $e$ fermion $\chi $. 
This 
would imply that the composite heavy electron  decays
into a neutral ``spinon''and a spinless charge e fermion 
$e^{-}_{\sigma }\rightleftharpoons s_{\sigma } + \chi ^{-}$.
The critical Lagrangian 
in this case would take the form
\begin{equation}\label{}
L=L_{F}[\psi ] + L_{F-M}[\psi ,\chi , z]+L_{M}[z,\psi ].
\end{equation}
We have to be cautious of course, because this is undoubtedly one of
many alternative ways we might begin to construct a new class of 
critical Lagrangians. What we do see quite clearly however, 
is this line of reasoning leads us to into the notion that 
the break-up of the heavy fermion QCP {\sl involves spin-charge separation}. 

Hall constant measurements may provide a good way
to discern between the spin density wave and composite quasiparticle
alternatives. In the former,
regions around the hot-line do not contribute
to the Hall conductivity, and the change in the Hall constant is 
expected to evolve as the staggered
magnetization\cite{questions2}.  By contrast, the composite fermion scenario leads
to a much more rapid evolution: provided that  the density
of spinless fermions is finite at the QCP the 
Hall constant will jump suddenly 
at the QCP \cite{questions}. 
\begin{equation}\label{hall}
\Delta R_{H}\propto  \left\{\begin{array}{cr
}
M_{\bf Q}, & \qquad \hbox{(vectorial )}
\cr
O (1)&(\hbox{spinorial})
\end{array} \right.
\end{equation}
The only available Hall measurement at a QCP to date shows a 
change in sign takes place in the close vicinity of the QCP
in critical $CeCu_{6-x} Au_{x}$, 
it is not yet clear whether there is a discontinuity 
at the transition\cite{onuki00}.
This is clearly an area where more experimental input is highly
desirable. 

Let me end with a few speculations.  If we are to construct a new
critical theory for the heavy electron quantum critical point, then
we will need  some new theoretical ideas.  A new critical theory
will require as a first step, a new kind of mean-field description
that permits us to understand why the magnetism  and the Kondo
effect die at a common critical point. At present,
we do not know how to construct a mean-field theory that contains
a heavy electron quantum critical point.  One interesting idea here, may
be the incorporation of \underline{supersymmetry}.  
This is an idea that I have tried to develop, as yet with only partial
success, with my graduate student
John Hopkinson and collaborators,
Catherine P{\'e}pin and Alexei Tsvelik.\cite{susy,susy2}
At a very naive level,  magnetism involves
the manifestation of spin as a bosonic excitation, whereas 
heavy electron behavior involves the manifestation of spin as a fermionic
object.   If the two phenomena share the same quantum critical point, then
is it possible that the spin manifests both types of behavior at a
quantum critical point- in otherwords, that it displays some kind of
supersymmetry? 
\fg{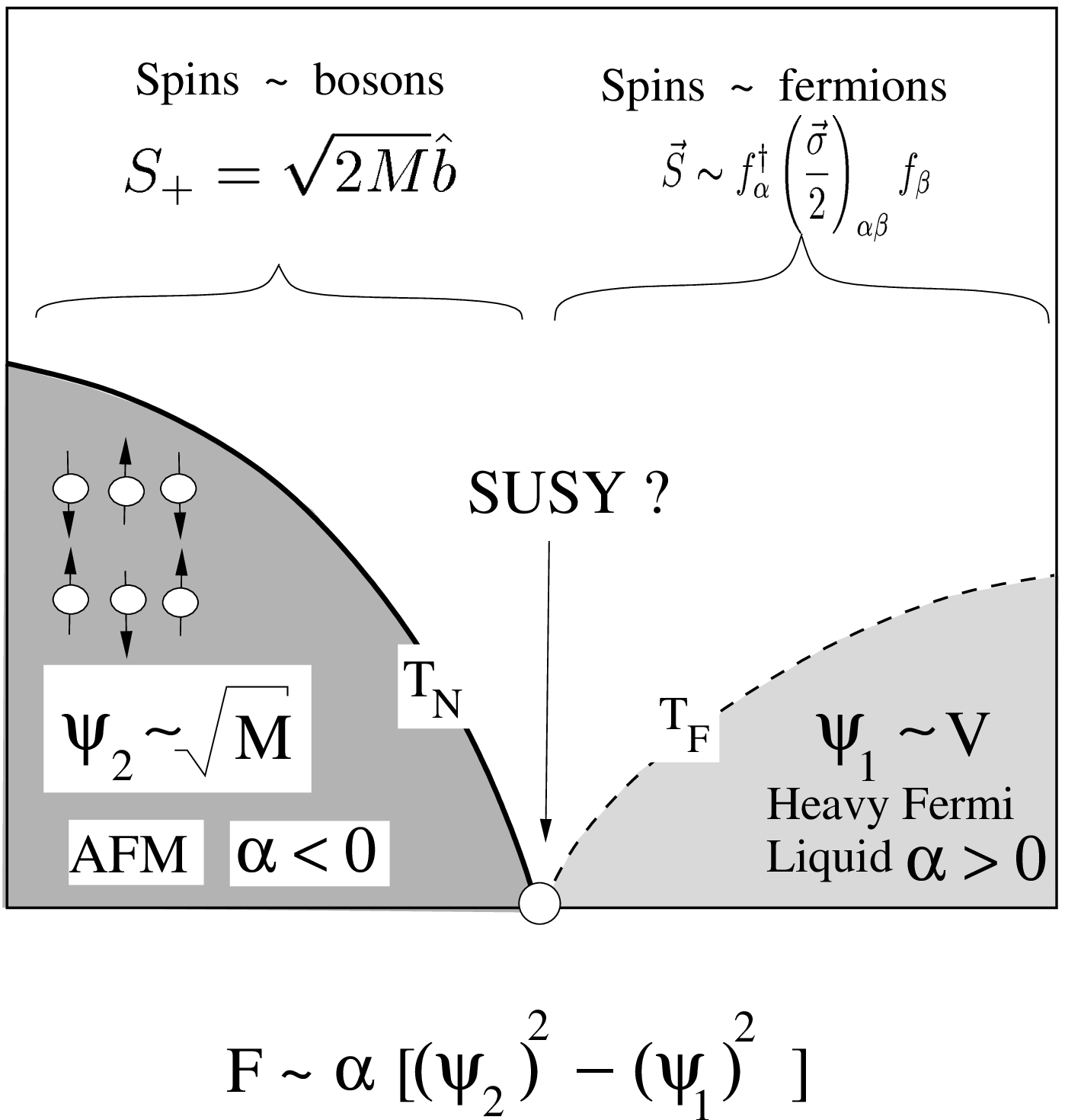}
{The heavy fermion QCP may involve a
supersymmetric gauge symmetry. 
 To understand the fact that
the magnetization ($M$) disappears at precisely the same point in the
phase diagram where the amplitude ($V$) for the formation of composite
fermions goes to zero, we need a special symmetry between the two
order parameters. Local moments behave as fermions in the
paramagnetic phase, but as bosons in the antiferromagnet. Does a
supersymmetry develop at the 
quantum critical point and is this responsible for the ``Minkowskii''
metric between the two order parameters that is required 
for them to vanish at the same point in the phase diagram?
}{susyfig}

It does prove possible to represent both spin and Hubbard operators
in a way that involves a locally supersymmetric gauge theory, but a
mean-field theory still eludes us at the current time. We do have an
idea about the structure of this mean-field theory, which I shall briefly
mention to you. In this putative mean-field theory, we require
\underline{two} order parameters- one for the formation of the
composite heavy electron corresponding to the amplitude for composite
fermion formation $\psi _{1}\sim V$ and one for the magnetism ($\psi
_{2}\sim \langle
z_{\sigma }\rangle \sim \sqrt{M}$). Suppose we can integrate out all of the
fermions in the theory, so that we are left with an effective theory for
$\psi _{1}$ and $\psi _{2}$, given by a Landau Ginzburg free energy 
$F[\psi _{1},\psi _{2}]$. Now here's the remarkable thing- for the two order
parameters to share a quantum critical point, then the expansion of
the Free energy near the QCP must take the form
\[
F\sim \alpha 
(\vert \psi _{2}\vert ^{2}-\vert \psi _{1}\vert ^{2}) 
+ \hbox{interactions}
\]
When $\alpha >0$, we have the heavy electron phase with $\psi _{2}=0,\qquad
\psi _{1}\neq 0$ but when $\alpha
<0$, we have the magnetic phase, where $\psi _{2}\neq 0, \qquad \psi
_{1}=0$.  At $\alpha =0$, both order parameters vanish simultaneously.
The two
must go to zero at the same point in the phase diagram, so they can
only
come together in the quadratic 
combination $(\vert \psi _{1}\vert ^{2}-\vert \psi _{2}\vert ^{2}) 
$.
This suggests that the negative definite metric 
\[
(\vert \psi _{2}\vert ^{2}-\vert \psi _{1}\vert ^{2}) 
\]
is a symmetry  invariant of the quantum critical point.
The appearance of a minus sign - a Minkowskii type metric- 
is required by the phenomenology, yet 
traditional invariant symmetry groups of a critical point involve a positive
metric associated with a trace over order parameter combinations.
The minus sign would occur if the residual symmetry 
between these two order parameters corresponded to symplectic group. 
One way for such minus signs might appear
is via the supertrace- an invariant
of a supergroup. This prompts the following conjecture, on which I
will end:  that the 
the above  invariance is  a residue 
of a supertrace in a supersymmetric Lagrangian
for quantum criticality.

\subsection{Summary}

This section has discussed the origin of the mass divergence
at the heavy fermion quantum critical point, emphasizing that a
quantum  spin density wave  picture can not explain the observed
properties. The proposal of fundamentally new kinds of
quantum critical points has been reviewed. This is clearly an area
with a huge potential for progress both on the experimental, and 
theoretical front. 


\subsection{Exercises }

\begin{problems}

\item  Consider the tree level scaling for the 
zero temperature Hertz-Millis Lagrangian at an
antiferromagnetic quantum critical point
\begin{equation}\label{}
S= S_{K}+S_{I}
\end{equation}
where
\begin{eqnarray}\label{}
S_{K}&=&\int^{\vert \omega \vert <\omega _{o},k<\Lambda 
}\frac{d\omega d^{d}k}{( 2\pi)^{d+1} }
\chi ^{-1} (\kappa )
M (\kappa )M (-\kappa )\cr
\chi ^{-1} (\kappa )&=& \left[(\vec{k}-\vec{Q})^{2}
+\frac{\vert \omega\vert ^{2/z}
}{\Gamma } \right]
\end{eqnarray}
describes the propagation of an overdamped spin fluctuation with
inverse susceptibility $\chi ^{-1} (\vec{k},\omega )$ at 
quantum criticality, with a critical wavevector $\vec{Q}$.  Here we have used the notation
$\kappa \equiv (\vec{k},\omega )$ to denote the wavevector and
frequency of a magnetization mode. 
The non-linear 
interaction term in the model takes the form
\begin{equation}\label{}
S_{I}[U]= U\int 
\left(\prod _{j=1,4}\frac{d^{(d+1)}\kappa _{j}}{( 2\pi)^{d+1} }
\right)M (\kappa _{1})\dots M (\kappa _{4})\delta ^{d+1}
(\sum_{j}\kappa _{j})
\end{equation}
\begin{enumerate}

\item Derive the tree level scaling that keeps the Kinetic term
$S_{K}$ invariant. First note that from the kinetic term, the scaling
dimension of frequency is $[\omega ]=[k]^{z}$. 
Show that if the wavevector and frequency cut-off are 
rescaled according to 
\begin{equation}\label{}
\tilde{\Lambda }= \frac{\Lambda }{b} ,\qquad 
\tilde{\omega _{0}}= 
\frac{\omega _{0}}{b^{z}}
\end{equation}
then one must rescale
\begin{equation}\label{}
k=\tilde{k}b,\qquad \omega =\tilde{\omega }b^{z}
\end{equation}
to keep the form of the Kinetic term invariant. Show that under this
scaling
\[
S_{K}\rightarrow  
b^{(z+d-2)}\int^{\vert \tilde{\omega }\vert <\omega _{o},\tilde{k}<\Lambda 
}\frac{d^{d+1}\tilde{\kappa }}{( 2\pi)^{d+1} }
\chi ^{-1} (\tilde{\kappa } )
M (\kappa )M (-\kappa )
\]
so that with scaling 
\[
M (\kappa )=\tilde{M} (\tilde{\kappa } )b^{- (z+d-2)/2}
\]
the kinetic energy remains invariant.

\item Using the tree level scaling derived above, show that the
interaction term transforms as
\[
S_{I} (U)\rightarrow S_{I} (U^{*})
\]
where
\[
U^{*}= \frac{U}{b^{(z+d-4)}}
\]
showing that the interaction term scales to zero for $d+z>4$, proving that
$d=4-z=2$ is the upper-critical dimension for the Hertz-Millis model.

\end{enumerate}

\end{problems}

\begin{theacknowledgments}
I should like to thank the students to whom I lectured this course,
for their patience and their wonderful questions, many of which helped
sharpen the notes. I am particularly indebted to Anna Posazhennikova,
for taking painstaking notes of the lectures.  The ideas on quantum
criticality in section four are largely the result of collaborative
work with theorists John Hopkinson, Catherine P{\'e}pin, Revaz
Ramazashvili, Qimiao Si and Alexei Tsvelik, and experimentalists
Gabriel Aeppli, Johann Custers, Philipp Gegenwart, Almut Schroeder,
Frank Steglich and Heribert Wilhelm.  Many people, particularly Natan
Andrei, Andrey Chubukov and Simon Kos deserve special thanks for
discussions related to the ideas in these notes.  Thanks to Adolfo
Avella and Anna Posazhennikova for proof reading the draft
version. This work was supported by the National Science Foundation
under grant DMR 9983156.
\end{theacknowledgments}


\bibliographystyle{aipproc}   


\IfFileExists{\jobname.bbl}{}
 {\typeout{}
  \typeout{******************************************}
  \typeout{** Please run "bibtex \jobname" to optain}
  \typeout{** the bibliography and then re-run LaTeX}
  \typeout{** twice to fix the references!}
  \typeout{******************************************}
  \typeout{}
 }
\label{}

\end{document}

%% file: parallels.table
\begin{table}

\caption{\bf Parallels between Superconductivity and the Kondo effect .}
{\small\begin{tabular}{|l||c|c|}
\hline
&&\\
& Superconductivity & Kondo effect\\
&&\\
\hline
\hline
&&\\
Bound State & $\psi_{\uparrow}(x)\psi _{\downarrow}(x')=F(x-x')$&
$\bigl({\vec \sigma}_{\alpha \beta} \cdot {\vec S} (t')\bigr)\psi_{\beta}(t)=
\Delta (t-t')f_{\alpha }(t')$\\
&&\\
& Bosonic& Fermionic\\
&&\\
\hline
&&\\
Characteristic energy&$T_{c}=\omega _{D}e^{-1/g\rho }$&
$T_{K}=D\sqrt{J\rho }e^{-1/J\rho }$\\
&&\\
Energy range contributing&$E\in [T_{c},\omega _{D}]$&$E\in 
[T_{K} , D]$\\
to bound state
&&\\
&&\\
\hline
&&\\
Extended in & space& time\\
& $\xi \sim v_{F}/T_{c}$& $\tau \sim \hbar /T_{K}$\\
&&\\
\hline
\hline
&&\\
Conserved Quantity&Total electron charge& 
Charge of local moment\\
&&\\
\hline
&&\\
Long Range Order & LRO $d>2$ & Powerlaw in time\\
&Powerlaw in space $ d \le 2$&\\
&&\\
\hline
&&\\
Phase stiffness & $\rho _{s}$ & $\rho _{\phi }$ \\
&&\\
\hline
&&\\
Consequences of & Meissner effect & Formation of charged\\
Phase stiffness & 
& heavy electron \\
(Anderson- Higgs)&& $2\frac{\Delta V_{F}}{(2\pi )^{3}}=n_{e}+n_{spins}$\\
&&\\
Quantity related &$\frac{1}{\lambda _{L}^{2}}\propto \rho _{s}$&
$
\frac{1}{U^{*}}=\rho _{\phi }$
\\
to phase stiffness &&\\
\hline
\end{tabular}
}
\noindent 
\end{table}

%% file: how.table
{\bf Table. 1. Selected Heavy Fermion compounds with quantum critical points.}
{\small
\vskip 0.1 truein
\begin{tabular}{|c||c|c|c|c|}
\hline
Compound & 
$x_c$/$P_c$/$H_c$ & $\frac{C_v}{T}\rarrow \infty $? & $\rho\sim T^a $ & Ref.\\
\hline
$YbRh_2Si_{2-x}Ge_x$ & $x_c=0.1$ &
$Log \left( \frac{T_o}{T}\right)$
$^{*}$ & $ T$ & \cite{steglich} \\
\hline
$CeCu_{6-x}Au_x $ & $x_c=0.1$ & $Log \left( \frac{T_o}{T}\right)$ & $ T + c $ & \cite{hilbert1,hilbert2,schroeder} \\
\hline
$CeCu_{6-x}Ag_x$ & $x_c=0.09$ & $Log \left( \frac{T_o}{T}\right)$ & $T^{1.1}$ & \cite{heuser}\\
\hline
\hline
$CeNi_2Ge_2$ & $P_c =0$  & $Log \left( \frac{T_o}{T}\right)$$^{*}$
& $T^{1.2}$ & \cite{grosche,gegenwart,julian,braithwaite,newgegen} \\
\hline
$U_2Pt_2In$ & $P_c=0$  & $Log \left( \frac{T_o}{T}\right)$
 & $T$ & \cite{devisser}\\
\hline
$U_2Pd_2 In$ & $P_c<0 $ &  ? & $T + c $ & \cite{devisser} \\
\hline
$CePd_2Si_2$ & $ P_c>0$  & ? & $ T^{1.2}$ & \cite{grosche} \\
\hline
$CeCoIn_5$ & $P_c\sim 1.6GPa$  & ?
 & $T$ & \cite{sarrao}\\
\hline
$CeIn_3$ & $P_c>0$ & ? & $T^{1.5} $ & \cite{grosche,mathur} \\
\hline
$U_3Ni_3Sn_4$ & $P_c >0$ & no & ? & \cite{devisser2} \\
\hline
$CeCu_2Si_2$ & $P_c=0$ & no & $T^{1.5}$ & \cite{cecu2si2}\\
\hline
\hline
$CeRu_2Si_2$ & $H^{\parallel}_c=7.7T$ &  $Log \left( \frac{T_o}{T}\right)$
 &  & \cite{aoki2}\\
\hline
$UPt_3$ & $H^{\perp}_c=20T$ &  $Log \left( \frac{T_o}{T}\right)$
 & $T^{1.2}$ & \cite{stewartupt3}\\
\hline
$Sr_3Ru_2O_7$ & $H^{\parallel}_c=7.7T$ &  $Log \left( \frac{T_o}{T}\right)$
 & $T^{1.1}$ & \cite{mackenzie}\\
\hline
\hline
\end{tabular}
}